%% file: TASI.tex
\documentclass[12pt]{article}
\pdfoutput=1
\usepackage[nosort]{cite}
\usepackage[height=8.85in,width=6.45in]{geometry}
\usepackage{pifont}

\usepackage{times}
 \newcommand{\CD}{{\cal D}}
  \newcommand{\CC}{{\cal C}}
  \newcommand{\mD}{\mathsf{D}}
  \newcommand{\CH}{{\cal H}}
  \newcommand{\HNS}{{\cal H}_\text{NSNS}}
    \newcommand{\HR}{{\cal H}_\text{RR}}
  \newcommand{\CN}{{\cal N}}
    \newcommand{\CL}{{\cal L}}
    \def\ee{\mathrm{e}}
\def\mm{\mathrm{m}}
\def\bZ{\mathbb{Z}}

\usepackage{booktabs}
\usepackage[utf8]{inputenc}
\usepackage{amsmath}
\usepackage{amssymb}
\usepackage{mathtools}
\numberwithin{equation}{section}
\usepackage{slashed}
\usepackage{braket}
\usepackage{url}
\usepackage[colorlinks,citecolor=teal,linkcolor=purple,linktocpage,pagebackref]{hyperref}
\usepackage{cite}
\usepackage{graphicx}
\usepackage{tikz}
\usepackage{tikz-cd}
\usepackage{tikzit}
\input{symbound.tikzstyles}

\usepackage{courier}
\usepackage{bm}

\usepackage{dashbox}
\usepackage{subcaption}
\usepackage{enumitem}

\usepackage{mdframed}

\usepackage{blkarray}
\usepackage{arydshln}
\usepackage{dsfont}
\usetikzlibrary{patterns}
\usetikzlibrary{arrows.meta}

\usepackage{calc}

\usepackage{accents}

\newcommand{\ie}{\begin{equation}\begin{aligned}}
\newcommand{\fe}{\end{aligned}\end{equation}}

\renewcommand{\title}[1]{\vbox{\center\LARGE{#1}}\vspace{5mm}}
\renewcommand{\author}[1]{\vbox{\center#1}\vspace{5mm}}
\newcommand{\address}[1]{\vbox{\center\em#1}}

\makeatletter
\newsavebox{\@brx}
\newcommand{\llangle}[1][]{\savebox{\@brx}{\(\m@th{#1\langle}\)}
  \mathopen{\copy\@brx\kern-0.5\wd\@brx\usebox{\@brx}}}
\newcommand{\rrangle}[1][]{\savebox{\@brx}{\(\m@th{#1\rangle}\)}
  \mathclose{\copy\@brx\kern-0.5\wd\@brx\usebox{\@brx}}}
\makeatother

\usetikzlibrary{decorations.markings}
\tikzset{middlearrow/.style={
	decoration={markings,
		mark=at position 0.55 with {\arrow[>=stealth]{#1}}},
	postaction={decorate}
	}
}

\tikzset{line/.style={line width=0.5mm},
curve/.style={line,smooth,tension=1},
->-/.style={decoration={
  markings,
  mark=at position #1 with {\arrow[>=stealth]{>}}},postaction={decorate}},
-<-/.style={decoration={
  markings,
  mark=at position #1 with {\arrow[>=stealth]{<}}},postaction={decorate}},
}

\usetikzlibrary{snakes}
\usetikzlibrary{shapes.misc}

\newcommand{\Zpre}[5]{
\tikzset{line/.style={line width=0.25mm},
curve/.style={line,smooth,tension=1}}
\draw [line] (0,0) -- (2,0) -- (2,2) -- (0,2) -- (0,0);
\ifthenelse{\equal{#4}{}}{
	\ifthenelse{\equal{#2}{1}}{\draw [line,dashed] (0,1) -- (2,1)}{
		\ifthenelse{\equal{#2}{2}}{\draw [line,middlearrow={>},line width=0.5mm] (0,1) -- (2,1)}{}
	};
	\ifthenelse{\equal{#3}{1}}{\draw [line,dashed] (1,0) -- (1,2)}{
		\ifthenelse{\equal{#3}{2}}{\draw [line,middlearrow={>},line width=0.5mm] (1,0) -- (1,2)}{}
	};
}{
	\ifthenelse{\equal{#3}{1}}{
			\draw [line,dashed] (1,0) -- (1,0.7);
			\draw [line,dashed] (1,2) -- (1,1.3);
			\draw [line] (1,0.7) -- (1,1.3);
			
		}{
		\ifthenelse{\equal{#3}{2} \AND \not\equal{#2}{2}}{\draw [line] (1,0) -- (1,2)}{
			\ifthenelse{\equal{#3}{2} \AND \equal{#2}{2}}{
				\draw [line] (1,0) -- (1,0.7);
				\draw [line] (1,2) -- (1,1.3);
				\ifthenelse{\equal{#5}{1}}{
					\draw [line,dashed] (1,0.7) -- (1,1.3);
				}{};
			}{};
		};
	};
	\ifthenelse{\equal{#4}{1}}{
		\ifthenelse{\equal{#2}{1}}{
			\draw [curve,dashed] plot coordinates {(0,1) (0.7,1) (1,1.3)};
			\draw [curve,dashed] plot coordinates {(2,1) (1.3,1) (1,0.7)};
			}{
			\ifthenelse{\equal{#2}{2}}{
			\draw [curve] plot coordinates {(0,1) (0.7,1) (1,1.3)};
			\draw [curve] plot coordinates {(2,1) (1.3,1) (1,0.7)};
			}{};
		};
	}{
		\ifthenelse{\equal{#4}{0}}{
			\ifthenelse{\equal{#2}{1}}{
				\draw [curve,dashed] plot coordinates {(0,1) (0.7,1) (1,0.7)};
				\draw [curve,dashed] plot coordinates {(2,1) (1.3,1) (1,1.3)};
				}{
				\ifthenelse{\equal{#2}{2}}{
				\draw [curve] plot coordinates {(0,1) (0.7,1) (1,0.7)};
				\draw [curve] plot coordinates {(2,1) (1.3,1) (1,1.3)};
				}{};
			};
		};
	};
};
}

\newcommand{\Z}[5]{
\begin{gathered}
\begin{tikzpicture}[scale = #1]
\Zpre{#1}{#2}{#3}{#4}{#5}
\end{tikzpicture}
\end{gathered}
}

\begin{document}
    
\begin{titlepage}
    \hfill      YITP-SB-2023-19
    \\

\title{\textit{What's Done Cannot Be Undone:}\\
TASI Lectures on Non-Invertible Symmetries}

\author{Shu-Heng Shao}

        \address{C.\ N.\ Yang Institute for Theoretical Physics, Stony Brook University}
        ~~\\

\abstract

  We survey recent developments in a novel kind of generalized global symmetry, the non-invertible symmetry, in diverse spacetime dimensions. 
   We start with several different but related constructions of the non-invertible Kramers-Wannier duality symmetry in the Ising model, and conclude with a new interpretation for the neutral pion decay and other applications. 
  These notes are based on lectures given at the TASI 2023 summer school ``Aspects of Symmetry."

\end{titlepage}

\eject

\tableofcontents

\section{Introduction}

\noindent\textit{``To bed, to bed. There’s knocking at the
gate. \\
Come, come, come, come. Give me your
hand. \\
What’s done cannot be undone. \\
To bed, to
bed, to bed."\\
--- Lady Macbeth; Act 5, Scene 1, Macbeth}

~~\\

Symmetry has long been a guiding principle in physics.  
Recently, there has been a transformative development in our understanding of global symmetries, stimulated by progress in high energy physics, condensed matter physics, quantum information theory, and mathematics. 
The notion of  symmetry has been broadened in several different directions, including the higher-form symmetry, non-invertible symmetry, subsystem symmetry, fractal symmetry, and many more.  
New examples of generalized global symmetries are found both in continuum quantum field theory (QFT) and lattice models of all kinds. 
These new symmetries and their anomalies have applications in a variety of quantum systems, ranging from the Ising model, to topological phases of matter of anyons and fractons, to gauge theory, and to string and M-theory.

Global symmetry provides an invariant characterization of the physical system. 
It acts nontrivially on the states and operators, and serves as an ID number that helps distinguish between distinct quantum systems. 
This is to be contrasted with the role of gauge ``symmetry", which is more of a redundancy in our description for the system. 
It can exist in one description of the model, but not in another dual description, such as in the particle-vortex duality.\footnote{Historically, the word ``global" is used to mean that the symmetry transformation parameter is a constant in the spacetime coordinates. 
For the more general symmetries discussed below, sometimes the symmetry parameters can have nontrivial dependence  in spacetime. Nonetheless, they are still true symmetries that act nontrivially on the Hilbert space, rather than redundancies in the description. 
We will use the word ``global" in the sense that it is not a gauge ``symmetry"; it is a true symmetry that acts faithfully on the physical configurations.}

Wigner's theorem states that ordinary global symmetries in quantum mechanics are implemented by (anti-)unitary operators, which in particular, have inverses. 
In higher spacetime dimensions, the story is however different. Symmetries can be non-invertible -- they are implemented by conserved operators without an inverse. 
Nonetheless, these \textit{non-invertible symmetries} lead to new conservation laws, selection rules, and dynamical constraints on renormalization group (RG) flows.

Non-invertible symmetries have a long history in theoretical physics.  
In integrable systems, there are infinity many conserved charges that do not lead to unitary operators. 
In 1+1d, non-invertible symmetries are implemented by topological line operators. 
Building on the seminal work of \cite{Verlinde:1988sn,Moore:1988qv,Moore:1989yh}, topological lines in rational conformal field theory (RCFT)  have been systematically studied   through a series of developments \cite{Oshikawa:1996dj,Petkova:2000ip,Fuchs:2002cm,Fuchs:2003id,Fuchs:2004dz,Frohlich:2004ef,Fuchs:2004xi,Fjelstad:2005ua,Frohlich:2006ch,Feiguin:2006ydp,Fuchs:2007tx,Fredenhagen:2009tn,Frohlich:2009gb,Davydov:2010rm,Carqueville:2012dk,Bachas:2013ora,Brunner:2013ota,Brunner:2013xna,Brunner:2014lua,Ho:2014vla,Hauru:2015abi,Aasen:2016dop,Aasen:2020jwb}. 
In recent years, it has been advocated that these non-invertible topological defects should be viewed
as a generalization of ordinary global symmetries 
\cite{Bhardwaj:2017xup,Chang:2018iay}. See also \cite{Tachikawa:2017gyf,Ji:2019ugf,Cordova:2019wpi,Lin:2019hks,Thorngren:2019iar,Konechny:2019wff,Huang:2020lox,Gaiotto:2020fdr,Pal:2020wwd,Brunner:2020miu,Gaiotto:2020iye,Komargodski:2020mxz,Chang:2020imq,Huang:2021ytb,Inamura:2021wuo,Thorngren:2021yso,Sharpe:2021srf,Huang:2021zvu,Huang:2021nvb,Vanhove:2021zop,Inamura:2021szw,Burbano:2021loy,Inamura:2022lun,Chang:2022hud,Lu:2022ver,Lin:2022dhv,Robbins:2022wlr,Li:2023mmw,Lin:2023uvm,Zhang:2023wlu,Cao:2023doz,Jacobsen:2023isq,Choi:2023xjw,Haghighat:2023sax,Seiberg:2023cdc}.  
Even more recently,  a large class of non-invertible symmetries  were found in general spacetime dimensions, which was built on an interesting interplay with the higher-form symmetries.  
Furthermore, these non-invertible symmetries exist in realistic QFTs such as the 3+1d pure Maxwell gauge theory, QED, QCD, and axion models, with new dynamical consequences. 
See \cite{Koide:2021zxj,Choi:2021kmx,Kaidi:2021xfk,Roumpedakis:2022aik,Bhardwaj:2022yxj,Arias-Tamargo:2022nlf,Hayashi:2022fkw,Choi:2022zal,Kaidi:2022uux,Choi:2022jqy,Cordova:2022ieu,Antinucci:2022eat,Bashmakov:2022jtl,Damia:2022seq,Damia:2022bcd,Moradi:2022lqp,Choi:2022rfe,Bhardwaj:2022lsg,Bartsch:2022mpm,Lin:2022xod,GarciaEtxebarria:2022vzq,Apruzzi:2022rei,Heckman:2022muc,Freed:2022qnc,Niro:2022ctq,Kaidi:2022cpf,Mekareeya:2022spm,
Antinucci:2022vyk,Chen:2022cyw,Bashmakov:2022uek,Karasik:2022kkq,Cordova:2022fhg,Decoppet:2022dnz,GarciaEtxebarria:2022jky,Choi:2022fgx,Yokokura:2022alv,Bhardwaj:2022kot,Bhardwaj:2022maz,Bartsch:2022ytj,Hsin:2022heo,Heckman:2022xgu,Antinucci:2022cdi,Apte:2022xtu,Garcia-Valdecasas:2023mis,Delcamp:2023kew,Kaidi:2023maf,Etheredge:2023ler,Putrov:2023jqi,Amariti:2023hev,Carta:2023bqn,Koide:2023rqd,Yamamoto:2023uzq,Inamura:2023qzl,Chen:2023qnv,Cvetic:2023plv,Bashmakov:2023kwo,Anber:2023pny,Bhardwaj:2023ayw,Bartsch:2023wvv,Damia:2023ses,Pasquarella:2023deo,Copetti:2023mcq,Argurio:2023lwl,Zhang:2023wai,vanBeest:2023dbu,Decoppet:2023bay,Lawrie:2023tdz,Yu:2023dfx,Bah:2023ymy,Moradi:2023dan,Chen:2023czk,Apruzzi:2023uma,Cordova:2023ent,Sun:2023xxv,Carqueville:2023qrk} for a partial list of recent advancements in non-invertible symmetries in higher spacetime dimensions.

In these notes we review aspects of non-invertible symmetries and their dynamical applications.  
Most of the discussions are formulated in QFT, while some others are lattice examples.   
In Section \ref{sec:generalities} we give a general discussion of global symmetries in QFT, and emphasize the relation between operators and defects. 
In Section \ref{sec:Ising}, we discuss several different (but related) constructions of the simplest nontrivial non-invertible symmetry: the Kramers-Wannier duality symmetry of the 1+1d Ising model.  
We review approaches both in the continuum and on the lattice. 
Section \ref{sec:interlude} comments on the relation between higher-form and non-invertible symmetries in general spacetime dimensions. 
 Section \ref{sec:higher} discusses the most basic non-invertible symmetries in higher than 1+1d, the condensation defect from higher gauging.  
 In Section \ref{sec:half}, we review a powerful construction of non-invertible symmetries from half gauging, and apply it to the Ising model, $c=1$ compact boson CFT, the 3+1d Maxwell theory, and the 3+1d ${\cal N}=4$ super Yang-Mills gauge theory. 
In Section \ref{sec:pion} we discuss the non-invertible symmetries in the real-world QED and QCD, and provide an alternative interpretation for the neutral pion decay.  
Section \ref{sec:application} covers dynamical applications from non-invertible symmetries. These include the universal bounds on the axion string tension and monopole mass in axion physics. 
We conclude in Section \ref{sec:conclusion}.

Unfortunately, there are many fascinating topics on non-invertible symmetries, and more generally, on generalized global symmetries, not covered in the current notes. 
In particular, we will not provide a comprehensive discussion of the mathematical framework behind these new symmetries, but focus more on the physical examples. 
We refer the readers to the recent reviews \cite{McGreevy:2022oyu,Cordova:2022ruw,Gomes:2023ahz,Brennan:2023mmt,Schafer-Nameki:2023jdn,Bhardwaj:2023kri,Luo:2023ive} for complementary discussions.

\section{Generalities on global symmetries}\label{sec:generalities}

What is symmetry? 
In quantum mechanics, the minimum requirement for any sort of symmetry is the existence of  an operator $U$ that is conserved under time evolution, i.e., $U$ commutes with the Hamiltonian, $[U,H]=0$. 
In QFT, the appropriate generalization of this condition  is that  the symmetry operator commutes with the stress-energy tensor $T_{\mu\nu}$. 
However, this is not sufficient. 
There are more constraints on symmetries from spacetime locality, as we discuss below.

\subsection{Conservation and topology}

For relativistic QFTs in Euclidean signature, time is on the same footing as any other spatial directions. 
Therefore, the conservation under time evolution $[U,H]=0$ should be upgraded to the invariance under any deformation in spacetime, i.e., the operator $U$ should be topological. 
``Topology" is the suit-up version of ``conservation"  in relativistic QFT. 

Let us illustrate this concept of topological operators using a familiar example. 
Consider a general QFT in  $d$ spacetime dimensions with a continuous $U(1)$ global symmetry.  
It is associated with a conserved Noehter current $j_\mu(x)$ satisfying the conservation equation, which in Lorentzian signature is
\ie
\partial^\mu j_\mu = -\partial_0 j_0 +\partial_i j_i= 0
\fe
where $i$ runs over the spatial coordinates. 
We define a charge operator $Q$ as
\ie
Q  =  \oint d^{d-1} x j_0\,.
\fe
For simplicity, we assume our space has no boundary, or equivalently, we assume appropriate fall-off conditions on the fields at infinity. 
The charge $Q$ is conserved because of the conservation equation
\ie
\partial_0 Q = \oint d^{d-1} x\,  \partial_0 j_0 
= \oint d^{d-1}x \, \partial_i j_i = 0 \,.
\fe
The conserved, unitary operator $U_\theta$ that implements a $U(1)$ rotation with angle $\theta$ is
\ie\label{unitaryU}
U_\theta =e^{i\theta Q}=  \exp\left[  i \theta \oint d^{d-1}x j_0 \right]\,.
\fe
We   refer to $U_\theta$ as the symmetry operator, while $Q$ as the charge operator. 

The unitary symmetry operator $U_\theta$ can be generalized in a covariant way in Euclidean signature as follows. 
Let $M^{(d-1)}$ by a closed $(d-1)$-manifold with no boundary in the $d$-dimensional Euclidean spacetime. 
We define
\ie
U_\theta (M^{(d-1)} ) 
=\exp\left[ i\theta\oint_{M^{(d-1)} } j_\mu dn^\mu\right]
= \exp\left[ i\theta \oint_{M^{(d-1)}} \star j\right]
\fe
where $\star$ is the Hodge dual of a differential form. 
Since $\star j$ is a closed form, i.e., $d\star j=0$, by Stokes' theorem, correlation functions of $U_\theta(M^{(d-1)})$ is independent of small deformations of $M^{(d-1)}$. 
Therefore, we see that a conserved current operator $j_\mu(x)$ in a relativistic QFT leads to a topological object $U_\theta (M^{(d-1)})$ supported on a  codimension-1 manifold in spacetime.

\subsection{Operators versus defects}\label{sec:opdefect}

 What is $U_\theta (M^{(d-1)})$? 
 When $M^{(d-1)}$ is the whole space at a fixed time, $U_\theta(M^{(d-1)})$ is the conserved, unitary operator that acts on the Hilbert space ${\cal H}(M^{(d-1)})$. 
When $M^{(d-1)}$ is extended in the time direction and localized in one of the spatial directions, say at $x=0$,  $U_\theta(M^{(d-1)})$ is a defect that modifies the quantization. The modified quantization gives rise to a twisted Hilbert space ${\cal H}_\theta(M^{(d-1)})$, labeled by the rotation angle $\theta$.

For instance, consider a free complex scalar field in 1+1d (i.e., $d=2$), 
\ie
{\cal L} =   \partial_\mu \Phi \partial^\mu \Phi^\dagger\,.
\fe
There is a $U(1)$ global symmetry $\Phi \to e^{i\theta}\Phi$ whose Noether current is  
\ie
j_\mu = i (\partial_\mu \Phi^\dagger) \Phi -i \Phi^\dagger\partial_\mu \Phi \,.
\fe
Let the space be a circle $S^1_x$ parametrized by $x\sim x+2\pi$. 
The Hilbert space ${\cal H}(S^1_x)$ on a circle is obtained by the canonical quantization of the free scalar field subject to the periodic boundary condition
\ie
\Phi(\tau, x+2\pi ) =\Phi(\tau,x)\,.
\fe
The conserved current leads to a unitary operator $U_\theta(S^1_x) = \exp(i \theta \oint dx j_0)$ that acts on this Hilbert space ${\cal H}(S^1_x)$. 
Alternatively, in Euclidean signature, we can insert a defect $U_\theta(S^1_\tau) = \exp( i \theta \oint d\tau j_x)$ along the Euclidean time direction at $x=0$. 
This defect changes the boundary condition  of the scalar field to 
\ie
\Phi(\tau,x+2\pi) =  e^{i\theta} \Phi(\tau,x)\,.
\fe
Canonical quantization subject to the above twisted boundary condition leads to a twisted Hilbert space ${\cal H}_\theta(S^1_x)$ labeled by the $U(1)$ group element $\theta\in [0,2\pi)$.

For a discrete symmetry $G$ (such as a $\mathbb{Z}_N$ symmetry), we do not have a conserved Noether current or a charge operator. 
Nonetheless, a discrete symmetry in a general QFT can be defined in terms of the existence of a conserved unitary operator $U_g$ for every group element $g\in G$.
 In relativistic QFTs,  these conserved operators lead to  topological operators/defects $U_g(M^{(d-1)})$ in Euclidean spacetime, with their correlation functions subject to various consistency conditions.

These examples point to  a general principle of global symmetries in relativistic QFTs: Spacetime locality requires that every global symmetry can be interpreted either as an operator or as a defect.  
More specifically, every global symmetry $g\in G$ should serve two purposes in life:
\begin{enumerate}
\item It leads to a conserved \textit{operator}  that acts on the Hilbert space $\cal H$.
\item It leads to a topological \textit{defect} that modifies the quantization and gives a twisted Hilbert space ${\cal H}_g$.\footnote{Traditionally, ``topological defects"  refer to solitons or other extended objects (such as strings or domain walls) that arise from the nontrivial topology of the field space. These defects typically have nonzero tension, so it costs energy to move them in spacetime.  
In contrast, in this review, ``topological defects" refer to  defects whose infinitesimal deformations  in spacetime do not change any physical observables. 
Euclidean correlation functions of these topological defects do not depend on the detailed shape and location of the defect insertions, but only the topology (e.g., the homological cycles). 
In particular, these defects have zero tension. 
This is similar to the use of the term ``topological QFTs", which refers to QFTs that depend only on the topology of the spacetime but not on the detailed shape. 
(The traditional, finite-tension ``topological defects" are referred to  as ``homotopy defects" in \cite{Pace:2023kyi} to avoid this confusion in terminology.) } 
\end{enumerate}
This \textit{operator/defect principle} imposes strong constraints on symmetries in QFT and is essential for the consistency of Euclidean correlation functions.  
There are instances where a conserved operator does not lead to a well-defined defect upon Wick rotation, and therefore cannot be inserted in a Euclidean correlation function. 
We will keep coming back to the constraints from this principle in later sections.

To summarize, both the operator and defect are captured in a single object $U_g(M^{(d-1)})$, which is the invariant way to characterize a global symmetry $G$ in relativistic QFT in  Euclidean spacetime \cite{Gaiotto:2014kfa}.\footnote{In the rest of this paper, we will sometimes use the term ``operator" and ``defect" interchangeably when there is no potential confusion.}  
Correlation functions involving $U_g(M^{(d-1)})$ are invariant under infinitesimal deformation of $M^{(d-1)}$. 
In particular, when $M^{(d-1)}$ is the whole space, the topological nature implies the conservation under time evolution.

\subsection{Higher-form symmetries}

Compared to quantum mechanics, another bonus in QFT  is that there can be conserved operators that have support only along some higher codimensional submanifolds. 
In Euclidean signature, they correspond to  topological defects of higher codimensions. 
These are called \textit{higher-form global symmetries} \cite{Gaiotto:2014kfa}. See \cite{Nussinov:2009zz,Kapustin:2013uxa,Kapustin:2014gua} for earlier works. 
More specifically, a $q$-form global symmetry is associated with a $(d-q-1)$-dimensional topological defect $U_g(M^{
(d-q-1)})$ in $d$ spacetime dimensions. 
In particular, an ordinary global symmetry is an invertible 0-form symmetry. 

A $q$-form global symmetry acts on a $q$-dimensional object $W$ as
\ie
U_g(M^{(d-q-1)} )\cdot W(N^{(q)})  = g(W) W(N^{(q)})
\fe
where $M^{(d-q-1)}$ and $N^{(q)}$ are linked in spacetime.  Here $g(W)$ is a representation of $g$, which is a phase.

Arguably the simplest higher-form symmetry is the Gauss law operator in free Maxwell theory with no charged matter. 
One defines a topological operator as
\ie
U_\theta(M^{(d-2)}) = \exp\left[- {\theta\over e^2} \oint_{M^{(d-2)}} \star F\right]\,,
\fe
where $e$ is the electric coupling constant in the Maxwell action. 
The exponent is nothing but the electric flux. 
It is topological thanks to the Gauss law $d\star F=0$, and implements a $U(1)^{(1)}$ 1-form global symmetry.\footnote{The superscript of a group denotes the form degree of the higher-form global symmetry. For an ordinary, 0-form symmetry, we sometimes suppress the superscript $(0)$ when there is no potential confusion. }
The charged objects are the (non-topological) Wilson lines $\exp(i n\oint_{N^{(1)}} A)$. 
The 1-form symmetry operator acts on the Wilson line by a phase:
\ie
U_\theta (M^{(d-2)}) \cdot \exp(i n \oint_{N^{(1)}} A) = e^{ in\theta } \exp(i n \oint_{N^{(1)}} A)\,.
\fe

There are also discrete higher-form symmetries. For instance, the center symmetry of a pure $SU(N)$ gauge theory with no matter fields is a $\mathbb{Z}_N^{(1)}$ 1-form global symmetry that measures the $N$-ality of a  Wilson line (i.e., the number of boxes mod $N$ in the Young tableau for its $SU(N)$ representation).

\subsection{The space of topological defects}\label{sec:space}

We have seen that global symmetries in a relativistic QFT are invariantly characterized in terms of the topological symmetry operators/defects. 
Given a QFT, what is  the space of topological operators/defects $\CD_a$?  
The complete structure in general spacetime dimensions is complicated, and requires the machinery of a full-fledged higher fusion category theory. 
Below we briefly discuss a subset of  the structure from a physics point of view, which allows us to take a first look into non-invertible symmetries.

\subsubsection{A first look into non-invertible symmetries}

Given two $(d-q-1)$-dimensional topological operators $\CD_1,\CD_2$ for a $q$-form global symmetry,  we can act them successively on the Hilbert space. 
More generally in Euclidean spacetime, 
we can insert two parallel topological operators/defects  near each other and bring them together parallelly. 
Specifically, we place $\CD_1$ and $\CD_2$ at the two boundaries of  $I\times M^{(d-q-1)}$  with no other operator insertion in between. Here  $I$ is an interval and $M^{(d-q-1)}$ is a closed manifold,
Since they are topological, the correlation function does not depend on the distance between them along the interval. 
Hence, this configuration defines a fusion product $\times$ between topological operators/defects. 
For an ordinary invertible global symmetry $G$, the fusion product takes the form of group multiplications:
\ie\label{groupmul}
U_{g_1} \times U_{g_2} = U_{g_3} \,,~~~~g_1 \times g_2 = g_3\,,~~~~g_i \in G\,.
\fe
In particular, every ordinary symmetry defect $U_g$ has an inverse $U_{g^{-1}}$, labeled by the inverse group element $g^{-1}\in G$, so that $U_g \times U_{g^{-1}} = U_{g^{-1}}\times U_g=1$.

If $q\ge 1$, one can move one higher-form symmetry defect past another in the ambient spacetime without intersecting the second one. This means that the fusion product for a higher-form global symmetry is always commutative, i.e., $U_{g_1} \times U_{g_2} = U_{g_2} \times U_{g_1}$ if $q\ge 1$.  (However,  it need not be invertible.) 
Indeed, the center symmetry groups  in gauge theory are always abelian. 
In contrast, for the ordinary global symmetry $q=0$, there isn't enough space to move one defect past another without intersection, and the fusion product is generally non-commutative. 
Indeed, there are non-abelian ordinary global symmetries such as $SU(2)$ or $S_3$.

In addition to the multiplication, we can also define a sum. 
Given any two topological defects $\CD_1$ and $\CD_2$,   the sum $\CD_1 + \CD_2$ is defined  so that its correlation function is the sum of those for the constituents:
\ie\label{group}
\langle (\CD_1+\CD_2) (M^{(d-q-1)}) \cdots \rangle
=\langle \CD_1 (M^{(d-q-1)}) \cdots \rangle+\langle \CD_2(M^{(d-q-1)}) \cdots \rangle 
\fe
where $\cdots$ represent the other operator insertions. 
The twisted Hilbert space associated with $\CD_1+\CD_2$ is  ${\cal H} _{\CD_1} \oplus {\cal H}_{\CD_2}$. 
More generally, we can take linear combinations of topological defects with non-negative integer coefficients. 
In contrast, while $\CD_1 - \CD_2$ or $\frac 17 \CD_1$ serve as well-defined conserved operators (and are therefore symmetries in the context of quantum mechanics), there are no Hilbert spaces  associated with them.  Therefore, they are not valid topological defects.  
This is one constraint from the operator/defect principle.

There is one exception to this constraint from the operator/defect principle. 
When the topological operator is a point in spacetime, i.e., if it is a $(d-1)$-form global symmetry operator,  we cannot use it as a defect to twist the Hilbert space. 
Hence, we cannot associate a Hilbert space to such a topological local operator.  
For topological local operators, we are allowed to consider general linear combinations of them with complex coefficients. 
For instance, when $d=1$, these are the ordinary global symmetries in quantum mechanics, where there is no notion of defects.

In relativistic QFT, an ordinary global symmetry is associated with a topological defect.  
Is the converse true?  
Interestingly, the answer is no: there are many topological defects that are not associated with an ordinary, invertible  symmetry. 
These topological defects do not obey a group multiplication law \eqref{groupmul}.  
In 1+1d, their fusion rule takes the form
\ie
\CL_a \times \CL_b = \sum_c N^c_{ab} \CL_c
\fe
with $N^c_{ab}\in \mathbb{Z}_{\ge0}$. 
In particular, they generally do not have an inverse. That is, given $\CL$, there isn't another topological defect $\CL^{-1}$ such that $\CL\times \CL^{-1}=1$.  
For this reason, they are called the non-invertible defects.

Higher-form symmetries can also  be non-invertible (but are necessarily commutative).  There are  topological operators/defects of general codimensions which do not obey a group-like fusion rule.  
Most generally,  topological defects of all dimensionalities should be viewed as generalized global symmetries of a relativistic QFT.

\subsubsection{Simple defects and boundaries}

A $p$-dimensional topological defect with $p>0$ is called \textit{simple} if there is a unique  topological point operator (which is the restriction of the bulk identity operator) living on it.
The sum of two defects $\CD_1+\CD_2$  has  at least two topological point operators coming from the two constituent defects. 
Therefore, a simple $p$-dimensional defect cannot be written as a sum of other $p$-dimensional defects. 
A simple (or elementary) boundary condition is defined in the same way.

An interesting question is to find simple non-invertible topological operators/defects that cannot be written as linear combinations of other defects of the same dimensionality. 
This rules out trivial examples like the projection operator, $\CC=I+\eta$ with $\eta^2=1$.\footnote{More precisely, $\CC$ is twice the projection. The projection operator on the other hand $1+\eta\over2$ is not a valid defect because of the division by 2.} 

There is one exception to this definition.  For $(d-1)$-form symmetries generated by topological local operators (i.e., $p=0$), one cannot define a notion of simpleness.  There is no preferred integral basis in the space of topological local operators, and one can take arbitrary linear combinations of them with complex coefficients.  For this reason there is no interesting non-invertible $(d-1)$-form symmetry in $d$ spacetime dimensions. In particular, there is no interesting non-invertible symmetry in quantum mechanics.\footnote{This is not to say that they are not useful; it is just that the notion of non-invertibility is unclear for $(d-1)$-form symmetry. See \cite{Nguyen:2021naa} for an application of such symmetries.}

\subsubsection{Stacking with TQFTs}

Let us discuss some rather trivial topological defects. 
In a $d$-dimensional QFT, we can insert  a decoupled $p$-dimensional TQFT, which can be viewed as a trivial topological defect of the ambient QFT.\footnote{While adding additional localized degrees of freedom to a given QFT   might appear unorthodox in the context of QFT, it is actually very common in condensed matter systems where people consider impurities localized at a point in space. 
Some of these impurities flow to the low energy to a topological (or conformal) defects.} 
This defect is simple if the TQFT has a unique topological local operator.  
When the TQFT is non-invertible (meaning that there does not exist another TQFT such that the tensor product of the two is a trivial TQFT), such as a general Chern-Simons gauge theory, the resulting defect is also non-invertible. 
This is a very trivial way to construct a simple non-invertible symmetry.

Next, given a $p$-dimensional topological defect $\CD$, we can stack a $p$-dimensional TQFT $\cal T$ on top, and obtain another defect ${\cal T}\otimes \CD$.  
More generally, given a set of $p$-dimensional topological defect $\CD_a$, the combination
\ie\label{TQFTc}
\sum_a {\cal T}_a \otimes \CD_a\,.
\fe
gives another well-defined defect. 
That is, we can take linear combinations of topological defects with ``coefficients" ${\cal T}_a$ valued in TQFT. 
The fusion of two general topological defects takes the   form
\ie
\CD_a \times \CD_b = \sum_c {\cal T}_c \otimes \CD_c\,,
\fe
where ${\cal T}_c$ are TQFT-valued coefficients.

In the special case of line defects $\CL_a$, the  allowed coefficients are the 0+1d topological quantum mechanics, which is nothing but a free $n$-dimensional qunit completely characterized by a positive integer $n$  (also known as the Chan-Paton factor). 
Therefore, for line defects, \eqref{TQFTc} reduces to
\ie
\sum_a  n_a \CL_a \,,~~~
\fe
with non-negative integer coefficients $n_a\in \mathbb{Z}_{\ge0}$.

In general, given a $d$-dimensional QFT, the space of $p$-dimensional simple topological defects, and therefore the full set of generalized global symmetries, is very wild. 
The classification is at least as complicated as the classification of $p$-dimensional TQFT.  
For $p=1$, since the only 0+1d TQFTs are just free qunits, the ambiguity is less severe; it's just an overall multiplicity factor $n_a$ for each line defect $\CL_a$. 
For $p=2$, since every 1+1d nontrivial TQFT (without symmetry) has more than one topological local operator, the stacking procedure above necessarily gives a non-simple surface defect. 
(See for instance \cite{Moore:2006dw}.) 
Hence, the classification of simple topological surface defects is under control. However, once we are at $p=3$ or higher, there are infinity many $p$-dimensional TQFTs with a unique topological local operator, such as the Chern-Simons gauge theory. This makes the classification of simple $p\ge3$-dimensional topological defects an infinite problem.\footnote{To make any progress, one needs to introduce a notion of equivalence classes between topological defects. See \cite{Copetti:2023mcq} for  a proposal.}

Similar comments apply to boundary conditions as well.  
Given a $d-1$-dimensional boundary condition of a $d$-dimensional QFT, one can stack a $d-1$-dimensional QFT to obtain another boundary. 
For 1-dimensional conformal boundary of a 1+1d CFT, since all the possible 0+1d CFTs are again just free qunits, the classification of conformal boundaries is not subject to the above ambiguity apart from the multiplicity factor.  
This is the reason why the classification of conformal boundary conditions is meaningful in 1+1d CFT \cite{Cardy:1989ir}. 
Similarly, for 1+1d topological boundaries of a 2+1d TQFT, one can stack a  1+1d TQFT on such a boundary to obtain another one. Since every  nontrivial TQFT has more than one topological local operator, the classification of topological boundaries of a 2+1d TQFT is again under better control. 
 However, the classification of topological boundaries for $d\ge 4$ TQFTs is much more challenging because of the freedom of stacking.

\section{Ising model}\label{sec:Ising}

In this section we discuss the simplest nontrivial example of the non-invertible global symmetry --- the Kramers-Wannier duality defect in the 1+1d critical Ising model, which is arguably the simplest quantum system beyond quantum mechanics. 
There are at least six constructions of the non-invertible symmetry in the Ising model:
\begin{itemize}
\item Topological defect line in Ising CFT \cite{Oshikawa:1996dj,Petkova:2000ip,Frohlich:2004ef,Chang:2018iay}.
\item From the chiral symmetry of the Majorana CFT via bosonization \cite{Thorngren:2018bhj,Ji:2019ugf,Kaidi:2021xfk}.
\item Half gauging and the Kramers-Wannier duality  \cite{Choi:2021kmx}.
\item Non-invertible lattice  operator  \cite{Seiberg:2023cdc,Seiberg:2024gek} and the duality defect \cite{Grimm:1992ni} in the transverse-field Ising lattice model.
\item Aasen-Fendley-Mong  (AFM) construction in the statistical Ising model \cite{Aasen:2016dop,Aasen:2020jwb}.
\item Anyonic chain and its generalization \cite{Feiguin:2006ydp,2009arXiv0902.3275T,2013PhRvB..87w5120G,Buican:2017rxc,Aasen:2020jwb}.
\end{itemize}
The first three constructions are in the continuum, and the latter three are  on the lattice. 
In addition, in Section \ref{sec:bulk}, we discuss a bulk perspective  from the 2+1d $\bZ_2$ gauge theory, i.e., the low energy limit of the toric code \cite{Kitaev:1997wr}. 
Some of these constructions are related. 
For instance, the anyonic chain, which is a Hamiltonian lattice model, is obtained from an anisotropic limit of the statistical AFM model. 
Also, the fourth construction is the lattice counterpart of the second.

We discuss some of these constructions in the rest of this section, and defer the half gauging construction   to Section \ref{sec:half} where we introduce it in general spacetime dimensions.  
 Instead of a chronological order, the presentation below is ordered to coordinate with the discussions in other sections.
 For readers who are familiar with 1+1d CFT, we recommend Sections \ref{sec:IsingCFT} and \ref{sec:bosonization} to start. 
 For readers from a more general background, we recommend Section \ref{sec:TFIM}, which only requires basic knowledge in quantum mechanics.

\subsection{Topological lines in Ising CFT}\label{sec:IsingCFT}

In the first approach, we view the Ising CFT as an abstract set of conformal data, including the set of local operators and their OPE coefficients, which are subject to the consistency conditions including crossing symmetry and unitarity.
This viewpoint, commonly adopted by the conformal bootstrap community in recent years, is completely universal and does not depend on the particular microscopic or Lagrangian realization of the CFT. 

We now extend these local CFT data by those for the line defects, in particular, the topological line defects. 
 Euclidean correlation functions involving  topological defect lines obey a general set of axioms, including various  topological manipulations of the lines. 
The mathematical structure behind these axioms for a finite non-invertible global symmetry is described by the \textit{fusion category}.  
Each simple topological line corresponds to a simple object of the fusion category.  
See \cite{etingof2016tensor} for a mathematical  exposition on this subject. 
We will not review the full set of consistency conditions obeyed by the topological lines. 
In particular, we will not have an extensive discussion of the $F$-symbols, which   capture the crossing relations between the lines and the anomalies. 
Comprehensive reviews can be found in \cite{Bhardwaj:2017xup,Chang:2018iay}, which build on the earlier works of \cite{Verlinde:1988sn,Petkova:2000ip,Fuchs:2002cm,Frohlich:2004ef,Frohlich:2006ch,Feiguin:2006ydp,Fuchs:2007tx,Fredenhagen:2009tn,Frohlich:2009gb,Carqueville:2012dk}. Below we will only use one particular consistency condition of the Euclidean correlation function in Ising CFT to derive its non-invertible topological line.

The Ising CFT has three local Virasoro primary operators: the identity 1, the thermal operator $\varepsilon$, and the order operator $\sigma$ (also known as the spin operator).  
Their conformal weights are $(0,0), (\frac 12,\frac 12), ({1\over 16},{1\over 16})$, respectively.  
The fusion rule between these local primary operators is
\ie\label{Isinglocal}
&\varepsilon\times \varepsilon =1\,,\\
&\sigma \times \varepsilon =  \varepsilon \times \sigma = \sigma\,,\\
&\sigma\times \sigma = 1+\varepsilon\,.
\fe
In 1+1d  CFT, there is an operator-state correspondence which states that the local operators are in one-to-one correspondence with the states in the Hilbert space $\cal H$ quantized on $S^1$. 
This is implemented by a conformal map that maps a plane to a cylinder.   
We denote the corresponding three Virasoro primary states as $\ket{1}, \ket{\varepsilon} ,\ket{\sigma}$, respectively.

\subsubsection{Modular covariance}

The torus partition function of the Ising CFT is
\ie\label{Z}
Z(\tau,\bar\tau )  =\text{Tr}_{\cal H} [q^{L_0 -{c\over 24}  } \bar q^{ \bar L_0 -{c\over24}}]
=| \chi_0(\tau)|^2+ |\chi_{1\over2}(\tau)|^2 + |\chi_{1\over 16}(\tau)|^2
\,,
\fe
with $c=\frac 12$, $q=e^{2\pi i \tau}, \bar q= e^{-2\pi i \bar \tau}$, and $\tau,\bar \tau$ being the complex structure moduli of the spacetime torus.  
$L_0,\bar L_0$ are the zero modes for the left- and right-moving Virasoro algebras, whose eigenvalues are $h,\bar h$. 
Here $\chi_h(\tau)$ is the Virasoro character that captures the contribution of the descendants in a Virasoro module whose primary state has conformal weight $h$. 
We will not need the explicit expressions for the characters (which can be found in every standard CFT textbook), but only their  modular S transformations under  $\tau \to -1/\tau$:
\ie\label{modularS}
\left(\begin{array}{c}\chi_0 (-1/\tau) \\ \chi_{1\over2}(-1/\tau) \\\chi_{1\over16} (-1/\tau)\end{array}\right)
= \frac12 
\left(\begin{array}{ccc}1 & 1 & \sqrt{2} \\1 & 1 & -\sqrt{2} \\\sqrt{2} & -\sqrt{2} & 0\end{array}\right)
\left(\begin{array}{c}\chi_0 (\tau) \\ \chi_{1\over2}(\tau) \\\chi_{1\over16} (\tau)\end{array}\right) = \sum_{j=0,\frac12, {1\over16}}S_{ij} \chi_j(\tau)\,.
\fe
Note that each $|\chi_h(\tau)|^2$ is invariant under the modular T transformation $\tau\to \tau+1$.

\begin{figure}[h!]
\begin{center}
\includegraphics[width=.4\textwidth]{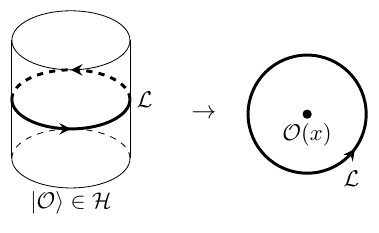}
\end{center}
\caption{Via the operator-state map, the action of a topological line $\CL$ on a state $\ket{\cal O}$ in the Hilbert space $\cal H$ is mapped to the Euclidean configuration of the line encircling the corresponding local operator ${\cal O}(x)$.}\label{fig:action}
\end{figure}

What are the global symmetries of the Ising CFT? 
Let $\CL$ be such a symmetry operator. 
Being  a symmetry in a relativistic CFT, we require $[\CL, T_{\mu\nu}]=0$. 
Therefore, the action of the operator $\CL$ on the Virasoro descendants is completely determined by that on the primary operator. 
Furthermore, $\CL$ does not change the conformal weights $(h,\bar h)$ of  a state, and therefore has to act on the primary states as eigenstates.\footnote{Here we used the fact that no two primary states have the same $(h,\bar h)$ in the Ising CFT.} 
Let $\lambda_0, \lambda_{1\over2}, \lambda_{1\over16}$ be the eigenvalues of $\CL$ acting on the three Virasoro primary states:
\ie
\CL \ket{1} = \lambda_0  \ket{1} \,,~~~~
\CL\ket{\varepsilon} = \lambda_{1\over2} \ket{\varepsilon}\,,~~~~
\CL\ket{\sigma} = \lambda_{1\over 16}  \ket{\sigma}\,.
\fe
Via the operator-state map, the above action can be translated into the Euclidean configuration of a topological line encircling a local operator as in Figure \ref{fig:action}.

The torus partition function of the Ising CFT with the \textit{operator} $\CL$ inserted inside the trace is: 
\ie
\Z{1}{2}{}{}{} =
Z^\CL (\tau,\bar\tau) =  \text{Tr}_{\cal H} [
\CL \, q^{L_0-{c\over24}} \bar q^{\bar L_0-{c\over 24}}
] 
= \lambda_0 |\chi_0(\tau)|^2 
+\lambda_{1\over2} |\chi_{1\over2}(\tau)|^2 
+\lambda _{1\over 16} |\chi_{1\over 16}(\tau)|^2\,.
\fe
In the figure above,  the vertical and horizontal directions are the time and space directions, respectively. 
The opposite sides of the square are identified so the spacetime is a two-torus. 
The topological line $\CL$, shown as the dark, oriented line,  is inserted at a fixed time and extends in the spatial direction.

If we view the Ising CFT on a circle as a quantum mechanics, then there is no additional constraint on the three real eigenvalues $\lambda_0, \lambda_{1\over2} ,\lambda_{1\over16}$.  
However, as emphasized in Section \ref{sec:opdefect}, every symmetry should serve as a conserved operator and as a defect associated with a well-defined Hilbert space. 
Demanding that $\CL$ leads to such a well-defined defect imposes additional constraints on the three eigenvalues.  
This is one application of the operator/defect principle.

\begin{figure}[h!]
\begin{center}
\includegraphics[width=.4\textwidth]{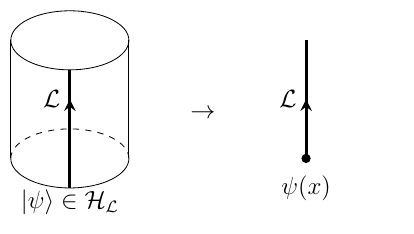}
\end{center}
\caption{Under the operator-state map, a state $\ket{\psi}$ in the Hilbert space ${\cal H}_{\cal L}$ twisted by the topological line $\cal L$ is mapped to a point operator $\psi(x)$ attached to the topological line ${\cal L}$.  Here $\psi(x)$ need not have integer spin $h-\bar h$.}\label{fig:HL}
\end{figure}

Concretely, consider another torus partition function with    a \textit{defect} $\CL$ located at a fixed point in space and extends   in the time direction. 
This partition function can be written as a trace over the twisted Hilbert space ${\cal H}_{\cal L}$ associated with the defect:
\ie
\Z{1}{}{2}{}{} =Z_\CL ( \tau,\bar \tau) = 
\text{Tr}_{{\cal H}_{\cal L}} [ \,  q^{L_0-{c\over 24}} \, \bar q^{\bar L_0 -{c\over24}}\, ] \,,
\fe
A priori, we do not know much about the twisted Hilbert space ${\cal H}_{\cal L}$. 
One thing we do know is that since the topological line $\CL$ commutes with the stress tensor,  the states in ${\cal H}_\CL$ are organized by the left- and right-moving Virasoro algebras. 
Hence, the partition function $Z_\CL$ can be expanded into the Virasoro characters with \textit{non-negative integer coefficients}, which are the degeneracies of the Virarsoro primaries in ${\cal H}_{\cal L}$:
\ie
Z_\CL(\tau,\bar\tau) 
= \sum_{i,j=0,\frac 12,{1\over16}} n_{ij} \chi_i(\tau ) \bar\chi_j(\bar \tau)\,,~~~~n_{ij}\in \mathbb{Z}_{\ge 0}\,.
\fe
Note that the Lorentz spin $h-\bar h$ (which is the spatial momentum eigenvalue on a circle) of a state in  the twisted Hilbert space need not be an integer, in contrast to the untwisted Hilbert space $\cal H$. 
That is, $n_{ij}$ need not be a diagonal matrix. 
This is because the states in ${\cal H}_{\cal L}$ do not correspond to local operators under the cylinder-plane conformal map. 
Rather, they correspond to point operators $\psi(x)$ attached to a topological line ${\cal L}$.\footnote{In these notes, a ``local operator" is an operator whose correlation functions depend only on the position of the point. In contrast, a ``point operator" is an operator whose correlation depends not only on the position of a point, but also on the topological line attached to it. For instance in the Ising CFT, the order operator $\sigma$ is a local operator, while the disorder operator $\mu$ is a point operator attached to the $\bZ_2$ topological line. In the literature, some other authors define ``local operators" and ``point operators" oppositely.} 
This generalizes the conventional operator-state correspondence to a one-to-one correspondence between non-local operators attached to a topological line $\CL$ with the states in the twisted Hilbert space ${\cal H}_\CL$. 
See Figure \ref{fig:HL}.  
In fact, the Lorentz  spin is constrained by the property of the topological line (such as the 't Hooft anomaly). See \cite{Aasen:2016dop,Chang:2018iay,Lin:2019kpn,Lin:2021udi,Lin:2023uvm} for discussions on these spin selection rules.

As Euclidean partition functions, $Z^\CL$ with an operator insertion is related to $Z_\CL$ with a defect insertion by a modular S transformation:
\ie
Z^\CL(\tau,\bar\tau) = Z_\CL(-1/\tau, -1/\bar\tau)\,.
\fe
This exemplifies  how an operator is related to a defect under the Wick rotation in Euclidean QFT.  
By applying the modular S transformation \eqref{modularS}, we find that not every set of real eigenvalues $\lambda_0,\lambda_{1\over2},\lambda_{1\over16}$  leads to a consistent twisted Hilbert space with non-negative integer degeneracies $n_{ij}\in \mathbb{Z}_{\ge0}$. 
It is straightforward to show that all valid solutions of the $\lambda_i$'s are generated by the following three:
\ie\label{Isingaction}
\left.\begin{array}{cc|ccc} && \lambda_0 & \lambda_{1\over2} & \lambda_{1\over16} \\
\hline
\text{identity}~~~&I & 1 &1 & 1 \\
\mathbb{Z}_2~~~&\eta & 1 & 1 & -1 \\
\text{non-invertible}~~~&\CD & \sqrt{2} & -\sqrt{2} & 0
\end{array}\right.
\fe
Any other valid solutions are obtained by taking non-negative integer linear combinations of the above three. 
The first one corresponds to the trivial identity line $I$. 
We discuss the other two below.

The $\eta$ line generates the invertible $\mathbb{Z}_2$ global symmetry, under which $\sigma$ is odd and $1,\varepsilon$ are even. 
We can grade the untwisted Hilbert space by this $\mathbb{Z}_2$ symmetry as $\CH=\CH^+\oplus\CH^-$. The Virasoro primaries in $\CH^\pm$ are
\ie\label{CHpm}
& \CH^+: ~~~~1, ~(h,\bar h)=(0,0)\,,\qquad \varepsilon ,(h,\bar h)=\left(\frac 12, \frac 12\right)\,,\\
&\CH^- :~~~~\sigma,~(h,\bar h)=\left({1\over 16},{1\over 16}\right)\,.
\fe
From \eqref{Isingaction}, we find the torus partition function with the $\mathbb{Z}_2$ defect $\eta$ extended along the time direction:
\ie\label{Zeta}
&Z_\eta(\tau,\bar\tau) = \chi_{1\over16}(\tau)\bar \chi_{1\over 16}(\bar\tau) + \chi_{1\over2}(\tau)\bar \chi_{0}(\bar\tau) + \chi_{0}(\tau)\bar \chi_{1\over 2}(\bar\tau)  \,.
\fe
 We again grade the $\bZ_2$-twisted Hilbert space $\CH_\eta$ by the $\bZ_2$ symmetry as $\CH_\eta =\CH^+_\eta \oplus\CH^-_\eta$. 
 The Virasoro primaries in $\CH^\pm_\eta$ are
 \ie\label{CHetapm}
& \CH_\eta^+: ~~~~ \mu ,(h,\bar h)=\left({1\over16},{ 1\over16} \right)\,,\\
&\CH_\eta^- :~~~~\psi_\text{L},~(h,\bar h)=\left({1\over 2},0\right)\,,\qquad \psi_\text{R},~(h,\bar h)=\left(0,\frac12\right)\,.
\fe
They are the disorder operator $\mu$ and the left- and right-moving fermions $\psi_\text{L},\psi_\text{R}$. 
These states in $\CH_\eta$ do not correspond to local operators; rather, they correspond to point operators attached to the $\bZ_2$ line $\eta$ (see Figure \ref{fig:HL}). 
In particular, the free fermion fields $\psi_\text{L},\psi_\text{R}$ are not local operators in the bosonic Ising CFT, but are attached to the $\bZ_2$ line.

The more interesting symmetry is $\CD$. It is a non-invertible symmetry because it has a kernel:  it annihilates the state $\ket{\sigma}$.  
From the action on the states, we find the following algebra in the untwisted Hilbert space $\CH$:
\ie\label{Isingcat}
&\eta^2 =I\,,\\
&\eta\times  \CD= \CD\times\eta=\CD\,,\\
&\CD\times \CD = I+\eta\,.
\fe
We see that there is no line defect $\CD^{-1}$ such that $\CD \times \CD^{-1} = \CD^{-1}\times \CD=I$. 
This is our first encounter of a (non-trivial) non-invertible global symmetry in QFT.  
These three topological lines do not generate a group, but the \textit{Ising fusion category}, which is one of the  $\mathbb{Z}_2$ \textit{Tambara-Yamagami  fusion categories} TY$_+$\cite{TY}.\footnote{There is another $\mathbb{Z}_2$ Tamabara-Yamagami fusion category TY$_-$ obeying the same fusion rule as in \eqref{Isingcat}. It is realized, for instance, in the $\mathfrak{su}(2)_2$ WZW model.   While they share the same fusion rule, their $F$-symbols (more specifically, their Frobenius–Schur indicators)
 are different. Consequently,  the Lorentz spin  in ${\cal H}_\CD$ obeys $h-\bar h \in \pm {1\over 16}+{\mathbb{Z}\over2}$ in  the Ising fusion category TY$_+$ , but $h-\bar h \in \pm {3\over16} +{\mathbb{Z}\over2}$ in TY$_-$  \cite{Aasen:2016dop,Chang:2018iay,Lin:2023uvm}.}

The torus partition function with a duality defect $\CD$ extended in the time direction is
\ie\label{ZD}
&Z_\CD(\tau,\bar\tau) = \chi_{1\over16}(\tau)\bar \chi_{0}(\bar\tau) + \chi_{1\over16}(\tau)\bar \chi_{1\over2}(\bar\tau) + \chi_{0}(\tau)\bar \chi_{1\over 16}(\bar\tau) 
+\chi_{1\over2}(\tau)\bar \chi_{1\over 16}(\bar\tau)\,.
\fe
There are four Virasoro primaries with $(h,\bar h) = \left({1\over 16}, 0 \right),\left({1\over 16}, \frac12\right), \left(0, {1\over 16}\right), \left({1\over 2}, {1\over 16} \right)$.
(They can be graded by the $\bZ_2$ and the non-invertible operators in the twisted Hilbert space. We will not discuss this grading here, but refer the readers to \cite{Chang:2018iay,Lin:2019hks}.)

In addition to the lack of an inverse, there is another unusual feature of $\CD$: its eigenvalue  on the ground state $\ket{1}$ is $\sqrt{2}$. 
In a compact, unitary CFT with a unique ground state, the eigenvalue $\langle \CL\rangle$ of a symmetry operator $\CL$ on the identity state
\ie\label{qdim}
\CL \ket{1} = \langle \CL\rangle \ket{1}
\fe
is called the \textit{quantum dimension} of $\CL$. 
A symmetry operator $\CL$ of a finite (non-invertible) symmetry always has $\langle \CL\rangle \ge1$. Furthermore, $\langle \CL\rangle =1$ if and only if it is invertible. 
(See \cite{Chang:2018iay} for a physics argument.) 
Using the map in Figure \ref{fig:action}, we see that the circle expectation value of the non-invertible line $\CD$ is $\sqrt{2}$.\footnote{The phase of the circle expectation value of a topological line on the plane can be changed by a topological counterterm associated with the extrinsic curvature. 
Therefore, the circle expectation value on the plane generally differs from the quantum dimension $\langle \CL\rangle$ (defined on a cylinder) by a phase. We refer the readers to \cite{Chang:2018iay,Cordova:2019wpi} for discussions. In this paper,  we   make the choice so that the circle expectation value on the plane is positive and equals $\langle \CL\rangle$.}

Notice that the algebra \eqref{Isingcat} of the topological lines $I,\eta,\CD$ is isomorphic to that of the local Virasoro primary operator \eqref{Isinglocal}. 
This is not an accident. 
It is shown in \cite{Petkova:2000ip} that in any diagonal RCFT, there is a finite set of topological lines that commute with the extended chiral algebra, which are in one-to-one correspondence with the local chiral algebra primary operators. 
These lines are sometimes referred to as the \textit{Verlinde lines} \cite{Verlinde:1988sn}, and they obey the same algebra as the fusion rule for the local primary operators.\footnote{In a  RCFT, there are generally many other topological lines, and therefore generalized global symmetries, that do not commute with the chiral algebra. For instance, the $\mathfrak{su}(2)_1$ WZW model has one nontrivial Verlinde line, corresponding to the $\mathbb{Z}_2$ center symmetry, but it has a continuous $(SU(2)\times SU(2))/\mathbb{Z}_2$ global symmetry.}  
Mathematically, the Verlinde lines form a fusion category, which is obtained from the modular tensor category associated with the RCFT by forgetting the braiding structure.

\subsubsection{Topological lines in Euclidean correlation functions}\label{sec:euclidean}

We have seen that $\CD$ when viewed as an operator acting on the (untwisted) Hilbert space $\cal H$ is non-invertible.  
This operator action can be mapped to the Euclidean configuration in Figure \ref{fig:action}.
But this is not the full story.  
There is another Euclidean process one can consider. 
Consider locally in the Euclidean correlation function, there is a local operator ${\cal O}(x)$ insertion and a topological line $\CL$ nearby.  
Since $\CL$ is topological, we can continuously deform it and bring it past ${\cal O}(x)$. 
This produces another Euclidean configuration of $\CL$ and another point ${\cal O}'(x)$. The latter may be a local operator, or a point operator attached to another topological line. 
The statement that these two configurations are equivalent means that any Euclidean correlation  function containing this local patch is invariant under this local deformation.

\begin{figure}[h!]
\begin{center}
\includegraphics[width=.3\textwidth]{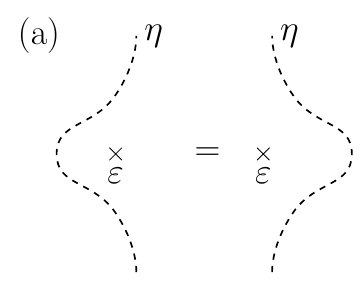}
\qquad\qquad
\includegraphics[width=.3\textwidth]{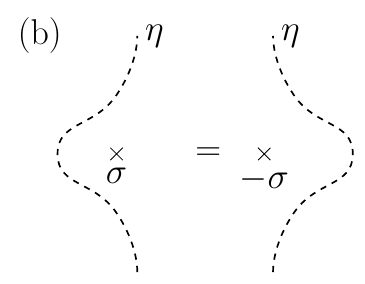}
\end{center}
\caption{As we sweep the $\mathbb{Z}_2$ line $\eta$ past a local operator, we produce a sign for the $\mathbb{Z}_2$ odd operator $\sigma$, while the correlation function is invariant for the $\mathbb{Z}_2$ even local operator $\varepsilon$.}\label{fig:eta}
\end{figure}

Let us demonstrate this Euclidean process for the invertible $\mathbb{Z}_2$ line $\eta$ in the Ising CFT.  
As the $\eta$ line is deformed past the local operators $\varepsilon$ and $\sigma$, it picks up a sign in the latter case since $\sigma$ is odd. See Figure \ref{fig:eta}.\footnote{A topological line $\CL$ is generally oriented. $\eta$ and $\CD$  are their own orientation reversals,  therefore we do not draw an arrow for them.}

\begin{figure}[h!]
\begin{center}
\includegraphics[width=.3\textwidth]{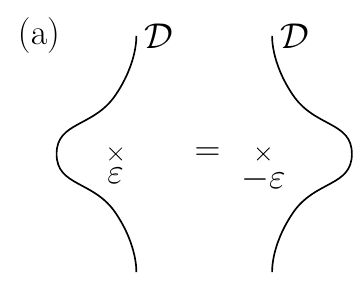}
\qquad\qquad
\includegraphics[width=.3\textwidth]{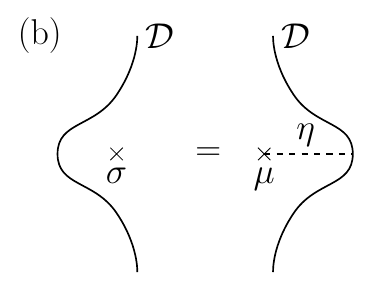}
\end{center}
\caption{Left: The line $\CD$ flips the sign of the thermal operator $\varepsilon$. Right: As we sweep the non-invertible duality  line $\CD$ past the local, order operator $\sigma$, it becomes the non-local, disorder operator $\mu$ attached to the $\mathbb{Z}_2$ line $\eta$.}\label{fig:duality}
\end{figure}

Next, we move on to the non-invertible duality line $\CD$. 
As we bring $\CD$ past $\varepsilon$, the latter receives a sign. 
Hence, as far as $\varepsilon$ is concerned, $\CD$ is trying to be an invertible $\mathbb{Z}_2$ symmetry.  
However, there cannot possibly be such an invertible $\bZ_2$ symmetry since it's incompatible with the fusion rule \eqref{Isinglocal} of the local operators.  
Indeed, as we bring $\CD$ past the order operator $\sigma$, the latter does not just change by a sign; rather, it becomes a disorder operator.  
The disorder operator $\mu$ is not a local operator; it is attached to the $\mathbb{Z}_2$ topological line $\eta$. 
Via the operator-state map, the disorder operator $\mu$ is mapped to a state $\ket{\mu}$ in the $\mathbb{Z}_2$-twisted Hilbert space ${\cal H}_\eta$. 
The fact that this Euclidean process involving $\CD$ maps a local operator to a non-local operator is another hallmark of the non-invertible symmetry. 
We also see that the line $\CD$ implements the Kramers-Wannier duality, which exchanges the order with the disorder operator. 
See Figure \ref{fig:duality}.

\begin{figure}[h!]
\begin{center}
\includegraphics[width=.3\textwidth]{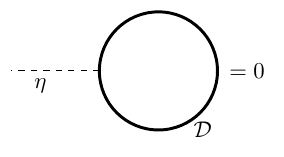}
\end{center}
\caption{Shrinking the $\CD$ loop  in this tadpole diagram creates a $(h,\bar h)=(0,0)$ operator attached to the $\eta$ line. However, there is no such a state in ${\cal H}_\eta$, and hence this Euclidean configuration  must vanish.  }\label{fig:tadpole}
\end{figure}

The Euclidean processes in Figures \ref{fig:eta} and \ref{fig:duality} can also be thought of as the action of the  topological line on the operators. 
How are they related to  the action \eqref{Isingaction} on the Hilbert space ${\cal H}$?   
In each figure, we close the topological line to the right and  make a loop enclosing the local operator on the lefthand side as in Figure \ref{fig:action}.  
For Figure \ref{fig:duality}(a),   we reproduce the action $\CD\ket{\varepsilon }=- \sqrt{2}\ket{\varepsilon}$, where the $\sqrt{2}$ arises from the quantum dimension of  $\CD$.  
For Figure \ref{fig:duality}(b), we end up with a ``tadpole" diagram of the $\eta$ line connecting to an empty bubble of $\CD$. 
Shrinking the $\CD$ loop, we end up with a topological endpoint of the $\eta$ line. 
However, as we computed in \eqref{Zeta}, there is no $(h,\bar h)=(0,0)$ state in the $\mathbb{Z}_2$-twisted Hilbert space ${\cal H}_\eta$. Therefore, this Euclidean correlation function must vanish. 
We hence reproduce the non-invertible action  $\CD \ket{\sigma}=0$. See Figure \ref{fig:tadpole} and \cite{Chang:2018iay} for more general discussions on the vanishing tadpole condition..

More generally, a topological line gives rise to not only an operator on the untwisted Hilbert space, but also maps between different twisted Hilbert spaces. 
These maps can be described by certain ``lasso"  diagrams \cite{Chang:2018iay}. In the mathematical literature, they are related to the tube algebra of the fusion category. See \cite{Lin:2022dhv} for a physicist friendly discussion of the tube algebra.

\subsubsection{Selection rules}\label{sec:selection}

Ordinary global symmetries lead to selection rules of correlation functions, so do non-invertible symmetries. 

\begin{figure}
\centering
\includegraphics[width=1\textwidth]{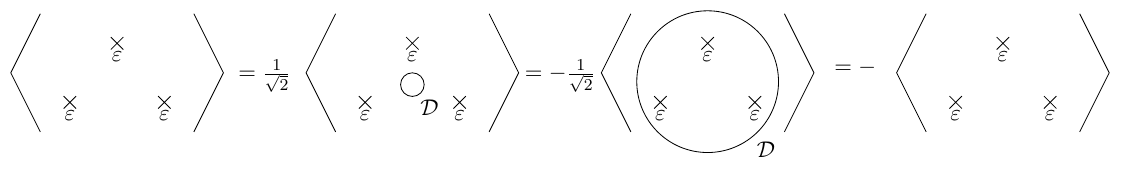}
\caption{Selection rule $\langle \varepsilon \varepsilon \varepsilon\rangle=0$ from the non-invertible symmetry $\CD$. We first nucleate a circle of the non-invertible line $\CD$, and then bring it past the three thermal operators $\varepsilon$. 
Each one of them gives a minus sign according to Figure \ref{fig:duality}(a). Finally, we shrink the non-invertible line $\CD$ on the other side of the two-sphere. The factor $\sqrt{2}$ comes from the quantum dimension of $\CD$. }\label{fig:3e}
\end{figure}

In the Ising CFT, the fusion rule \eqref{Isinglocal} implies that the three-point function  of the thermal operator on the two-sphere  vanishes:
\ie\label{3e}
\langle \varepsilon (z_1, \bar z_1) \varepsilon(z_2,\bar z_2) \varepsilon(z_3,\bar z_3) \rangle = 0 \,.
\fe
Can we understand this from a global symmetry principle? 
It does not follow from the invertible $\bZ_2$ symmetry $\eta$ since $\varepsilon$ is $\bZ_2$-even.  
It turns out that \eqref{3e} is a consequence of the non-invertible symmetry $\CD$, which flips the sign of $\varepsilon$. 
See Figure \ref{fig:3e} for the derivation of \eqref{3e}.

\begin{figure}
\centering
\includegraphics[width=1\textwidth]{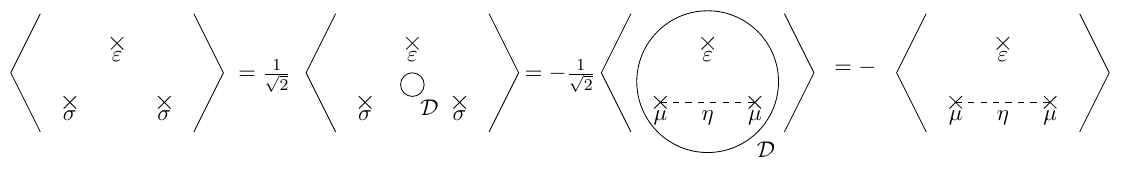}
\caption{Selection rule $\langle \varepsilon \sigma\sigma\rangle=- \langle \varepsilon \mu\mu\rangle$ from the non-invertible symmetry $\CD$. We first nucleate a circle of the non-invertible line $\CD$, and then bring it past the three local operators.  
Using \ref{fig:duality}(b), the two  order operators $\sigma$ turn into a pair of disorder operators $\mu$ connected by a $\bZ_2$ line $\eta$ . Finally, we shrink the non-invertible line $\CD$ on the other side of the two-sphere. }\label{fig:ees}
\end{figure}

As far as $\varepsilon$ is concerned, $\CD$ is trying to be a $\bZ_2$ symmetry. 
However,  $\CD$ cannot be an ordinary, invertible $\bZ_2$ symmetry in the full Ising CFT  because there is a nontrivial three-point function $\langle \sigma\sigma\varepsilon\rangle$ from \eqref{Isinglocal}.  
(Suppose it were, then $\sigma$ is either even or odd. But  either way $\langle \sigma\sigma\varepsilon\rangle$ would have to vanish.) 
Indeed, the action of $\CD$ on the order operator is unconventional; it maps a local operator to a non-local operator as in Figure \ref{fig:duality}.
Following the steps  in Figure \ref{fig:ees}, the non-invertible symmetry leads to  a relation between the local operator correlation function and a correlation function involving the disorder operators:\footnote{To be more precise, in addition to Figure \ref{fig:duality}, we also need to perform an $F$-move to derive Figure \ref{fig:ees}. See \cite{Chang:2018iay} for more discussions on the $F$-moves. }
\ie
\langle \varepsilon (z_1,\bar z_1) \sigma(z_2,\bar z_2) \sigma(z_3,\bar z_3)\rangle
=- \langle \varepsilon (z_1,\bar z_1) \mu(z_2,\bar z_2) \mu(z_3,\bar z_3)\rangle\,.
\fe

We conclude that non-invertible symmetries not only  lead to selection rules  on correlators for local operators, but they also relate local operator correlation functions to those involving non-local ones. 
We refer the readers to \cite{Lin:2022dhv} for a more complete discussion of selection rules from non-invertible symmetries in 1+1d. 

\subsection{Bosonization and  chiral symmetry of the Majorana CFT}\label{sec:bosonization}

It is commonly stated that the Majorana CFT is the same as the Ising CFT. 
This is  incorrect. 
The Majorana CFT is a fermionic CFT (or a spin CFT), which can only be defined on a spin manifold. 
Furthermore, its correlation functions (such as the torus partition function) depend on a choice of the spin structures of the spacetime manifold. 
In contrast, the Ising CFT is a bosonic CFT (or non-spin CFT), which does not depend on the spin structure and can be defined on a general Riemannian manifold.  
In a bosonic CFT, all local operators have integer Lorentz spins,  $h-\bar h\in \mathbb{Z}$, while local operators in a fermionic CFT can have integer or half-integer spins, $h-\bar h \in \mathbb{Z}/2$.

As we discuss below, the Majorana CFT is related to the Ising CFT by gauging. 
This specific gauging procedure which relates a bosonic theory to a fermionic one is   known as the  bosonization/fermionzation.  
In this section we review  these bosonization and fermionization procedures, following the recent discussions in \cite{Kapustin:2017jrc,Karch:2019lnn,yujitasi,Ji:2019ugf,Lin:2019hks,Hsieh:2020uwb,Tan:2022vaz}, which build on the classic paper \cite{Witten:1983ar}.
This discussion leads to another realization of the non-invertible symmetry $\CD$ of the Ising CFT from the anomaly of the $\mathbb{Z}_2\times \mathbb{Z}_2^f$ global symmetry in the Majorana CFT under bosonization. 

\subsubsection{Symmetries and anomalies of the Majorana CFT}

Consider a non-chiral, free, massless Majorana fermion in 1+1d:
\ie
{\cal L} = i \psi_\text{L} (\partial_t-\partial_x )\psi_\text{L} 
+i \psi_\text{R} (\partial_t+\partial_x) \psi_\text{R}\,,
\fe
where $\psi_\text{L},\psi_\text{R}$ are the left- and right-moving Majorana-Weyl fermions, respectively.

The Majorana CFT has a $\mathbb{Z}_2\times \mathbb{Z}_2^f$ global symmetry, which is generated by the fermion parity $(-1)^F$ and the chiral fermion parity $(-1)^{F_\text{L}}$:\footnote{An invertible symmetry operator $U$ acts on a local operator ${\cal O}(x)$ by conjugation, i.e., $U {\cal O}(x) U^{-1} = {\cal O}'(x)$. We  sometimes abbreviate this action as $U:~{\cal O}(x)\to {\cal O}'(x)$. The invertible global symmetry of a QFT is defined as the group formed by  the conjugation action of the symmetry operators on the local operators.  This is generally different from the operator algebra realized on the states in a Hilbert space.  Indeed, as we will see later, the operator algebra in the (twisted) Hilbert space can be projective, which is related to  't Hooft anomalies.}
\ie\label{Z2Z2}
&(-1)^F: ~~\psi_\text{L}\to -\psi_\text{L} \,,\qquad \psi_\text{R} \to -\psi_\text{R}\,,\\
&(-1)^{F_\text{L}}: ~~\psi_\text{L}\to -\psi_\text{L} \,,\qquad \psi_\text{R} \to \psi_\text{R}\,.
\fe
We can compose the above two generators to obtain $(-1)^{F_\text{R}}$ which only flips the sign of $\psi_\text{R}$. 
There are also parity and time-reversal symmetries, but we will not discuss them here.

This gives another way to see that Majorana CFT is different from the Ising CFT: they have different internal global symmetries. The former has the (invertible) $\mathbb{Z}_2\times \mathbb{Z}_2^f$ global symmetry, while the latter has the (non-invertible) Ising fusion category TY$_+$.  See Table \ref{table:MajIsing} for comparisons.

\begin{table}
\begin{align*}
\left.\begin{array}{|c|c|c|}
\hline  & \text{Majorana CFT} & \text{Ising CFT} \\
\hline &\text{fermionic CFT}&\text{bosonic CFT}\\
  & ~\text{depends on  spin structures}~ &~ \text{independent of spin structures} ~\\
  \hline ~\text{internal}~ & \mathbb{Z}_2\times \mathbb{Z}_2^f & \text{Tambara-Yamagami category TY}_+ \\
  ~~\text{global symmetry} ~~& (-1)^{F_\text{L}},(-1)^F & I,\eta, \CD \\
   \hline\text{~local primary operators~} & 1,\psi_\text{L},\psi_\text{R},\psi_\text{L}\psi_\text{R}& 1,\varepsilon,\sigma \\
   \hline \end{array}\right.
\end{align*}
\caption{The Majorana CFT versus the  Ising CFT. }\label{table:MajIsing}
\end{table}

Below we review the quantization of  the Majorana CFT on a circle with $x\sim x+2\pi$.  
(See, for example, \cite{Delmastro:2021xox,Seiberg:2023cdc} for recent discussions.) 
We can impose the periodic or the antiperiodic boundary condition on the left and the right fermions:
\ie
\psi_\text{L}(t,x+2\pi)  = e^{2\pi i \nu_\text{L}} \psi_\text{L} (t,x)\,,\qquad \psi_\text{R} (t,x+2\pi )  =e^{2\pi i \nu_\text{R}} \psi_\text{R}\,,
\fe
with $\nu_\text{L,R}=0,\frac 12$.
Following the standard terminology in string theory, we refer to the periodic boundary condition as Ramond (R), while the anti-periodic boundary condition as Neveu-Schwarz (NS). 
Combining the left with the right, this leads to four different boundary conditions, NSNS, RR, NSR, and RNS. 

We start with the NSNS boundary condition, i.e., $\nu_\text{L}=\nu_\text{R}=\frac 12$.  
In this case, there is no fermion zero mode, and we have a unique ground state $\ket{1}$, which is symmetric under the $\mathbb{Z}_2\times \mathbb{Z}_2^f$ symmetry:
\ie
(-1)^F \ket{1}= \ket{1}\,,\qquad (-1)^{F_\text{L}}\ket{1}= \ket{1}\,.
\fe
The symmetry action on the excited states follow from \eqref{Z2Z2}. 
We find that the $\mathbb{Z}_2\times \mathbb{Z}_2^f$ symmetry is realized linearly on the NSNS Hilbert space $\HNS$:
\ie\label{NSNSlinear}
\text{NSNS}:~~~(-1)^F (-1)^{F_\text{L}} =  (-1)^{F_\text{L}}(-1)^F\,.
\fe
Via the operator-state correspondence, the states in the NSNS Hilbert space $\HNS$ are in one-to-one correspondence with the local operators in the Majorana CFT. 
In particular, the NSNS ground state $\ket{1}$ is mapped to the identity operator 1, while the excited states are mapped to the fermion fields  $\psi_\text{L},\psi_\text{R}$ and their composites. 
There are four  Virasoro primaries, $1, \psi_\text{L},\psi_\text{R},\psi_\text{L}\psi_\text{R}$ in $\HNS$. 
We can further grade NSNS Hilbert space by the $(-1)^F$ charge as  $\HNS = \HNS^+  \oplus\HNS^-$. 
The Virasoro primaries and their conformal weights in $\HNS^\pm$ are
\ie
&\HNS^+ :~~1,~(h,\bar h)=(0,0)\,,\qquad 
&&\psi_\text{L}\psi_\text{R},~(h,\bar h) = \left(\frac 12,\frac 12\right)\,,\\
&\HNS^- :~~\psi_\text{L},~(h,\bar h)=\left(\frac12,0\right)\,,\qquad 
&&\psi_\text{R},~(h,\bar h) = \left(0,\frac 12\right)\,.
\fe
Note that local operators in a fermionic CFT are allowed to have half-integer Lorentz spins.

Next, we consider the RR boundary condition, i.e., $\nu_\text{L}=\nu_\text{R}=0$.  
The RR Hilbert space $\HR$ can be viewed as the $(-1)^F$-twisted Hilbert space from $\HNS$.  
There are two fermion zero modes $\psi_\text{L,0},\psi_\text{R,0}$ which obey the following anti-commutation relations:
\ie
\{ \psi_\text{L,0} , \psi_\text{L,0} \} =
\{ \psi_\text{R,0} , \psi_\text{R,0} \} =2\,,\qquad
\{ \psi_\text{L,0} , \psi_\text{R,0} \} =0\,.
\fe
Since this anti-commutation relation cannot be realized on a one-dimensional representation, there must be degenerate ground states in $\HR$.  
The minimal representation is 2-dimensional. 
We can choose a basis $\{ \ket{\sigma},\ket{\mu}\}$ for the ground space
\ie\label{RRgs}
\ket{\sigma} = \left(\begin{array}{c}1 \\0\end{array}\right) \,,\qquad \ket{\mu} = \left(\begin{array}{c}0 \\1\end{array}\right)\,.
\fe
so that the fermion zero modes and the symmetry operators $(-1)^F,(-1)^{F_\text{L}}$ are represented by the following Pauli matrices:
\ie\label{RRrep}
~& \psi_\text{L,0}  = \sigma^y\,,\qquad && \psi_\text{R,0} = \sigma^x \,,\\
&(-1)^F = \sigma^z \,,\qquad &&(-1)^{F_\text{L}} = \sigma^x\,.
\fe
(As we will discuss momentarily, there is still freedom in redefining these zero modes and symmetry operators.) 
Interestingly, the $\mathbb{Z}_2\times \mathbb{Z}_2^f$ global symmetry is realized projectively on the RR Hilbert space $\HR$:
\ie\label{RRproj}
\text{RR}:~~~ (-1)^F (-1)^{F_\text{L}}  = -  (-1)^{F_\text{L}}(-1)^F\,.
\fe
Note that this sign cannot be removed by redefining the symmetry operators $(-1)^F,(-1)^{F_\text{L}}$. 
This signals an 't Hooft anomaly of the $\mathbb{Z}_2\times \mathbb{Z}_2^f$ symmetry. 
Indeed, it is known that this symmetry of the Majorana CFT has a mod 8 anomaly  classified by the spin cobordism group \cite{Qi:2012gjs,Ryu:2012he,Gu:2013azn,Kapustin:2014dxa,Kaidi:2019pzj}
\ie\label{mod8}
\text{Hom(Tors}\,\Omega^\text{Spin}_3(B\mathbb{Z}_2),U(1))=\mathbb{Z}_8 \,.
\fe  
In fact, this anomaly is related to the superstring spacetime dimensions \cite{Seiberg:1986by,Kaidi:2019pzj}. 
It was argued in  \cite{Delmastro:2021xox} (see also \cite{Thorngren:2018bhj,Grigoletto:2021zyv}) that the minus sign in \eqref{RRproj} reflects a quotient of the full mod 8 anomaly.

Via the operator-state map, the states in $\HR$ are mapped to (non-local) operators attached to a $(-1)^F$ line on the plane. 
More specifically, the two ground states \eqref{RRgs} are mapped to the order and disorder operators $\sigma$ and $\mu$.  
The RR Hilbert space can be graded by the $(-1)^F$ charge as $\HR = \HR^+ \oplus\HR^-$. The Virasoro primaries and their conformal weights in $\HR^\pm$ are:
\ie
\HR^+ :~\sigma,~~(h,\bar h) = \left({1\over16} ,{1\over 16}\right)\,,\\
\HR^- :~\mu,~~(h,\bar h) = \left({1\over16} ,{1\over 16}\right)\,.
\fe
The conformal weights are computed from the Casimir energy in the RR Hilbert space, which can be found in every standard string theory textbook.

In the NSNS Hilbert space $\HNS$, there is a canonical way to determine the normalization of the symmetry operator $(-1)^F$ so that the unique ground state $\ket{1}$ is even. 
This is not the case in the RR Hilbert space $\HR$, where the two ground states $\ket{\sigma},\ket{\mu}$ are on the same footing and are exchanged by another symmetry operator $(-1)^{F_\text{L}}$ (which is represented as $\sigma^x$ in \eqref{RRrep}). 
One can redefine the operator $(-1)^F$ in $\HR$ by an overall sign, so that the two sectors $\HR^+$ and $\HR^-$ are exchanged. 
This corresponds  to changing $(-1)^F=\sigma^z$ to $(-1)^F =-  \sigma^z$ in \eqref{RRrep}. 
More invariantly, this freedom corresponds to multiplying the Majorana CFT by a 1+1d local fermionic counterterm. The latter is known as the Arf invariant in mathematics, and as (the low-energy limit of) the Kitaev chain in condensed matter physics.  
The Arf invariant Arf$[\rho]$ is an invertible fermionic TQFT whose partition function is given by
\ie
(-1)^{\text{Arf}[\rho] }= \begin{cases}
+1\,,\qquad \text{if}~~\rho ~~\text{is even}\,,\\
-1\,,\qquad \text{if}~~\rho ~~\text{is odd}\,,
\end{cases}
\fe
where $\rho$ is the spin structure of the spacetime manifold.

\subsubsection{Bosonization and fermionization}

We now discuss the relation between the Ising and the Majorana CFT. 
While the $\mathbb{Z}_2\times \mathbb{Z}_2^f$ global symmetry has an 't Hooft anomaly and cannot be gauged, the $\mathbb{Z}_2^f$ subgroup generated by $(-1)^F$ is free of anomaly. 
The Ising CFT can be obtained by gauging the $(-1)^F$ global symmetry of the Majorana CFT.  
In this sense, the difference between the Majorana CFT and the Ising CFT is similar to that between the $S^1/\mathbb{Z}_2$ orbifold CFT and the $c=1$ compact boson. 
However, there is an important distinction: bosonization maps a fermionic CFT to a bosonic one, while gauging an ordinary $\mathbb{Z}_2$ global symmetry (sometimes known as orbifolding) maps a bosonic CFT to another bosonic CFT. Below we discuss in detail the difference between bosonization and the conventional orbifold.

We start with a review on gauging a non-anomalous $\mathbb{Z}_2$ global symmetry of a bosonic CFT $\cal B$, such as the Ising CFT.\footnote{Every bosonic CFT has chiral central charge $c_\text{L}- c_\text{R}\in 8\mathbb{Z}$. We will assume this condition throughout.} 
We denote its partition function on a genus $g$ Riemann surface $\Sigma_g$ with $\mathbb{Z}_2$ background gauge field $A\in H^1(\Sigma_g,\mathbb{Z}_2)$ by $Z_{\cal B}[A]$.  
Gauging the $\mathbb{Z}_2$ global symmetry gives another bosonic CFT, denoted by ${\cal B}/\mathbb{Z}_2$. 
When a global symmetry is gauged, it is by definition no longer a global symmetry in the gauged CFT.  
Instead, in 1+1d, the gauged CFT ${\cal B}/\mathbb{Z}_2$ has a new $\mathbb{Z}_2$ global symmetry, which is known as the dual (or quantum) $\mathbb{Z}_2$ symmetry \cite{Vafa:1989ih}. 
(See also \cite{Bhardwaj:2017xup,Tachikawa:2017gyf} for modern perspectives.) 
The symmetry operator/defect of the dual $\mathbb{Z}_2$ symmetry in ${\cal B}/\mathbb{Z}_2$ is the topological Wilson line  for the $\mathbb{Z}_2$ gauge symmetry. 
One important property of the dual $\mathbb{Z}_2$ symmetry is that gauging it in ${\cal B}/\mathbb{Z}_2$ returns the original bosonic CFT $\cal B$, which can be checked explicitly using the above expression.   
More explicitly, the partition function of   the gauged CFT ${\cal B}/\mathbb{Z}_2$ coupled to the background gauge field $A$ of the dual $\mathbb{Z}_2$ symmetry is:
\ie\label{orbifold}
Z_{{\cal B}/\mathbb{Z}_2} [A] 
={1\over 2^g} \sum_{a\in H^1(\Sigma_g, \mathbb{Z}_2)} 
Z_{\cal B}[a]  \, (-1)^{ \oint a\cup A}\,.
\fe
The sum in $H^1(\Sigma_g, \mathbb{Z}_2)$ is the discrete analog of the path integral over the dynamical gauge field when the gauge group is continuous. 

The formula \eqref{orbifold} takes the form of a generalized Fourier transformation. 
In Fourier transformation, we never lose information. 
The same applies to discrete gauging.  
No operator is thrown away under discrete gauging. Rather, some local operators become non-local, and some non-local operators become local. 
We will discuss this more below.

In the special case when $\cal B$ is the Ising CFT, ${\cal B}/\mathbb{Z}_2$ turns out to be isomorphic to the Ising CFT itself. 
One way to see it is that there is a unique $c=\frac12$ unitary CFT by the BPZ classification. 
The invariance of the Ising CFT under gauging the $\mathbb{Z}_2$ is related to the non-invertible symmetry $\CD$ and will be important in Section \ref{sec:half}.

Next, let $Z_{\cal F}[\rho]$ be the partition function of  a fermionic CFT $\cal F$ such as the Majorana CFT on  $\Sigma_g$ with spin structure $\rho$.  
While there is no canonical zero in the space of spin structure, the difference between any two spin structures is a $\mathbb{Z}_2$ gauge field $A \in H^1(\Sigma_g,\mathbb{Z}_2)$. 
Therefore, given a reference spin structure $\rho$, the partition function of the Majorana CFT on any other spin structure $\rho'$ can be written as $Z_{\cal F}[\rho+A]$ for an appropriate choice of $A$. 
Every fermionic CFT has a $(-1)^F$ global symmetry (in its untwisted Hilbert space), which couples to the background gauge field $A$ and shifts the spin structure. 
To obtain a bosonic CFT from  a fermionic CFT, we gauge  $(-1)^F$ to sum over the spin structures, so that the resulting theory is independent of the spin structures. 
In the string theory literature, this is known as the GSO projection \cite{Gliozzi:1976qd,Seiberg:1986by}.\footnote{More precisely, the GSO projection can be understood as bosonization in the BRST quantization. In the lightcone quantization, there are 8 Majorana-Weyl fermions and the chiral central charge is 4. Summing over the spin structure in the lightcone quantization does not give a bosonic CFT, because there isn't any bosonic CFT with chiral central charge 4.  Rather, it again gives a fermionic CFT.  }

After we gauge the $(-1)^F$ symmetry, it is no longer a global symmetry in the resulting bosonic CFT. 
Instead, we have a fermionic version of the dual $\mathbb{Z}_2$ symmetry \cite{Karch:2019lnn,Ji:2019ugf,Lin:2019hks}. 
The partition function of the resulting bosonic CFT $\cal B$ coupled to the  background gauge field $A$ for the  dual $\mathbb{Z}_2$ global symmetry is
\ie\label{bosonization}
Z_{\cal B}[A]  = {1\over 2^g} (-1)^{ \text{Arf}[A+\rho] +\text{Arf}[\rho] }
\sum_{a\in H^1(\Sigma_g,\mathbb{Z}_2) }  
Z_{\cal F}[a+\rho] (-1)^{\oint a\cup A} \,.
\fe
Note that this fermionic dual $\mathbb{Z}_2$ symmetry (which is free of anomaly) is different from the bosonic version in \eqref{orbifold} because of the counterterm $(-1)^{ \text{Arf}[A+\rho] +\text{Arf}[\rho] }$. 
This term is added to ensure that the lefthand side is independent of the choice of the reference spin structure $\rho$ on the righthand side. 
Gauging $(-1)^F$ as in \eqref{bosonization} therefore maps a fermionic CFT to a bosonic one, which is sometimes referred to as the bosonization. 

The inverse map of \eqref{bosonization} is known as the fermionization. 
Starting with any bosonic CFT $\cal B$ with a non-anomalous $\mathbb{Z}_2$ global symmetry, we can couple it to the invertible fermionic TQFT (i.e., the Arf invariant) in the following way to produce a fermionic CFT $\cal F$:
\ie\label{fermionization}
Z_{\cal F}  [A+\rho ] =  {1\over 2^g} (-1)^{\text{Arf}[\rho]} \sum_{a\in H^1(\Sigma_g,\mathbb{Z}_2)}Z_{\cal B}[a] (-1)^{\text{Arf}[a+\rho] + \oint a\cup A} \,.
\fe

\begin{figure}[h!]
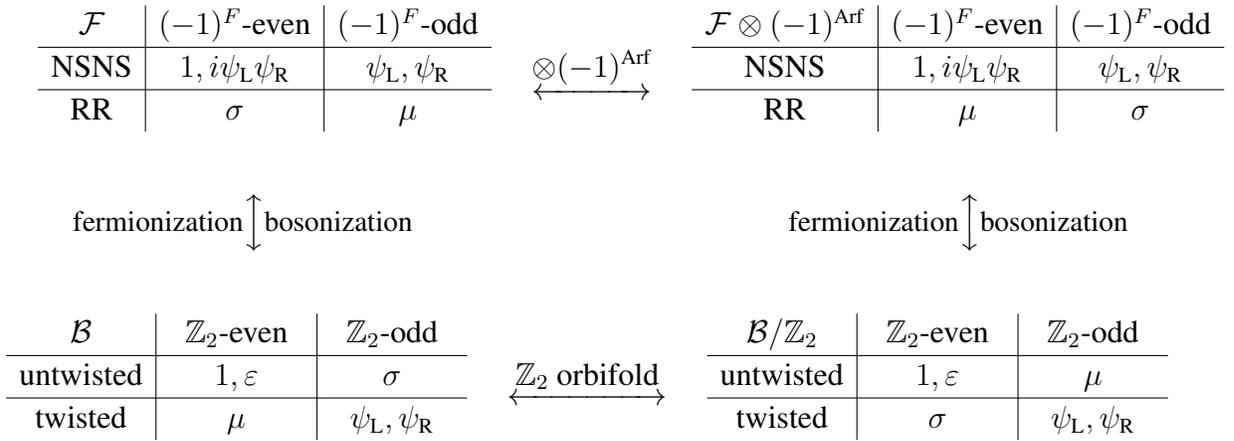

\begin{align*}
~&\quad
\left.\begin{array}{c|c|c}{\cal F} & (-1)^F\text{-even}&(-1)^F\text{-odd}\\\hline   \text{NSNS}& 1, i \psi_\text{L}\psi_\text{R} & \psi_\text{L}, \psi_\text{R} \\\hline    \text{RR} & \sigma & \mu\end{array}\right.
\quad
\underleftrightarrow{\otimes (-1)^\text{Arf}} \quad
\left.\begin{array}{c|c|c}{\cal F}\otimes (-1)^\text{Arf} & (-1)^F\text{-even}&(-1)^F\text{-odd}\\\hline   \text{NSNS}& 1, i\psi_\text{L}\psi_\text{R} & \psi_\text{L},\psi_\text{R} \\\hline    \text{RR} & \mu & \sigma\end{array}\right.\\
&~~\\
&\hspace{1cm}\text{\small fermionization}\Big\updownarrow \text{\small bosonization}\hspace{5cm}
\text{\small fermionization}\Big\updownarrow \text{\small bosonization}\\
&~~\\
&\left.\begin{array}{c|c|c}{\cal B} & ~~\mathbb{Z}_2\text{-even}~~&~~\mathbb{Z}_2\text{-odd}~~\\\hline   \text{untwisted}& 1,\varepsilon & \sigma \\\hline    \text{twisted} & \mu & \psi_\text{L}, \psi_\text{R}\end{array}\right.
\quad
\underleftrightarrow{~\mathbb{Z}_2 \text{~orbifold}~} \quad
\left.\begin{array}{c|c|c}{\cal B}/\mathbb{Z}_2 &~~ \mathbb{Z}_2\text{-even}~~&~~\mathbb{Z}_2\text{-odd}~~\\\hline   \text{untwisted}& 1,\varepsilon & \mu \\\hline    \text{twisted} & \sigma & \psi_\text{L}, \psi_\text{R}\end{array}\right.\\
~~
\end{align*}
\caption{The maps between the Hilbert spaces of the bosonic and fermionic CFTs under the $\mathbb{Z}_2$ orbifold \eqref{orbifold}, bosonization \eqref{bosonization}, and fermionization \eqref{fermionization}. Here ``untwisted" stands for the untwisted Hilbert space  $\CH$ of the bosonic CFT, whose states are in one-to-one correspondence with the local operators. On the other hand, ``twisted" stands for the Hilbert space $\CH_\eta$ twisted by the $\mathbb{Z}_2$ global symmetry, whose states are in correspondence with point operators attached to the $\mathbb{Z}_2$ topological line. For the fermionic CFTs, ``NSNS" stand for the Hilbert space $\HNS$ where all fermions are anti-periodic along the spatial circle, while ``RR" stands for the Hilbert space  $\HR$ twisted by $(-1)^F$. In each entry we write the corresponding (local or non-local) operators in the Hilbert space for the special case when $\cal F$ is the Majorana CFT and $\cal B$ is the Ising CFT. The thermal operator $\varepsilon$ is identified with $i\psi_\text{L}\psi_\text{R}$. }\label{fig:bosonization}
\end{figure}

The three maps, the $\mathbb{Z}_2$ orbifold \eqref{orbifold}, bosonization \eqref{bosonization}, fermionization \eqref{fermionization}, relate two bosonic CFTs $\cal B$, ${\cal B}/\mathbb{Z}_2$, and two fermionic CFTs $\cal F$, ${\cal F}\otimes \text{Arf}$ together.\footnote{One can view the two fermionic CFTs, $\cal F$ and ${\cal F}\otimes \text{Arf}$, equivalent since they only differ by a 1+1d fermionic counterterm, the  Arf invariant.} 
On a spacetime torus, the maps above relate the Hilbert spaces of the bosonic CFTs to the those of the fermionic CFTs. More specifically, the bosonic CFT $\cal B$ has four sectors of Hilbert spaces: the $\mathbb{Z}_2$-even/odd sectors of the untwisted Hilbert spaces, $\CH=\CH^+\oplus \CH^-$ in \eqref{CHpm}, and the $\mathbb{Z}_2$-even/odd sectors of the $\mathbb{Z}_2$-twisted Hilbert spaces $\CH_\eta=  \CH_\eta^+ \oplus\CH_\eta^-$ in \eqref{CHetapm}. 
The fermionic CFT $\cal F$ on the other hand also has four sectors of their Hilbert spaces: the $(-1)^F$-even/odd sectors of the NSNS and RR Hilbert spaces, $\HNS=\HNS^+ \oplus\HNS^-$, $\HR= \HR^+\oplus\HR^-$. 
These  sectors of Hilbert spaces of the bosonic $\cal B$ and fermionic CFTs $\cal F$ are mapped into each under bosonization as:
\ie\label{Hmap}
\CH^+  = \HNS^+ \,,\qquad \CH^-  = \HR^+\,,\qquad \CH_\eta^+ = \HR^- \,,\qquad \CH_\eta^- = \HNS^-\,.
\fe 
See Figure \ref{fig:bosonization} for more details. 

\subsubsection{Non-invertible symmetry from the mixed anomaly}

Having discussed the general bosonization/fermionization in 1+1d, we now track the global symmetries in the bosonic and fermionic CFTs. 

The $(-1)^F$ global symmetry of Majorana CFT is gauged under bosonization, and is therefore no longer a global symmetry in the Ising CFT.  
The $\mathbb{Z}_2$ global symmetry $\eta$ is the (fermionic) dual $\mathbb{Z}_2$ symmetry of $(-1)^F$, but they never coexist in a 1+1d theory.  
The dual symmetry $\eta$ acts as $-1$ on the twisted sector states, which in this case include the order operator $\sigma$.

What happens to $(-1)^{F_\text{L}}$ when we gauge $(-1)^F$? 
It is trying to be a $\mathbb{Z}_2$ symmetry that flips the sign of $\varepsilon$ (which is identified with $i\psi_\text{L}\psi_\text{R}$ under bosonization). 
However, the Ising fusion rule \eqref{Isinglocal} forbids such a $\mathbb{Z}_2$ global symmetry.  
Therefore $(-1)^{F_\text{L}}$ cannot turn into an ordinary $\mathbb{Z}_2$ symmetry in the Ising CFT. 
Indeed, we know there is only one $\mathbb{Z}_2$ symmetry $\eta$ in the Ising CFT, not two. 

Furthermore, in the Majorana CFT, the $(-1)^{F_\text{L}}$ symmetry exchanges  the order  $\sigma$ and disorder operator $\mu$ since $(-1)^{F_\text{L}}=\sigma^x$ in \eqref{RRrep}. 
However, in the Ising CFT, while $\sigma$ is a local operator, $\mu$ is a non-local operator associated with the $\mathbb{Z}_2$ twisted Hilbert space.

The source of the problem is that there is a mixed anomaly between the $(-1)^F$ and the $(-1)^{F_\text{L}}$ symmetries. This can be seen  from the projective sign in \eqref{RRproj}. 
The mixed anomaly turns the invertible $(-1)^{F_\text{L}}$ of the Majorana CFT into the non-invertible duality symmetry $\CD$ under bosonization \cite{Thorngren:2018bhj,Ji:2019ugf,Kaidi:2021xfk}. 
Indeed, the non-invertible symmetry $\CD$ flips the sign of the thermal operator $\varepsilon$ under the Euclidean process in Figure \ref{fig:duality}(a). 
In addition, the Euclidean process of $\CD$ in Figure \ref{fig:duality}(b) exchanges the local, order operator $\sigma$ with the non-local, disorder operator $\mu$, consistent with the action of $(-1)^{F_\text{L}}=\left(\begin{array}{cc}0 & 1 \\1 & 0\end{array}\right)$ on the ground space in $\HR$ of the Majorana CFT before bosonization.  
We will discuss the lattice counterpart  of this relation between $(-1)^{F_\text{L}}$ and $\CD$ in Section \ref{sec:TFIM}.

To summarize, under bosonization, the $(-1)^F$ symmetry of the Majorana CFT is dual to the $\mathbb{Z}_2$ symmetry $\eta$ of the Ising CFT. 
On the other hand, the $(-1)^{F_\text{L}}$ symmetry of the Majorana CFT turns into the non-invertible symmetry $\CD$ of the Ising CFT.

\subsubsection{Duality twisted Hilbert space from NSR and RNS Hilbert spaces}

Under the bosonization/fermionization, we have related the untwisted  and $\mathbb{Z}_2$-twisted Hilbert spaces $\CH,\CH_\eta$ of the Ising CFT to the NSNS and RR Hilbert spaces $\HNS,\HR$ of the Majorana CFT.  
We now extend this discussion to the duality twisted Hilbert space $\CH_\CD$ of the Ising CFT and the NSR and RNS Hilbert spaces of the Majorana CFT.

In the RNS Hilbert space $\CH_\text{RNS}$ of the Majorana CFT (corresponding to $\nu_\text{L} = 0, \nu_\text{R}=\frac12$), there is  a single fermion zero mode $\psi_\text{L,0}$.  
We can canonically quantize the RNS theory by demanding that the single fermion zero mode acts on the unique ground state as
\ie\label{RNScanonical}
\psi_\text{L,0} \ket{\Omega}_\text{RNS} = \ket{\Omega}_\text{RNS}\,.
\fe
There are two Virasoro primary states, $\ket{\Omega}_\text{RNS}$ and $\chi_{\text{R}, -\frac12}\ket{\Omega}_\text{RNS}$, in the RNS Hilbert space with the following conformal weights:
\ie
\CH_\text{RNS}:~~\left({1\over 16},  0\right) \,,\qquad \left(\frac {1}{16}, {1\over 2}\right)\,.
\fe
Importantly, we  longer have the $(-1)^F$ operator in $\CH_\text{RNS}$ because it is incompatible with  \eqref{RNScanonical}.  
On the other hand, we still have $(-1)^{F_\text{R}}$, under which $\ket{\Omega}_\text{RNS}$ is even and $\chi_{\text{R}, -\frac12}\ket{\Omega}_\text{RNS}$ is odd.

The problem with an odd number of fermion zero mode is known to be subtle. 
For instance, the computation of the torus partition function of the RNS theory from the functional integral disagrees with the trace over the Hilbert space $\CH_\text{RNS}$ obtained from canonical quantization by a factor of $\sqrt{2}$. 
We will not discuss this here and refer the readers to \cite{Stanford:2019vob,Delmastro:2021xox,Witten:2023snr,Seiberg:2023cdc} for recent discussions.

The RNS Hilbert space can be obtained from a $(-1)^{F_\text{L}}$ twist of the NSNS Hilbert space $\HNS$. 
The fractional Lorentz spins 
\ie\label{1/16}
\text{RNS}:~~~h-\bar h \in {1\over 16} +{\mathbb{Z}\over2}
\fe
of the states in $\CH_\text{RNS}$ provide another way to see the mod 8 anomaly of $(-1)^{F_\text{L}}$ in \eqref{mod8} \cite{yujitasi,Delmastro:2021xox,Grigoletto:2021zyv}. (See \cite{Chang:2018iay,Lin:2019kpn,Lin:2021udi} for the corresponding discussions in bosonic CFT.) 
One way to see the obstruction in gauging $(-1)^{F_\text{L}}$ is that none of the states in the twisted Hilbert space $\CH_\text{RNS}$ can be ``promoted" to a local operator in either a bosonic or fermionic CFT because of their fractional spins \eqref{1/16}.  
However, if we have 8 copies of the system, then the RNS states have Lorentz spins in $\mathbb{Z}/2$, which are consistent with the spins of local operators in a fermionic CFT. 
Indeed, the combined chiral symmetry of 8 Majorana fermions is gaugeable. 
In fact, gauging it returns 8 Majorana fermions and implements a triality transformation.  
If, on the other hand, we take 16 Majorana fermions, then the RNS states have integer spins. 
Gauging the combined chiral symmetry then gives a bosonic CFT, which is the $(E_8)_1$ WZW model.

The NSR Hilbert space can be quantized in a similar way. It has two Virasoro primary states:
\ie
\CH_\text{NSR}:~~\left(0 , {1\over 16} \right) \,,\qquad \left({1\over 2}, {1\over 16}\right)\,.
\fe
Comparing with the duality-twisted Hilbert space $\CH_\CD$ \eqref{ZD} in the Ising CFT, we find 
\ie\label{HDNSR}
\CH_\CD = \CH_\text{NSR} \oplus\CH_\text{RNS}\,.
\fe

\subsection{Transverse-field Ising lattice model}\label{sec:TFIM}

We now discuss the microscopic origin of the non-invertible symmetry $\CD$ of the Ising CFT. 
In this subsection we discuss the transverse-field Ising model, which is a Hamiltonian lattice model. 
At the critical point, there is a second-order phase transition described by the Ising CFT in the thermodynamic limit. 
Our discussion follows \cite{Seiberg:2023cdc,Seiberg:2024gek} closely.

Consider a  periodic 1d chain of $N$ sites with a 2-dimensional qubit $\CH_j$ on every site $j=1,\cdots, N$. 
The total Hilbert space $\CH$ is $2^N$-dimensional and is a tensor product of the local Hilbert spaces, 
\ie
\CH = \bigotimes_{j=1}^N \CH_j\,.
\fe
On each site, the operators are generated by the Pauli matrices $X_j,Z_j$ obeying $X_j^2= Z_j^2=1, X_j Z_j = - Z_j X_j$, while $X_j,Z_j$ with different $j$'s commute with each other. 
In each local Hilbert space, our convention is that $X_j \ket{\pm }_j = \pm \ket{\pm}_j$.

The Hamiltonian of the ferromagnetic transverse-field Ising model is  
\ie\label{TFIM}
H  = - g \sum_{j=1}^N X_j -  \sum_{j=1}^N Z_j Z_{j+1}\,,
\fe
where we impose the periodic boundary condition and identify $X_{N+1}=X_1,Z_{N+1}=Z_1$. 
The sign of $g$ can be flipped by a field redefinition. Without of generality,   we assume $g>0$ below.   

The Ising model has a $\mathbb{Z}_2$ spin-flip symmetry for any $g$. 
On the one hand, it leads to a conserved, unitary operator
\ie\label{etaTFIM}
\eta = \prod_{j=1}^N X_j\,,
\fe
that commutes with the Hamiltonian $[\eta,H]=0$ and obeys a $\mathbb{Z}_2$ algebra, i.e., $\eta^2=1$. 
On the other hand, it leads to a defect which gives a  twisted Hamiltonian
\ie\label{Heta}
H_\eta =- g \sum_{j=1}^N X_j - \sum_{j=1}^{N-1}Z_j Z_{j+1} + Z_N Z_1 \,.
\fe
The sign flip of the interaction on the $(N,1)$-link  is a defect localized at  a point in space and extends in the time direction. 
The twisted Hamiltonian $H_\eta$ has a different energy spectrum compared to the original Hamiltonian $H$.

At a generic $g$, there is also a $\mathbb{Z}_N$ lattice translation symmetry $T_\text{Ising}$ obeying $T_\text{Ising}^N=1$ and $T_\text{Ising} H = HT_\text{Ising}$. 
Explicitly, it is given by \cite{Grimm:1992ni}\footnote{Throughout this paper, the symbol $\prod_{j=1}^L$ is defined as the ordered product.  That is, $\prod_{j=1}^L a_j = a_1a_2 \cdots a_L$. While this does not make any difference for products of $c$-number, it matters for products of matrices. }
\ie
&T_\text{Ising}  = \prod_{j=1}^{N-1} t_j^\text{Ising}\,,\\
&t^\text{Ising}_j= \frac12 \left( X_j X_{j+1} +Y_j Y_{j+1} +Z_j Z_{j+1} +1\right) \,.
\fe
The overall phase of this operator is chosen so that $T_\text{Ising}\ket{++\cdots+}=\ket{++\cdots+}$.

The spectrum of the original untwisted Hamiltonian $H$ is gapped with a non-degenerate ground state when $g>1$. This is the disordered phase where the $\mathbb{Z}_2$ is unbroken. 
It is gapped with a 2-fold degenerate ground space when $0<g<1$. This is the ordered phase where the $\mathbb{Z}_2$ is spontaneously broken. 
At $g=1$, there is a second order phase transition  described by the Ising CFT in the thermodynamic limit. 

In a Hamiltonian lattice model, a conserved operator and a topological defect are rather different. 
The former is an operator that commutes with the Hamiltonian and acts on the original Hilbert space $\CH$. 
For discussions on the relation between operators and defects on  the lattice, see, for example, \cite{Aasen:2016dop,Aasen:2020jwb,Cheng:2022sgb,Seiberg:2023cdc,Seifnashri:2023dpa}.
We will focus more on the operator and briefly discuss the associated defect.

\subsubsection{Non-invertible lattice operator}

At the critical point $g=1$, is there an additional operator in the transverse-field Ising model that obeys the following requirements?
\begin{enumerate}
\item It is an operator that acts on the $2^N$-dimensional Hilbert space $\CH$.
\item It commutes with the Hamiltonian $H$.
\item It flows to the non-invertible operator $\CD$ in the continuum Ising CFT.
\end{enumerate}

Naively, at $g=1$, there is a \textit{transformation} of the $\bZ_2$-even local operators that leaves the Hamiltonian invariant:
\ie\label{KWtransformation}
X_j \rightsquigarrow Z_j Z_{j+1} \,,~~~
Z_j Z_{j+1}\rightsquigarrow X_{j+1}\,,~~~j=1,\cdots,N\,.
\fe
This is known as the Kramers-Wannier transformation. 
Is this transformation implemented by an invertible operator $U$? 
Suppose it were, then we would have $U X_j U^{-1} =Z_jZ_{j+1}$. 
However, this immediately leads to 
$U\eta U^{-1} = U\prod_{j=1}^N X_j U^{-1} = \prod_{j=1}^N Z_jZ_{j+1}=1$, implying $\eta=1$, which is  a contradiction. 
Therefore, the arrow $\rightsquigarrow$ above cannot possibly be implemented by an invertible operator. 
This gives us a hint that the Kramers-Wannier transformation is associated with an unconventional symmetry at the critical point.
So what are the precise equations for \eqref{KWtransformation}? 

We first present the answer for this symmetry, and then derive it later. 
Define a unitary (and in particular, invertible) operator \cite{PhysRevB.91.195143,2019ScPP....6...29H,Tan:2022vaz,Chen:2023qst}
\ie\label{UKW}
U_\text{KW} =  e^{- {2\pi i N\over 8}}\left( \prod_{j=1}^{N-1} 
{1+iX_j \over \sqrt{2}} {1+iZ_jZ_{j+1}\over \sqrt{2}} \right)
{1+iX_N\over \sqrt{2}}\,.
\fe
This operator is not translationally invariant. 
It acts on the local terms of the Hamiltonian $H$ as
\ie
~&U_\text{KW} X_j U_\text{KW}^{-1} 
= Z_j Z_{j+1}  \,,~~~~j = 1,2,\cdots, N-1\,,\\
&U_\text{KW} X_N U_\text{KW}^{-1} =   \eta Z_N Z_1 \,,\\
&U_\text{KW} Z_j  Z_{j+1} U_\text{KW}^{-1}    = X_{j+1} \,,~~~~j = 1,2,\cdots, N-1\,,\\
&U_\text{KW} Z_N  Z_1 U_\text{KW}^{-1}  =\eta X_1\,.
\fe
Because of the term $\eta$ in the action around  site $j=N$, $U_\text{KW}$ does \textit{not} commute with $H$. It is not a symmetry. 

To fix this, we multiply $U_\text{KW}$ by a $\mathbb{Z}_2$ projector and define a new operator $\mD$ acting on $\CH$ \cite{Seiberg:2023cdc,Seiberg:2024gek}:
\ie\label{mD}
\mD  = U_\text{KW} {1+\eta\over2} \,.
\fe
Because of the projector, $\mD$ is not invertible; it annihilates all the $\mathbb{Z}_2$-odd states. 
What we gain is that $\mD$ is translationally invariant (i.e., $T\mD= \mD T$) and acts on the local operators uniformly as
\ie
\mD X_j  =  Z_j Z_{j+1} \mD \,,~~~~~
\mD Z_jZ_{j+1} = X_{j+1} \mD\,,~~~~j=1,\cdots, N\,,
\fe
with $X_{N+1}=X_1, Z_{N+1}=Z_1$. 
This gives the precise equations for $\rightsquigarrow$ in \eqref{KWtransformation}. 
It follows that
\ie
\mD H = H \mD \,,\qquad \text{for}~~g=1\,.
\fe

Thus, the critical Ising lattice model has the following list of conserved operators: the $\mathbb{Z}_2$ symmetry  $\eta$,  the lattice translation symmetry $T_\text{Ising}$, and the non-invertible operator $\mD$. Together they obey the following algebra \cite{Seiberg:2023cdc,Seiberg:2024gek}\footnote{The phase $e^{-{2\pi i N\over8}}$ in \eqref{UKW} is chosen so that there is no phase in the first line of \eqref{noninvfusion}.}
\ie\label{noninvfusion}
&\mD^2 =\frac12 (1 +\eta) \, T_\text{Ising} \,,\\
&\eta^2=1\,,~~~~\mD \, \eta = \eta\, \mD= \mD\,,\\
&T_\text{Ising}^N=1\,,~~~T_\text{Ising}  \,\mD = \mD\,T_\text{Ising} \,,~~~T_\text{Ising} \, \eta= \eta \,T_\text{Ising}\,.
\fe
Since $\mD^2$ involves the lattice translation  $T_\text{Ising}$ by one Ising site, $\mD$ appears as a ``half-translation" (see for instance \cite{PhysRevB.3.3918}).  
More precisely, it is non-invertible and only squares to the lattice translation on the $\mathbb{Z}_2$-even states.

Note the conserved symmetry operator $\mD$ cannot be written as a product of local operators.\footnote{Instead, $\sqrt{2}\mD$ is a \textit{matrix product operator} with bond dimension 2 \cite{Seiberg:2024gek}. See also \cite{Tantivasadakarn:2021vel,PhysRevX.5.011024,Lootens:2021tet,Lootens:2022avn}.} 
This is to be contrasted with the invertible $\bZ_2$ operator \eqref{etaTFIM}.
 Rather, $\mD$ is a sum of two products of local operators, but separately each term in the sum does not commute with the Hamiltonian.

What is the relation between this lattice algebra \eqref{noninvfusion} and the continuum one \eqref{Isingcat}? 
First, the normalization of $\mD$ as a conserved operator is undetermined without further inputs from the information of the defect. 
We could have redefined $\mD$ by a factor of $\sqrt{2}$, but here we chose a different normalization that is natural from the fermionic viewpoint as we will discuss later. 
Second, in addition to the normalization, the lattice algebra mixes with the lattice translation  symmetry $T_\text{Ising}$ in an interesting way.  
On the low-lying states in the large $N$ limit, the lattice operator $\mD$ is related to the continuum one $\CD$ as
\ie
\mD = {1\over \sqrt{2}} \CD e^{2\pi i (L_0 - \bar L_0 )\over 2N}\,.
\fe
For the low-lying states, this relation is exact.\footnote{In the terminology of \cite{Cheng:2022sgb,Seiberg:2023cdc}, $\CD$ of the Ising CFT is a non-invertible emanant symmetry. An emanant symmetry of the IR continuum field theory emanates from a fixed source from the microscopic model, such as the lattice translation symmetry. In contrast, an emergent symmetry of the IR continuum field theory emerges out of nowhere from the microscopic model and is typically violated by irrelevant local operators. } 
In the thermodynamic limit $N\to \infty$, $T_\text{Ising}\sim 1$ on the low-lying states, and the algebra \eqref{noninvfusion} reduces to the continuum one \eqref{Isingcat} (up to rescaling the operator $\mD$).

We emphasize that the operator $\mD$ acts on a single, $2^N$-dimensional Hilbert space. 
This is different from, but related to, the map considered in \cite{Aasen:2016dop,Tantivasadakarn:2021vel,Tantivasadakarn:2022hgp,Li:2023mmw,Cao:2023doz,Li:2023knf}. 
The authors of these references define a map from the Hilbert space on the sites to another Hilbert space on the link, and vice versa. 
Furthermore, the algebra of their maps is different from \eqref{noninvfusion} and does not involve the lattice translation.

\subsubsection{Bosonization on the lattice}

We now derive the operator $\mD$ by imitating the  steps in the continuum bosonization  in Section \ref{sec:bosonization}.  
Locally, the lattice bosonization is achieved by the famous Jordan-Wigner transformation. But globally, this is not the full story. 
Below we will pay special attentions to the global issues and track the global symmetries carefully.  
Again, our presentation follows \cite{Seiberg:2023cdc}, which is closely related to the earlier discussions in \cite{Baake:1988vg,Grimm:1992ni,Grimm:2001dr,Grimm:2002rd,Hsieh:2020uwb}.

Consider a 1d closed chain of $L$ sites indexed by $\ell=1,\cdots, L$.  
We start with the even $L=2N$ case so the Hilbert space is $2^N$-dimensional. 
On each site there is a real fermion $\chi_\ell$ with the following anti-commutation relation
\ie
\{ \chi_\ell ,\chi_{\ell'} \} = 2\delta_{\ell, \ell'} \,.
\fe
The Majorana chain has been discussed extensively in the literature \cite{Kitaev:2000nmw,2010PhRvB..81m4509F,2011PhRvB..83g5103F,2015PhRvB..92w5123R,Hsieh:2016emq,OBrien:2017wmx}. Here we focus on two Hamiltonians:
\ie
H_\pm = i  \sum_{\ell=1}^{N-1} \chi_{\ell+1}\chi_\ell \pm i\chi_1\chi_N\,.
\fe
But we emphasize that the construction below applies to more general Hamiltonians with the same symmetry. For instance, one can add to $H_+$   a four-Fermi interaction $\sum_\ell \chi_\ell \chi_{\ell+1}\chi_{\ell+2}\chi_{\ell+3}$. 

Let us now discuss the global symmetry of these Hamiltonians. 
Both $H_\pm$ have a fermion parity $(-1)^F: \chi_\ell \to-\chi_\ell$ symmetry:
\ie\label{latfp}
(-1)^F = i^N \chi_1 \cdots \chi_{2N}\,.
\fe
The phase $i^N$ is chosen so that $((-1)^F)^2=1$. 
They also have   translation symmetries $T_\pm$ that map $\ell\to \ell+1$.  More precisely, $T_+ :\chi_\ell \to \chi_{\ell+1}$ and $T_-: \chi_\ell \to \chi_{\ell+1}$ for $\ell=1,\cdots, L-1$ and $\chi_L \to -\chi_1$.  
They can be written in terms of the Majorana fermions as
\ie
~&T_ + = {e^{2\pi i (N-1)\over8}\over 2^{2N-1\over 2}}\chi_1(1+\chi_1\chi_2)(1+\chi_2\chi_3)\cdots (1+\chi_{2N-1}\chi_{2N})\,,\\
&T_ - =  {e^{-{2\pi i N\over8}}\over 2^{2N-1\over 2}}(1-\chi_1\chi_2)(1-\chi_2\chi_3)\cdots (1-\chi_{2N-1}\chi_{2N})\,.
\fe
The phases are chosen so that $T_+^{2N}=1$ and $T_-^{2N} = (-1)^F$. 
From the expression in \eqref{latfp} and the way $T_\pm$ acts on the fermions, it is easy to find \cite{2015PhRvB..92w5123R,Hsieh:2016emq}
\ie\label{latticeproj}
&T_+ (-1)^F = -(-1)^F T_+\,,\\
&T_- (-1)^F =  (-1)^F T_-\,.
\fe
The minus sign in the first line will be crucial below. 

In the continuum limit, for even $L=2N$,  the Hamiltonians $H_+$ and $H_-$ flow to the Majorana CFT with RR and NSNS boundary conditions, respectively.  
The fermion parity $(-1)^F$ is realized manifestly on the lattice.  
On the other hand, it is well-known that the chiral fermion parity $(-1)^{F_\text{L}}$ (and similarly $(-1)^{F_\text{R}}$) of the Majorana CFT   arises from the lattice translation symmetry of the Majorana chain.  
To see this, we note that the left- and right-moving modes of the continuum Majorana CFT arise from the momentum modes on the lattice around $k=N$ and $k=0$, respectively. (Here we normalize the lattice momentum so that $k$ is identified with $k+2N$.) 
The lattice translation operator acts on the momentum mode labeled by $k$ with eigenvalue $e^{2\pi i k\over 2N}$. 
Therefore, the lattice translation symmetry acts with an extra minus sign on the modes around $k=N$ compared to those around $k=0$.

In the terminology of \cite{Cheng:2022sgb,Seiberg:2023cdc}, $(-1)^{F_\text{L}}$  
is an emanant symmetry that arises from the Majorana lattice translation $T_\pm$.   
In particular, we recognize that the lattice algebra in \eqref{latticeproj} agrees with the continuum ones in \eqref{RRproj} and \eqref{NSNSlinear}. 
In the large $N$ limit, the low-lying states obey the following exact relations between the lattice operator on the lefthand side and the continuum operators in the RR and NSNS theores:
\ie
& T_\pm = (-1)^{F_\text{L}}  e^{2\pi i (L_0 -\bar L_0) \over 2N} \,.
\fe

In the continuum, we have seen that the non-invertible operator $\CD$ of the Ising CFT arises from the chiral fermion parity $(-1)^{F_\text{L}}$ under bosonization. Below we imitate this on the lattice by tracking the Majorana translation symmetries $T_\pm$ under the lattice bosonization. 

The first step of the lattice bosonization is  the Jordan-Wigner transformation, which   maps the fermionic variables $\chi_\ell$ to the bosonic variables  in terms of the Pauli matrices $\sigma^a_j$:
\ie
&\chi_{2j-1} =  \left(\prod_{j'=1}^{j-1} \sigma_{j'}^x\right) \sigma^z_j\,,~~~~
\chi_{2j} =\left( \prod_{j'=1}^{j-1}\sigma_{j'}^x \right)\sigma^y_j\,,
\fe
for $j=1,\cdots, N$.  The free fermion Hamiltonians $H_\pm$ written in terms of the bosonic variables are
\ie
H_\pm = - \sum_{j=1}^N \sigma_j^x - \sum_{j=1}^{N-1} \sigma_{j}^z \sigma_{j+1}^z \pm  (-1)^F\sigma^z_N\sigma_1^z\,.
\fe
However, this is not the end of the story. 
While these Hamiltonians look almost like the transverse-field Ising Hamiltonian \eqref{TFIM}, there is a problem with the last link where there is a non-local term $(-1)^F = \prod_{j=1}^N \sigma^x_j$.

To obtain a local Hamiltonian, we take the direct sum of two copies of the $2^N$-dimensional Hilbert space for the Majorana chain, denoted as $\CH_\pm$, to obtain a bigger Hilbert space 
\ie
\widetilde \CH= \CH_- \oplus \CH_+
\fe
 of $2^{N+1}$-dimensional. 
We define the Hamiltonian on $\widetilde\CH$ as $\widetilde H  = \left(\begin{array}{cc}H_- & 0 \\0 & H_+\end{array}\right)$, where each entry is a $2^N\times 2^N$ matrix. 
In this bigger Hilbert space, there is a new $\mathbb{Z}_2$ symmetry\footnote{The $\mathbb{Z}_2\times \mathbb{Z}_2$ symmetry generated by $(-1)^F$ and $\widetilde\eta$ in $\widetilde\CH$ is the ``categorical symmetry" of \cite{Ji:2019jhk}. However, the system with the Hamiltonian $\widetilde H$ on the $2^{N+1}$-dimensional Hilbert space is not a local 1+1d system, but the boundary of a 2+1d system. See \cite{2019arXiv190205957J} for the latter perspective.}
\ie
\widetilde\eta =  \left(\begin{array}{cc}1& 0\\0 & -1\end{array}\right).
\fe 
To imitate the continuum relation $\CH = \HNS^+ \oplus \HR^-$, we project $\widetilde \CH$ to the 
\ie
\tilde \eta (-1)^F=+1
\fe
 sector.\footnote{Comparing with the convention in \eqref{Hmap} in the continuum, here we flip the $(-1)^F$ eigenvalues in the RR Hilbert space by stacking the Majorana CFT with an Arf invariant.}  
In this projected, $2^N$-dimensional Hilbert space 
\ie
\CH = \widetilde\CH\Big|_{(-1)^F\widetilde\eta=1},
\fe
 the Hamiltonian becomes that of the transverse-field Ising model \eqref{TFIM}, with  $Z_j = \left(\begin{array}{cc}0& \sigma_j^z\\\sigma_j^z & 0\end{array}\right), 
X_j = \left(\begin{array}{cc}\sigma^x_j& 0\\ 0 & \sigma^x_j\end{array}\right)$. 
This completes the lattice bosonization.

What happens to the Majorana translation operator $T_\pm$ after the lattice bosonization? First, one can show that the square of the Majorana translation becomes the translation of the Ising chain:
\ie
T_\text{Ising} = \left(\begin{array}{cc}T_-^2& 0\\0 & T_+^2\end{array}\right)\Big|_{\CH}\,,
\fe
where $|_\CH$ denotes the restriction to the projected Hilbert space $\CH$. 
However, the operator $\left(\begin{array}{cc}T_-& 0\\0 & T_+\end{array}\right)$ does not commute with $(-1)^F \widetilde\eta$ because of the minus sign in \eqref{latticeproj}. Therefore, it does not act in the Ising Hilbert space $\CH$.

To get rid of the problematic $T_+$, we define
\ie
\mD = \left(\begin{array}{cc}T_-& 0\\0 & 0\end{array}\right) \Big|_{\CH} \,.
\fe
Written in terms of the bosonic variables $X_j,Z_j$, this is precisely the non-invertible lattice  operator introduced in \eqref{mD}.

Having discussed the conserved operator $\mD$ of the transverse-field Ising model, we now briefly discuss the corresponding topological defect. 
The Ising Hamiltonian twisted by this non-invertible defect is given by \cite{Schutz:1992wy,Grimm:1992ni,Ho:2014vla,Hauru:2015abi,Aasen:2016dop,Seiberg:2024gek}
\ie\label{HD}
H_\mD = -\sum_{j=2}^{N} ( X_j + Z_{j-1}Z_{j})  -Z_N X_1\,.
\fe
This defect can also be obtained from a lattice version of the relation \eqref{HDNSR} by bosonizing the Majorana chain with an odd number of fermions \cite{Aasen:2016dop,Seiberg:2023cdc}.   
We refer the readers to \cite{Hauru:2015abi,Seiberg:2024gek} for the non-invertible fusion of the topological defect associated with $\mD$ on the lattice. 
In summary, we have three Hamiltonians $H,H_\eta,H_\mD$ in \eqref{TFIM}, \eqref{Heta}, \eqref{HD} twisted by the topological defects on the lattice. These three defects  flow to  $I,\eta, \CD$ in the continuum Ising CFT.

\subsection{Aasen-Fendley-Mong model and anyonic chain}

In this section we briefly review the statistical AFM model \cite{Aasen:2020jwb} and the anyonic chain, which is  a quantum Hamiltonian lattice model. 
These models realize a general fusion category symmetry on the lattice. 
We refer the readers to  Section 1 of \cite{Inamura:2023qzl} for a concise review of these models.

\subsubsection{General fusion category}

The statistical Ising model can be viewed as the simplest example of the  general AFM model \cite{Aasen:2016dop}. 
To realize  the duality defect $\CD$, it is crucial that the construction involves both the original 2d spacetime lattice and the dual lattice. 
The duality defect is introduced as a 1d locus where the Boltzmann weights are modified in the neighborhood of the locus. 
The defect is topological in the sense that the partition function is invariant under small deformation of the locus of the defect. 
More generally, given any fusion category $\CC$ (together with a choice of some objects and morphisms), there is   a statistical AFM model realizing the objects of $\CC$ as topological defects  \cite{Aasen:2020jwb}.

 \begin{figure}[h!]
 \centering
 \includegraphics[width=.5\textwidth]{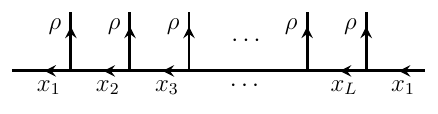}
 \caption{The fusion tree that defines the Hilbert space of a  closed periodic, anyonic chain.   }\label{fig:fusiontree}
 \end{figure}

Let us focus on the Hamiltonian lattice model obtained from taking an anisotropic limit of the statistical AFM model. 
The resulting 1+1d Hamiltonian lattice model is known as the  anyonic chain \cite{Feiguin:2006ydp}.  
To define the Hilbert space we consider the fusion tree in Figure \ref{fig:fusiontree}. 
The input data consists of a fusion category $\CC$ and a choice of a reference simple object $\rho\in \CC$.\footnote{The original AFM model and anyonic chain were defined only for those fusion categories with fusion coefficients $N^a_{bc}$ being either 0 or 1. It was later generalized in \cite{Inamura:2023qzl} to include fusion categories with $N^a_{bc}>1$.  Here for simplicity we assume all the $N^a_{bc}$ are 0 or 1.}  
The basis states are labeled by $|x_1x_2\cdots x_L\rangle$, with each $x_\ell$ taking values in the set of simple objects of $\CC$.  Moreover, a state is allowed only if $x_\ell\times \rho$ contains $x_{\ell+1}$. 
Because of the constraints, the Hilbert space of the anyonic chain is generally not a tensor product of local Hilbert spaces.

\begin{figure}
\centering
\includegraphics[width=.7\textwidth]{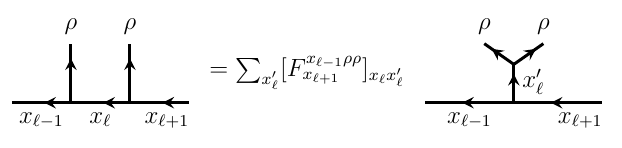}
\caption{A  change of basis of states using the $F$-symbols.}\label{fig:basis}
\end{figure}

There is a family of Hamiltonians acting on this Hilbert space that enjoy the non-invertible symmetry $\CC$. 
To define the Hamiltonian, it is convenient to introduce an alternative basis of states. 
Locally around link $\ell$, we define a change of basis as in Figure \ref{fig:basis}:
\ie
~|\cdots x_{\ell-1}x_\ell x_{\ell+1}\cdots \rangle
 = \sum_{x_\ell' }  [F^{x_{\ell-1}\rho\rho }_{x_{\ell+1}}]_{x_\ell x_\ell'}  \,
 |\cdots x_{\ell-1}x_\ell' x_{\ell+1}\cdots \rangle^{(\ell)}
\fe
where the sum is over simple objects of $\CC$ and the superscript $(\ell)$ on $|\cdots\rangle^{(\ell)}$ denotes the basis in which the $\ell$-th and the $\ell+1$-th vertical lines are fused.  
Here $F$ is the $F$-symbol (also known as the associator) of the fusion category $\CC$, which describes the crossing relation of the lines.  
For each simple object $\rho'$ of $\CC$, we then define a local Hamiltonian:
\ie
H_\ell^{(\rho')} \, |\cdots x_{\ell-1}x_\ell' x_{\ell+1}\cdots \rangle^{(\ell)}
=  - \delta_{x_\ell' , \rho'}|\cdots x_{\ell-1}x_\ell' x_{\ell+1}\cdots \rangle^{(\ell )}  \,.
\fe
It is  minus the projection operator onto the state where the fusion between the $\ell$-th and the $\ell+1$-th vertical lines is $x_\ell'=\rho'$. 
Obviously, $\sum_{\rho'} H_\ell ^{(\rho')} $ is minus the identity operator, where the sum is over simple objects $\rho'$ in the fusion channel $\rho\times \rho$. 
The general $\bZ_L$-translationally invariant Hamiltonian  takes the form 
\ie\label{anyonicH}
H =\sum_{\rho'} c_{\rho'}  \sum_\ell H_\ell^{(\rho')}\,.
\fe

The most prominent feature of the family of  Hamiltonians \eqref{anyonicH} is that they enjoy the non-invertible symmetry  $\CC$. 
On this Hilbert space, one can define a set of operators, each labeled by a simple object of $\CC$. 
The action of these  operators on the Hilbert space is pictorially represented by fusing  a line   with the anyonic chain from below using  a series of $F$-moves. 
Furthermore,  these operators obey the fusion rule of $\CC$.  
They commute with the Hamiltonian \eqref{anyonicH} because the operator action  is defined by fusion from below, while the Hamiltonian is defined using the change of basis from above as in Figure \ref{fig:basis}. 
These conserved operators are called the ``topological symmetries" in the literature.

The original ``golden chain" paper \cite{Feiguin:2006ydp} studies the case where $\CC$ is the Fibonacci fusion category, which has a unique non-trivial simple object $W$ obeying the fusion rule 
\ie\label{Fib}
W^2= I+W\,.
\fe
The reference line is $\rho=W$.  The model flows to the $c=7/10$ tricritical Ising CFT, which realizes  the non-invertible Fibonacci symmetry in the continuum.

 \begin{figure}[h!]
 \centering
 \includegraphics[width=.6\textwidth]{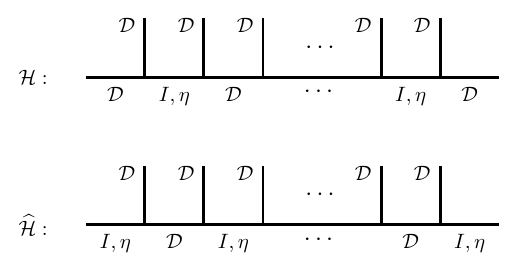}
 \caption{The fusion tree in the special case where the fusion category is the Ising fusion category TY$_+$. The fusion rule \eqref{Isingcat} constrains the fusion tree to be one of the above two cases.  Here we periodically identify the two ends and assume the number $L$ of the horizontal edges to be even. The Hilbert space is   a direct sum $\CH\oplus \widehat{\CH}$ of two tensor product Hilbert spaces, each of $2^{L\over2}$-dimensional. }\label{fig:isingtree}
 \end{figure}

\subsubsection{Ising fusion category}

Let us analyze  the case where the fusion category $\CC$ is   the Ising category TY$_+$ \cite{Aasen:2020jwb}, whose simple objects are $I,\eta,\CD$.  
The fusion rule is  in \eqref{Isingcat}. 
 The reference object $\rho$ will be chosen to be $\CD$.  
Using the fusion rule \eqref{Isingcat}, one finds  that the Hilbert space is empty if the total number $L$ of the horizontal edges in the periodic anyonic chain is odd. When $L$ is even, the fusion rule constrains the configuration to be one of the two cases in Figure \ref{fig:isingtree}: 
\begin{itemize}
\item $\CH$:~~ $x_\ell=\CD$ for all odd $\ell$, and $x_\ell=I,\eta$ for all even $\ell$.
\item $\widehat\CH$:~~ $x_\ell=\CD$ for all even $\ell$ and $x_\ell=I,\eta$ for all odd $\ell$.  
\end{itemize} 
The Hilbert space therefore decomposes into a direct sum of two tensor product Hilbert spaces $\CH \oplus \widehat\CH$, each of $2^{L\over2}$-dimensional.  
The two configurations $x_\ell =I,\eta$ correspond to the two possible values of the Ising spin. 
One can think of the two $2^{L\over2}$-dimensional Hilbert spaces as one for the sites $\CH$ and one  for the links $\widehat{\CH}$ of the original transverse-field Ising model on $L/2=N$ sites. 

\begin{figure}[h!]
\centering
\includegraphics[width=.8\textwidth]{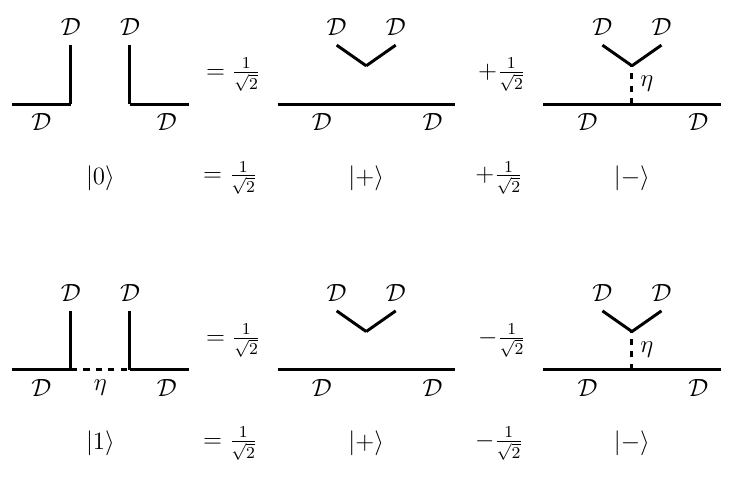}
\caption{At  link $\ell$, we write the states $x_\ell= I,\eta$  as $\ket{0},\ket{1}$, respectively. The links $\eta$ and $\CD$ as shown as dashed and the solid lines, whereas we do not draw the trivial line $I$. The righthand sides are written in terms of the states in the alternative basis $\ket{+}$ and $\ket{-}$, which stand for $x_\ell' =I,\eta$, respectively. Some $F$-symbols of the Ising fusion category TY$_+$ can be read off from this figure. (We suppress the subscripts $\ell$ in the figure.) }\label{fig:isingbasis}
\end{figure}

Let us focus on the subspace $\CH$ where all $x_\ell=\CD$ on all the odd links.  
The discussion for $\widehat{\CH}$ is similar. 
 On every even link, we write $\ket{0}_\ell$ for the $x_\ell= I$ state, and $\ket{1}_\ell$ for the $x_\ell=\eta$ state.   
 In the alternative basis, we write $\ket{+}_\ell$ for the $x_\ell'=I$ state, and $\ket{-}_\ell$ for the $x_\ell' =\eta$ state. 
 See Figure \ref{fig:isingbasis} for the relation between these two bases.  
 On every even link, we have the standard Pauli operators $X_\ell, Z_\ell$ which act on these states as 
 \ie
& X_\ell\ket{+} _\ell= \ket{+}_\ell ,&& X_\ell \ket{-} _\ell= -\ket{-}_\ell\,,\\
& Z_\ell\ket{0}_\ell = \ket{0}_\ell , &&Z_\ell \ket{1} _\ell= -\ket{1}_\ell\,.
\fe

Let us write down the Hamiltonian \eqref{anyonicH} on $\CH$. 
The local Hamiltonians\footnote{With $\rho=\CD$, there are only two possible $\rho'=I,\eta$. 
Furthermore, since $H_\ell^{(I)} +H_\ell^{(\eta)}$ is minus the identity operator, we can just focus on $H_\ell^{(I)}$ with $\rho'=I$.  } $H_\ell^{(I)}$ for $\ell$ even and odd are qualitatively different.  
When $\ell$ is even (i.e., the $I,\eta$-link), the vertical link $x_\ell'$ can be either $I$ or $\eta$. The local Hamiltonian $H_\ell^{(I)}$ favors the state $\ket{+}_\ell$ and penalizes $\ket{-}_\ell$. Therefore, $H_\ell^{(I)}$ is $-X_\ell$ for even $\ell$ (up to an immaterial constant shift). 
When $\ell$ is odd (i.e., the $\CD$-link), the vertical link $x_\ell'$ is $I$ if $x_{\ell-1}=x_{\ell+1}$, and is $\eta$ if $x_{\ell-1}\neq x_{\ell+1}$. 
It follows that $H_\ell^{(I)}$ penalizes the configuration where $x_{\ell-1}$ and $x_{\ell+1}$ are not aligned.
Therefore, $H_\ell^{(I)}$ is $-Z_{\ell-1} Z_{\ell+1}$ for odd $\ell$ (up to a constant shift).
We conclude that the Hamiltonian \eqref{anyonicH} in the special case when $\CC$ is  the Ising fusion category gives the critical  transverse-field Ising Hamiltonian \eqref{TFIM} for the even links.

Conversely, in the other half of the Hilbert space $\widehat{\CH}$, $H_\ell^{(I)}$ acts as $-X_\ell$ for odd $\ell$, and as $-Z_{\ell-1}Z_{\ell+1}$ for even $\ell$. We again find the  critical Ising Hamiltonian on $\widehat{\CH}$ for the odd links.

We have two conserved operators acting on $\CH\oplus \widehat{\CH}$, the $\bZ_2$ operator $\eta$ and the duality operator $\CD$. 
While the $\bZ_2$ line operator acts within each $2^{L\over2}$-dimensional  subspace, the duality operator  maps one to another. Together on this $2^{{L\over2}+1}$-dimensional Hilbert space, they obey the algebra \eqref{Isingcat}, and in particular, $\CD^2 = I+\eta$.  
This is to be contrasted with  the algebra $\mD^2=  {1+\eta\over2} T_\text{Ising}$ \eqref{noninvfusion}   of \cite{Seiberg:2023cdc} for the transverse-field Ising model on $L/2=N$ sites. 
There we only have a single $2^N$-dimensional Hilbert space for the  sites, and the non-invertible algebra mixes with the lattice translation. 
To summarize, we have the following two options:
\begin{itemize}
\item The algebra $\CD^2 = I+\eta$ does not mix with the lattice translation, but the Hilbert space is not a tensor product of local Hilbert spaces; rather, it is  a direct sum of two tensor product Hilbert spaces $\CH\oplus \widehat{\CH}$ \cite{Aasen:2016dop,Aasen:2020jwb}.
\item The  algebra $\mD^2 =  {1+\eta\over2} T_\text{Ising}$ mixes with the lattice translation, but the Hilbert space is a tensor product Hilbert space \cite{Seiberg:2023cdc}. 
\end{itemize}

Finally, we comment on the translation symmetry.  
The anyonic chain has a $\bZ_{2N}$ translation symmetry, which restricts  the coefficients of $H_\ell^{(I)}$ to be the same for all $\ell$. 
The resulting $\bZ_{2N}$-invariant Hamiltonian \eqref{anyonicH} is at the critical point.  
We can break the $\bZ_{2N}$ symmetry and only preserves the $\bZ_N$ lattice translation symmetry that shifts by two links $\ell\to \ell+2$. 
Imposing only the $\bZ_N$ symmetry, we can assign different coefficients for $\sum_{\ell:\text{even}}H_\ell^{(I)}$ and $\sum_{\ell:\text{odd}}H_\ell^{(I)}$ in \eqref{anyonicH}, deforming the system  away from the critical point. 
The resulting Ising Hamiltonian is in the high temperature phase in half of the Hilbert space, and in the low temperature phase in the other half. 
The full system is a direct sum of the Ising models in the high and low temperature phases. 
The duality operator $\CD$ maps between these two subspaces, implementing the Kramers-Wannier transformation.

\section{Interlude: non-invertible versus higher-form symmetries}\label{sec:interlude}

Non-invertible  symmetries are ubiquitous in 1+1d. 
They are implemented by codimension-1 topological defects in spacetime, and therefore are 0-form global symemtries. 
Every rational CFT has a set of  topological lines, the Verlinde lines, that commute with the extended chiral algebra.
They obey a (non-invertible) algebra that is isomorphic to the fusion rule of the local primary operators \cite{Petkova:2000ip}.  
So in almost all rational CFTs, there are non-invertible symmetries.  
There are also non-invertible symmetries in irrational CFT. For example, the  $c=1$ $S^1/\mathbb{Z}_2$ orbifold CFT has a rich spectrum of  finite and continuous non-invertible symmetries at every radius \cite{Chang:2020imq,Thorngren:2021yso}. 
There are also examples of non-invertible symmetries in non-conformal QFTs, such as the 1+1d adjoint QCD \cite{Komargodski:2020mxz}. 
These non-invertible global symmetries have interesting dynamical consequences and constrain the renormalization group flows \cite{Chang:2018iay,Thorngren:2019iar,Komargodski:2020mxz,Thorngren:2021yso}.  We will review some of these dynamical applications in Section \ref{sec:anomaly}.

Parallel to the progress  in 1+1d, there had been rapid developments in higher-form global symmetries \cite{Gaiotto:2014kfa} (as well as in the more general  higher groups \cite{Kapustin:2013uxa,Tachikawa:2017gyf,Cordova:2018cvg,Benini:2018reh}) in diverse spacetime dimensions.  
Most of these applications of higher-form symmetries are in higher than 1+1 spacetime dimensions.  
Indeed, in 1+1d,  1-form global symmetries are generated by   nontrivial topological local operators. They lead to different ``universes", which imply degenerate vacua for the Hilbert space on a circle.  
(See \cite{Sharpe:2022ene} for a review  of 1-form symmetries in 1+1d.) 
Roughly speaking, 1+1d is too crowded for higher-form symmetries.

For some years, the  developments of non-invertible symmetries in 1+1d and higher-form symmetries in higher dimensions appeared to be two orthogonal generalizations of the ordinary global symmetry.  

Having said that, people had discussed  various \textit{non-invertible higher-form symmetries} in diverse spacetime dimensions for some time.  
These are topological operators/defects supported higher codimensional manifolds which do not obey a group-like fusion rule. 
Below we mention a few examples:
\begin{itemize}
\item In 2+1d TQFT, such as the non-abelian Chern-Simons gauge theory, there are topological Wilson lines that do not obey a group multiplication law. They are the low-energy limit of the non-abelian anyons in the microscopic model.  
In the language of generalized global symmetry, the worldlines of the non-abelian anyons in the low-energy generate a non-invertible 1-form global symmetry.  See, for example, \cite{Kaidi:2021gbs,Yu:2021zmu,Benini:2022hzx,McGreevy:2022oyu} for this perspective. 
As such,  non-abelian topological orders can be viewed as spontaneously broken phases of  non-invertible 1-form symmetries, extending Landau's paradigm. (See \cite{Wen:2018zux} for related discussions for abelian topological orders.) 
\item Non-invertible topological surface operators in 2+1d TQFT \cite{Kapustin:2010if,Fuchs:2012dt,Lan:2014uaa,Carqueville:2017ono}. 
\item Starting with  a $d$-dimensional QFT with a non-anomalous, finite group $G$, we can gauge the latter to obtain another QFT. The gauged QFT has  Wilson lines $W_R$  associated with the $G$ gauge group. They are labeled by the irreducible representations  $R$ of the gauge group $G$.   
Since $G$ is a finite group, these Wilson lines are topological. 
Their fusion algebra is given by the representation ring Rep$(G)$ of $G$:
\ie
&W_{R_1} \times W_{R_2 }  = \sum_i W_{R_i}\,,\\
&R_1\otimes R_2 = \bigoplus_i R_i\,.
\fe
These topological Wilson lines generate a dual, $(d-2)$-form global symmetry \cite{Gaiotto:2014kfa,Bhardwaj:2017xup,Tachikawa:2017gyf}. When $G$ is abelian, Rep$(G)$ is isomorphic to $G$ itself, which is a group. The $(d-2)$-form symmetry Rep$(G)$ is invertible if $G$ is abelian. However, when $G$ is non-abelian, then Rep$(G)$ is not a group, and we hence have a non-invertible $(d-2)$-form symmetry. 
For instance, when $G=S_3$, there are three Wilson lines $W_\mathbf{1} = I, W_{\mathbf{1}_-}, W_\mathbf{2}$, where $\mathbf{1},\mathbf{1}_-, \mathbf{2}$ stand for the trivial, the sign, and the standard 2-dimensional irreducible representations of $S_3$. They generate a non-invertible $(d-2)$-form global symmetry in the $S_3$ gauge theory with the following algebra
\ie
&W_{\mathbf{1}_-} \times W_{\mathbf{1}_-}=I\,,~~
W_{\mathbf{1}_-}\times W_\mathbf{2} =W_\mathbf{2}\times W_{\mathbf{1}_-}=W_\mathbf{2}\,,\\
&W_\mathbf{2} \times W_\mathbf{2} = I  +W_{\mathbf{1}_-} +W_\mathbf{2}\,.
\fe
\item Cheshire charges \cite{Schwarz:1982ec,Alford:1990mk,Alford:1990fc,Alford:1991vr,Bucher:1991bc,Alford:1992yx}  measured by codimension-2 topological defects in gauge theory. 
These defect,  later also known as the topological Gukov-Witten defects \cite{Gukov:2006jk,Gukov:2008sn}, are labeled by the conjugacy classes,  are  defined by the nontrivial monodromy of the gauge field around the defect. 
They implement the electric 1-form symmetry which acts on the Wilson lines by linking. 
When the gauge group is a finite non-abelian group or a disconnected continuous group such as $O(2)$, the topological Gukov-Witten operators can be  non-invertible. 
See \cite{Rudelius:2020orz,Heidenreich:2020tzg,Nguyen:2021yld,Wang:2021vki,Cordova:2022rer,Arias-Tamargo:2022nlf,Antinucci:2022eat} for discussions. 
\item Consider a QFT  $\cal T$  of a   compact scalar field $\phi\sim \phi+2\pi$ with a $U(1)$ global shift symmetry $\phi\to \phi+\theta$ with $\theta\in[0,2\pi)$. 
The 1+1d $c=1$ compact boson CFT is one such an example, but our discussion also applies to the (non-conformal) compact boson in higher dimensions. 
The invertible $U(1)$ symmetry operator is $\exp( i \theta Q )$, with the conserved charge $Q$ linear in $\phi$. 
Now we gauge the $\mathbb{Z}_2$ symmetry $\phi\to -\phi$ in $\cal T$ to obtain another QFT, denoted as ${\cal T}/\mathbb{Z}_2$. 
In ${\cal T}/\mathbb{Z}_2$, $\exp(i\theta Q)$ is no longer a gauge-invariant operator because $Q\to -Q$ under the gauge symmetry. However, there is a continuous family of conserved operators that survive the $\mathbb{Z}_2$ gauging:
\ie\label{contnoninv}
\CN_\theta = 2\cos(\theta Q)\,.
\fe
It obeys the non-invertible fusion rule \cite{Chang:2020imq,Thorngren:2021yso}:
\ie
\CN_\theta \times \CN_{\theta'} = \CN_{\theta+\theta'} +\CN_{\theta-\theta'}\,.
\fe
\end{itemize}
 In addition to the above, there are various trivial ways to produce a non-invertible symmetry as discussed in Section \ref{sec:space}. 
 We can always add two invertible symmetry operators together to obtain a non-invertible symmetry.  
 Or we can stack a decoupled non-invertible TQFT to any (invertible) defect to create a non-invertible defect.

In some sense, it is easier to construct non-invertible higher-form symmetry in general spacetime dimensions. 
In contrast,  little was known about non-invertible 0-form symmetries (which are generated by codimension-1 topological operators in spacetime) in higher  dimensions beyond the TQFT examples.   
It was also unclear if they exist in realistic quantum systems.

The year 2021 brought many new developments. 
In \cite{Koide:2021zxj}, the authors generalized the model  in \cite{Aasen:2016dop,Aasen:2020jwb} to 4 spacetime dimensions. 
Their lattice $\mathbb{Z}_2$ gauge theory has a non-invertible 0-form global symmetry associated with the Kramers-Wannier-Wegner  duality \cite{Wegner:1971app}, and there is an interesting interplay between the $\mathbb{Z}_2$ 1-form global symmetry and the non-invertible symmetry. 
This construction was soon generalized to continuum field theory in \cite{Choi:2021kmx,Kaidi:2021xfk}, where examples of non-invertible symmetries were found in the 3+1d free Maxwell gauge theory, Yang-Mills theory, and  ${\cal N}=4$ super Yang-Mills theory. 
The techniques employed in these papers are very general and  were further generalized to a large class of quantum systems in diverse spacetime dimensions.
These include the non-invertible global symmetry in the 3+1d QED and QCD for the real world \cite{Choi:2022jqy,Cordova:2022ieu}.

In most of these constructions, higher-form symmetries play a central role. 
The two seemingly orthogonal developments of generalized global symmetries are no longer separable.

\begin{figure}
\centering
\includegraphics[width=.28\textwidth]{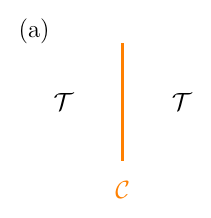}\qquad\qquad
\includegraphics[width=.4\textwidth]{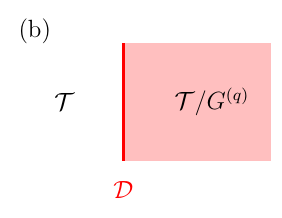}
\caption{(a) Higher gauging: the (non-invertible) condensation defect $\CC$ is obtained by gauging a finite higher-form global symmetry $G^{(q)}$ only along a higher codimensional manifold (shown in orange). 
(b) Half gauging: the topological interface $\CD$ is obtained from gauging  $G^{(q)}$ only in half of the spacetime, and impose a topological Dirichlet boundary condition along the interface.  While $\CD$ is generally an interface between two different systems $\cal T$ and ${\cal T}/G^{(q)}$, when it so happens that ${\cal T}$ is isomorphic to ${\cal T}/G^{(q)}$,  $\CD$ becomes a (non-invertible) topological defect in a single quantum system.}\label{fig:CD}
\end{figure}

Below we mention a few popular constructions for the non-invertible symmetries:

\paragraph{Higher gauging \cite{Roumpedakis:2022aik}} We gauge a discrete higher-form global symmetry $G^{(q)}$ of a QFT $\cal T$ along a higher-codimensional manifold in spacetime. Higher gauging does not change the bulk of the quantum system $\cal T$, but generates a topological defect, known as the condensation operator/defect $\CC$, which is  generally non-invertible. See Figure \ref{fig:CD}(a). 
 Condensation operators are  the most basic non-invertible symmetries, and are generalizations of the projection operator.
In some sense, they are like the identity element in the realm of non-invertible symmetries.

\begin{figure}
\centering
\includegraphics[width=.7\textwidth]{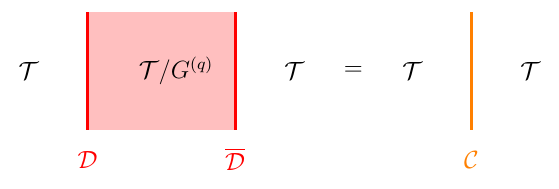}
\caption{The fusion of the interface $\CD$ and its orientation reversal $\overline{\CD}$ is equivalent to gauging the higher-form symmetry only along a higher codimensional manifold, and therefore gives the condensation defect $\CC$, i.e., $\CD\times \overline{\CD}=\CC$.  }\label{fig:DDC}
\end{figure}

\paragraph{Half gauging \cite{Choi:2021kmx}} We gauge a non-anomalous discrete higher-form global symmetry $G^{(q)}$ of a QFT $\cal T$ in half of the spacetime and impose a topological Dirichlet boundary condition at the interface. Half gauging produces a topological interface $\CD$ between two quantum systems, ${\cal T}$ and ${\cal T}/G^{(q)}$.  See Figure \ref{fig:CD}(b). 
The fusion of $\CD$ and its orientation reversal $\overline\CD$ correspond to gauging $G^{(q)}$ only in an interval. Since both $\CD$ and $\overline\CD$ are topological, shrinking the interval  results in gauging in a codimension-1 manifold, which is equivalent to higher gauging. 
Therefore, (see Figure \ref{fig:DDC})
\ie
\CD \times \overline\CD = \CC\,.
\fe 
In the special case when the system is invariant under gauging $G^{(q)}$, i.e.,  ${\cal T} \simeq {\cal T}/G^{(q)}$, 
 $\CD$ becomes a topological defect in a single quantum system.  This turns out to be a powerful way to construct non-trivial non-invertible symmetry in general spacetime dimensions.

\paragraph{From mixed anomalies \cite{Tachikawa:2017gyf,Kaidi:2021xfk}}
Starting with a QFT $\cal T$ with a mixed anomaly of two invertible (higher-form) global symmetries, gauging one of them sometimes turn the other into a non-invertible symmetry in the gauged QFT.\footnote{Some other times  gauging a symmetry with a mixed anomaly creates a higher group \cite{Tachikawa:2017gyf,Cordova:2018cvg,Benini:2018reh}. See Table \ref{table:triangle} for some examples in 3+1d.}  
For example, the non-invertible symmetry $\CD$ of the Ising CFT arises from the mixed anomaly between $(-1)^F$ and $(-1)^{F_\text{L}}$ of the Majorana CFT (see Section \ref{sec:bosonization}). 
When a non-invertible symmetry can be constructed from gauging a non-anomalous subgroup of an invertible, anomalous symmetry, it is called non-intrinsic \cite{Kaidi:2022uux}, or group-theoretical \cite{Sun:2023xxv}. 
Not all non-invertible symmetries are group-theoretical.

There are many other constructions of non-invertible symmetries that we did not mention here.  
In many cases (such as in the Ising model), a single non-invertible symmetry can be constructed from several different methods.

In the following sections we will review the higher and half gauging procedures,  connect these new ideas to the non-invertible symmetry of the Ising CFT in 1+1d, and provide examples in higher dimensions.

\section{Higher gauging and condensation defects}\label{sec:higher}

As discussed in Section \ref{sec:space}, a cheap way to construct a non-invertible symmetry is to take linear combination of invertible ones. For instance, the operator $\CC=1+\eta$ with $\eta$ a $\mathbb{Z}_2$ symmetry is non-invertible. 
However, this is not an interesting non-invertible symmetry since it is made out of the invertible ones; it's a derivative object. 
In other words, $1+\eta$ is not a simple operator/defect since it can be written as a non-negative integer combination of other operators/defects of the same dimensionality.

In higher spacetime dimensions, there are more creative operations one can perform other than just taking linear combinations of invertible symmetries. 
The rough idea is that one can sometimes sum over a finite set of topological defects of lower dimensions along the nontrivial cycles of a higher dimensional manifold. 
This creates a topological defect on the higher dimensional manifold, known as the \textit{condensation defect} \cite{Kong:2013aya,Kong:2014qka,Else:2017yqj,Gaiotto:2019xmp,Johnson-Freyd:2020twl,Roumpedakis:2022aik}.  
Condensation defects are simple defects because they cannot be written in terms of other defects of the same dimensionality; however, they are a mesh of topological defects of lower dimensions. In this sense, it is a categorical generalization of the projection operator. Nonetheless, it leads to interesting and familiar global symmetries, invertible or non-invertible. It is also the most basic part of the non-invertible fusion algebra in general spacetime dimensions.

\subsection{Higher gauging}

We start with a general introduction of higher gauging in diverse spacetime dimensions. 
The original idea of higher gauging came from anyon condensation along a 1d line in the 2-dimensional space \cite{Kong:2013aya}, and it was later interpreted in terms of  gauging  a higher-form global symmetry along a higher codimensional manifold in  \cite{Roumpedakis:2022aik}. 
Because it was introduced in the context of anyon condensation, the resulting defect was called a condensation defect. 
Another way to justify this name is that in many examples, the worldvolume Lagrangian of the condensation defect involves a Higgs field acquiring a vacuum expectation value \cite{Roumpedakis:2022aik}.

Let $\cal T$ be a QFT in $d$ spacetime dimensions with a discrete $q$-form global symmetry $G^{(q)}$. 
If $G^{(q)}$ is free of 't Hooft anomaly, then we can gauge it to obtain another QFT ${\cal T}/G^{(q)}$. 
Gauging a discrete $q$-form global symmetry is implemented by summing over network of the $(d-q-1)$-dimensional topological defects on the entire spacetime manifold. 
This is equivalent to summing over the discrete $q$-form gauge fields by the Poincare duality.  
There are generally different options in gauging $G^{(q)}$, which corresponds to choosing an invertible $G^{(q)}$-SPT in $d$ dimensions. 
For $q=0$ and $d=2$, i.e., ordinary gauging in 1+1d, these options are known as the discrete torsion, classified by $H^2(BG,U(1))$.

Given a  discrete $q$-form global symmetry, we can sometimes gauge it not in the entire spacetime, but only along a codimension $p$ submanifold $M^{(d-p)}$ in spacetime. 
This is known as \textit{higher gauging} \cite{Roumpedakis:2022aik}, or more precisely, $p$-gauging a $q$-form global symmetry. 
In this terminology, the ordinary gauging of $G^{(q)}$ corresponds to 0-gauging. 
Higher gauging does not change the bulk QFT $\cal T$, but generates a topological defect along $M^{(d-p)}$, known as the condensation defect.

Just like ordinary gauging, there can be obstruction to higher gauging. 
A $q$-form global symmetry is called \textit{$p$-gaugeable} if it can be gauged on a codimension $p$ submanifold. Otherwise, it is called \textit{$p$-anomalous}.  
A $p$ gaugeable symmetry is $p'$-gaugeable for all $p'>p$.  
There are also options in $p$-gauging coming from choosing different $(d-p)$-dimensional SPTs on $M^{(d-p)}$.

Since the $q$-form symmetry defects are $(d-q-1)$-dimensional, we can only $p$-gauge a $q$-form symmetry if $p\le q+1$. 
When $p=q+1$, higher gauging is rather trivial and corresponds to taking a linear combination of the $G^{(q)}$ topological defects. The resulting condensation operator is proportional to the  projection operator and is not simple. 
In particular, the ordinary projection operator $1+\eta$ (or more precisely, twice thereof) arises from 1-gauging an ordinary 0-form $\mathbb{Z}_2$ symmetry. 
When $p<q+1$, the condensation defect is simple in that it cannot be written in terms of other defects of the same dimensionality.

\begin{figure}[h!]
\centering
\includegraphics[width=.2\textwidth]{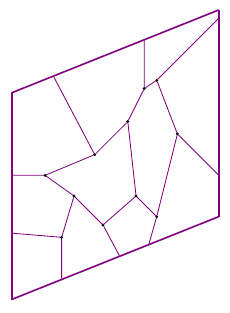}
\caption{Condensation defects arise from higher gauging a discrete higher-form symmetry along  a higher codimensional submanifold. They are defined by summing over networks of higher-form symmetry defects  along a submanifold. Therefore,  condensation defects are ``porous" and necessarily act trivially on the local operators. }\label{fig:porous}
\end{figure}

Condensation defects can be invertible or non-invertible, and they provide some of the simplest examples of non-invertible symmetries in diverse spacetime dimensions. 
Almost every time we have a higher-form global symmetry (with some assumptions on its $p$-anomaly), it leads to a non-invertible condensation defect. 

Since the condensation defect  is a mesh of lower dimensional defects, it is ``porous" to local operators. See Figure \ref{fig:porous}.
Therefore, it necessarily acts trivially on all the local operators. 
Indeed, topological  surface operators in 2+1d TQFT are examples of condensation defects, in which case there is simply no nontrivial local operator.

\subsection{Condensation defects in 2+1d}\label{sec:1gaugeZ2}

Let us illustrate the general idea of higher gauging in the simplest nontrivial case: 1-gauging a $\mathbb{Z}_2^{(1)}$ 1-form global symmetry in 2+1d. 
Our discussion will be brief and we refer the readers to \cite{Barkeshli:2014cna} for a comprehensive introduction to 2+1d TQFT, and to 
 \cite{Roumpedakis:2022aik} for more details of the condensation defects.

\begin{figure}[h!]
\centering
\includegraphics[width=.4\textwidth]{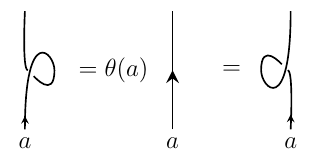}
\caption{The definition of the topological spin $\theta(a)=e^{2\pi i h(a)}$ of a topological line $a$.}\label{fig:topspin}
\end{figure}

In a 2+1d, a $\mathbb{Z}_2^{(1)}$ 1-form global symmetry is generated by a $\mathbb{Z}_2$ topological line $\eta$, with $\eta^2=1$.  
An important quantity associated with a topological line is its \textit{topological spin} $h$, which is defined modulo integers. We also define $\theta =e^{2\pi i h}$. 
The topological spin is defined by the relative phase when twisting the topological line as in Figure \ref{fig:topspin}. 
In a bosonic QFT, $\eta$ can have four possible topological spins, $h=0, \frac14, \frac 12, \frac 34$. 
In the case of a TQFT, they correspond to the low-energy limit of the worldline of a boson, a semion, a fermion, and an anti-semion, respectively.

The nontrivial spin (or more generally, the nontrivial $R$-symbols) of a topological line presents  obstructions to gauging the 1-form global symmetry \cite{Gaiotto:2014kfa,Gomis:2017ixy,Hsin:2018vcg}. 
This is because a topological move (such as in Figure \ref{fig:topspin}) of the lines  produces a nontrivial phase, making the partition function of the supposed gauged theory ambiguous. 
In the special case of a $\mathbb{Z}_2$ 1-form global symmetry, it is free of the 't Hooft anomaly (i.e., 0-gaugeable) if  $h=0$. 
This anomaly is classified by $H^4(B^2 \mathbb{Z}_2,U(1))=\mathbb{Z}_4$.

\begin{figure}[h!]
\centering
\includegraphics[width=.4\textwidth]{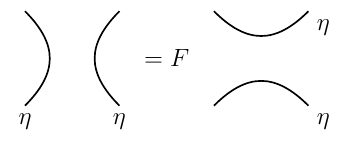}
\caption{The crossing relation of a $\mathbb{Z}_2^{(1)}$ anyon line on a 2-dimensional surface produces a nontrivial phase $F$.}\label{fig:Fsymbol}
\end{figure}

If we are less ambitious, and only wish to 1-gauge the $\mathbb{Z}_2^{(1)}$ 1-form symmetry along a codimension-1 surface $\Sigma$ in spacetime, then we need not be bothered by the nontrivial topological spin $h$. 
Indeed, there is no braiding (or $R$-symbols) on a 2-dimensional surface $\Sigma$: one cannot define the statistics of a particle in 1+1d. 
In contrast, there can be a nontrivial crossing move described by the $F$-symbol on a surface as in Figure \ref{fig:Fsymbol}. 
A nontrivial phase $F$ in the crossing relation presents an obstruction to 1-gauging the $\mathbb{Z}_2^{(1)}$ symmetry along a 2-dimensional surface in spacetime.  
This 1-anomaly  is classified by $H^3(B\mathbb{Z}_2,U(1))=\mathbb{Z}_2$. 
The phase $F$ in the crossing relation is related to the topological spin as
\ie
F = \theta(\eta)^2 = e^{4\pi i h}\,.
\fe
Therefore, we conclude that while only the boson line $h=0$ is 0-gaugeable, both the boson $h=0$ and the fermion  $h=\frac12$ lines are 1-gaugeable. 
The semion $h=\frac 14$ and anti-semion $h=\frac 34$ lines are 0- and 1-anomalous by contrast. 

When the $\mathbb{Z}_2^{(1)}$ symmetry is 1-gaugeable, i.e., if $F=e^{4\pi i h}= 1$,  we can 1-gauge the symmetry to obtain a topological condensation defect supported on a surface $\Sigma$:
\ie\label{CC}
\CC (\Sigma) = {1\over\sqrt{| H_1(\Sigma,\mathbb{Z}_2)|}} \sum_{\gamma\in H_1(\Sigma,\mathbb{Z}_2)} \eta(\gamma)\,.
\fe
Note that we can always redefine a topological surface operator by a topological Euler counterterm $\lambda^{\chi(\Sigma)}$ for any $\lambda\in \mathbb{R}$. As a result, the expectation value of a surface defect on a two-sphere is subject to this counterterm ambiguity. 
 Here we have made a choice for the Euler counterterm for later convenience.

Below we analyze the condensation defect $\CC(\Sigma)$ for the two cases when $\eta$ is a boson $h=0$ or a fermion $h=\frac12$.  
To proceed, we first note a simple relation obeyed by $\eta$ when restricted to the surface $\Sigma$. Let $\gamma, \gamma'$ be two 1-cycles on $\Sigma$. 
We view $\Sigma$ as our space and $\eta(\gamma)$ a conserved operator acting on the Hilbert space. Using Figure \ref{fig:topspin}, it is easy to show that the $\mathbb{Z}_2^{(1)}$ 1-form symmetry operators obey the following commutation relation:
\ie\label{qtorus}
\eta(\gamma) \eta(\gamma')  =\theta(\eta)^{\langle \gamma,\gamma'\rangle} \eta(\gamma+\gamma')\,,
\fe
where $\langle \gamma,\gamma'\rangle$ is the intersection number between the 1-cycles $\gamma, \gamma'\in H_1(\Sigma,\mathbb{Z}_2)$. 
For instance, when $\Sigma=T^2$ with $\{\mathbf{A,B}\}\in H_1(T^2,\mathbb{Z})$ a basis for the 1-cycles, we have $\eta(\mathbf{A}+\mathbf{B}) = \theta(\eta)\eta(\mathbf{A})\eta(\mathbf{B})=\theta(\eta)\eta(\mathbf{B})\eta(\mathbf{A})$. 
The condensation operator on $T^2$ is
\ie\label{CT2}
\CC(T^2)  = \frac12 \left[ 1+\eta(\mathbf{A}) +\eta(\mathbf{B}) + \theta(\eta) \eta(\mathbf{A})\eta(\mathbf{B})\right]\,,
\fe
with $\theta(\eta) = e^{2\pi i h}= \pm1$.

With \eqref{qtorus}, it is then straightforward to compute the algebra of the condensation operators $\CC(\Sigma)$ on a general Riemann surface:
\ie\label{Z2condfusion}
~&\CC(\Sigma) \times \CC(\Sigma) = \sqrt{|H_1(\Sigma,\mathbb{Z}_2)|} \, \CC(\Sigma)\,,~~&&\text{if}~~h=0\,,\\
&\CC(\Sigma) \times \CC(\Sigma) = 1\,,~~&&\text{if}~~h=\frac12\,.
\fe
(These equations can be easily verified when $\Sigma=T^2$ using \eqref{CT2}.) 
We find that $\CC(\Sigma)$ is non-invertible if $h=0$ and is an invertible $\mathbb{Z}_2$ 0-form symmetry if $h=\frac 12$.
The coefficient $\sqrt{|H_1(\Sigma,\mathbb{Z}_2)|}$ in the non-invertible fusion rule for $h=0$ is the partition function of a 1+1d $\mathbb{Z}_2$ gauge theory. 
We can therefore write the fusion rule as
\ie\label{cheshire}
\CC  \times \CC  = ({\cal Z}_2) \, \CC \,,~~&&\text{if}~~h=0\,,
\fe
where ${\cal Z}_2$ stands for the 1+1d $\mathbb{Z}_2$ gauge theory.  
This is a simple example of a TQFT-valued fusion rule that we discussed in Section \ref{sec:space}.

The above discussion can be generalized to any 2+1d bosonic QFT with a $\mathbb{Z}_N^{(1)}$ 1-form global symmetry. 
Let $\theta(\eta)=e^{2\pi ih}$ be the topological spin of the generator $\eta$ for $\mathbb{Z}_N^{(1)}$. 
The 1-form global symmetry is 1-gaugeable if $\theta(\eta)^N=1$, i.e., if $\theta(\eta)=e^{2\pi i k\over N}$ for some integer $k$ defined modulo $N$. 
We obtain a condensation defect $\CC_n$ for each $\mathbb{Z}_n$ subgroup of $\mathbb{Z}_N$, with $n|N$. 
Their fusion rule is \cite{Roumpedakis:2022aik}
\ie\label{generalC}
\CC_n\times\CC_{n'}
= ({\cal Z}_{\text{gcd}(n,n',k\ell)})  \, \CC_{ \text{gcd}(n,n',k\ell )nn'\over\text{gcd}(n,n')^2}\,,~~~~\ell = {N\over \text{lcm}(n,n')}\,.
\fe
In particular, $\CC_1$ is the trivial surface defect. 
When $k=1$, this reduces to the algebra in \cite{Kapustin:2010if}.

Mathematically, the topological line $\eta$, the condensation surface $\CC$, and their composites form a fusion 2-category \cite{2018arXiv181211933D}. See also \cite{2021arXiv210315150D,2022arXiv220310331D} for  the  mathematical formulation of the condensation surfaces discussed above.   
See \cite{Inamura:2023qzl} for 2+1d lattice models realizing a general fusion 2-category.  
See also \cite{Carqueville:2018sld,Mulevicius:2020bat,Mulevicius:2020tgg,Carqueville:2021edn,Yu:2021zmu,Carqueville:2021dbv,Mulevicius:2022gce} for the gauging of these topological surfaces.

\subsection{Examples in 2+1d TQFT}

In this subsection we discuss some examples of condensation defects in 2+1d TQFT. However, we emphasize that the higher gauging construction is not restricted to TQFT, but also applies to more general QFT. 
For the condensation defects in the 2+1d free Maxwell theory, see \cite{Roumpedakis:2022aik}. 
For condensation defects in 3+1d, see \cite{Choi:2022zal,Choi:2022jqy}.

\subsubsection{$U(1)_{2N}$ Chern-Simons theory and charge conjugation symmetry}

The Lagrangian for the $U(1)_{2N}$ Chern-Simons theory at level $2N$ is\footnote{Chern-Simons theories with odd levels are fermionic TQFT and can only be defined on spin manifolds. Here we restrict our discussions to bosonic QFT and therefore we assume the level to be even. }
\ie
{\cal L} =   {2Ni\over 4\pi } ada\,,
\fe
where $a$ is a dynamical 1-form gauge field. 
The equation of motion $da=0$ sets all the nontrivial local operators to be zero on-shell. The Chern-Simons theory therefore has a unique local operator, the identity. Correspondingly, the $S^2\times S^1$ partition function is 1 via the operator-state correspondence. 
(Note that the gauge field $a$ is not gauge invariant and does not qualify as a local operator.)

While the local operator data is completely trivial, there are  nontrivial topological line operators, which are the Wilson lines:
\ie
W^s = \exp\left( i s \oint a\right)\,,~~~s=1,\cdots, 2N\,.
\fe
The Wilson lines $W^s$ obey a $\mathbb{Z}_{2N}$ fusion rule and generate a $\mathbb{Z}_{2N}^{(1)}$ 1-form global symmetry. 
The topological spin of $W^s$ is
\ie
h(s) =  {s^2\over 4N}  ~\text{mod}~1\,.
\fe
Because of the nontrivial topological spin, the $\mathbb{Z}_{2N}^{(1)}$  has an 't Hooft anomaly.  
In fact, since $\theta(\eta)^{2N}\neq 1$, it is also 1-anomalous. 
In contrast, the $\mathbb{Z}_N^{(1)}$ subgroup, which is generated by $W^2$ with spin $h(\eta^2) = 1/N$, is 1-gaugeable (but still 0-anomalous).  

Let us focus on the condensation defect $\CC_N$ from 1-gauging the $\mathbb{Z}_N^{(1)}$ subgroup. 
Its fusion rule can be read off from \eqref{generalC} with $k=1$ and $n=n'=N$, which turns out to be an invertible $\mathbb{Z}_2$ algebra:
\ie
\CC_N \times\CC_N =  1\,.
\fe
In fact, the $\CC_N$ generates the charge conjugation $\mathbb{Z}_2$ 0-form symmetry, which acts on the gauge field as
\ie\label{chargeconj}
\CC_N:~~a\to -a\,.
\fe

The charge conjugation symmetry \eqref{chargeconj} in Chern-Simons theory is a rather subtle symmetry. 
It acts trivially on all the local operators, since there isn't any nontrivial one.  
Therefore, the operator $\CC(S^2)$ on a two-sphere is a trivial operator and does not act faithfully. 
However, it permutes the Wilson line $W^s$ with $W^{2N-s}$, so the operator $\CC(T^2)$ on a two-torus acts nontrivially on the states. 
It is an unusual 0-form symmetry since it doesn't act on the local operators, but it is also not a 1-form symmetry. The latter acts on the line defects by a phase, rather than permuting them. 
In some sense, the charge conjugation symmetry is something in between a 0-form and a 1-form symmetry. 
The notion of higher gauging makes this intuition precise: it is a condensation surface defect made of lines. 
In fact, all topological surface defects in 2+1d TQFT are condensation defects \cite{Fuchs:2012dt,Roumpedakis:2022aik}. 

In addition to the invertible charge conjugation symmetry $\CC_N$, the $U(1)_{2N}$ Chern-Simons theory has other non-invertible topological surfaces $\CC_n$ \cite{Kapustin:2010if},  arising from 1-gauging a $\mathbb{Z}_n^{(1)}$ subgroup of the 1-gaugeable $\mathbb{Z}_N^{(1)}$ symmetry. Their fusion algebra is given by \eqref{generalC} with $k=1$.

\subsubsection{$\mathbb{Z}_2$ gauge theory}\label{sec:Z2gauge}

The Lagrangian of the 2+1d continuum $\mathbb{Z}_2$ gauge theory takes the form of a Chern-Simons theory with two $U(1)$ gauge fields $a,b$ \cite{Maldacena:2001ss,Banks:2010zn,Kapustin:2014gua}:
\ie\label{Z2Lag}
{\cal L} = {2i\over 2\pi } adb\,.
\fe
This is the low-energy limit of the toric code \cite{Kitaev:1997wr}. 
There are four topological line operators:
\ie
1, ~~e=\exp\left(i\oint a\right),~~m=\exp\left(i\oint b\right)\,,~~f= \exp\left(i\oint (a+b)\right)\,.
\fe
They generate a $\mathbb{Z}_2^{(1)}\times \mathbb{Z}_2^{(1)}$ 1-form global symmetry
\ie
e\times m=m\times e= f ,~~e\times f=f\times e=m,~~m\times f= f\times m =e\,,~~e^2=m^2=f^2=1\,.
\fe
The $1,e,m$ lines are bosons ($h=0$) and the $f$ line is a fermion $(h=\frac12)$.

The $\mathbb{Z}_2^{(1)}\times \mathbb{Z}_2^{(1)}$ 1-form symmetry is 0-anomalous because of the nontrivial spin of fermion  $f$. Nevertheless, it is 1-gaugeable. 
We can 1-gauge the three $\mathbb{Z}_2^{(1)}$ subgroups generated by $e,m,f$ to obtain three condensation defects. 
We denote the corresponding condensation defects by $\CC_\ee, \CC_\mm,\CC_f$, respectively, defined by \eqref{CC} with $\eta$ replaced by $e,m,f$. 
Following our general discussion in Section \ref{sec:1gaugeZ2}, the $\CC_\ee,\CC_\mm$ defects obey the non-invertible fusion algebra \eqref{cheshire} because $e, m$ are bosons. 
On the other hand, $\CC_f$ is an invertible $\mathbb{Z}_2$ surface because $f$ is a fermion. 
$\CC_f$ generates a 0-form $\mathbb{Z}_2^\text{em}$ global symmetry that exchanges $e$ with $m$.

In  addition,  there are two ways to gauge the entire $\mathbb{Z}_2^{(1)}\times \mathbb{Z}_2^{(1)}$ which differed by a choice of the discrete torsion $H^2(B(\mathbb{Z}_2\times \mathbb{Z}_2),U(1))=\mathbb{Z}_2$. 
We denote the corresponding condensation defects by $\CC_{\ee\mm},\CC_{\mm\ee}$:
\ie
	\CC_{\ee\mm}(\Sigma) &= \frac{1}{\sqrt{|H_1(\Sigma,\mathbb{Z}_2 \times \mathbb{Z}_2)|}} \sum_{\gamma,\gamma' \in H_1(\Sigma,\mathbb{Z}_2)} e(\gamma)m(\gamma')  \, , \\
	\CC_{\mm\ee}(\Sigma) &= \frac{1}{\sqrt{|H_1(\Sigma,\mathbb{Z}_2 \times \mathbb{Z}_2)|}} \sum_{\gamma,\gamma' \in H_1(\Sigma,\mathbb{Z}_2)} (-1)^{ \langle \gamma, \gamma' \rangle }e(\gamma)m(\gamma')  \, .
\fe

Together, 1-gauging  leads to 6 condensation defects  with the following fusion rule \cite{Roumpedakis:2022aik}:
\begin{equation} \label{z2.gauge.theory.fusion}
\begin{array}{lllll}
	\CC_\ee \times \CC_\ee = (\mathcal{Z}_2) \CC_\ee \,, &~~~~& \CC_\ee \times \CC_\mm = \CC_{\ee\mm} \,, &~~~~& \CC_{\ee\mm} \times \CC_{\ee\mm} = \CC_{\ee\mm} \,,\\
	\CC_\mm \times \CC_\mm = (\mathcal{Z}_2) \, \CC_\mm \,, &~~~~& \CC_\ee \times \CC_f = \CC_{\ee\mm} \,, &~~~~& \CC_{\ee\mm} \times \CC_{\mm\ee} = (\mathcal{Z}_2) \,  \CC_{\ee} \,, \\
	\CC_f \times \CC_f = 1 \,, &~~~~& \CC_f \times \CC_\mm = \CC_{\ee\mm} \,, &~~~~& \CC_{\mm\ee} \times \CC_{\ee\mm} = (\mathcal{Z}_2) \,  \CC_{\mm} \,,\\
	\CC_\ee \times \CC_{\ee\mm} = (\mathcal{Z}_2) \, \CC_{\ee\mm} \,, &~~~~& \CC_\mm \times \CC_{\ee\mm} =  \CC_{\mm} \,, &~~~~& \CC_f \times \CC_{\ee\mm} = \CC_\mm \,, \\
	\CC_\mm \times \CC_{\mm\ee} = (\mathcal{Z}_2) \, \CC_{\mm\ee} \,, &~~~~& \CC_\ee \times \CC_{\mm\ee} =  \CC_{\ee} \,, &~~~~& \CC_f \times \CC_{\mm\ee} = \CC_\ee \,.
\end{array}
\end{equation}
These six condensation defects are also discussed in the context of gapped domain walls of the toric code in \cite{Lan:2014uaa}. 
The four non-invertible condensation defects $\CC_\ee,\CC_\mm,\CC_{\ee\mm} ,\CC_{\mm\ee}$ can be factorized into pairs of the topological boundary conditions of the $\bZ_2$ gauge theory.

\subsubsection{A bulk perspective on the non-invertible symmetry   of the  Ising CFT}\label{sec:bulk}

Here we discuss a bulk perspective of the non-invertible global symmetry $\CD$ of the Ising CFT from the 2+1d $\bZ_2$ gauge theory. 

In the 1+1d Ising CFT, the untwisted Hilbert space $\CH$ can be graded by the $\mathbb{Z}_2$ global symmetry as $\CH=\CH^+ \oplus\CH^-$ as in \eqref{CHpm}. 
Similarly, the $\mathbb{Z}_2$ twisted Hilbert space $\CH_\eta$ can be graded by the $\mathbb{Z}_2$ global symmetry $\eta$ as $\CH_\eta = \CH_\eta^+ \oplus\CH_\eta^-$ as in  \eqref{CHetapm}. 

These states in the 1+1d system have a natural embedding into a 2+1d system \cite{Witten:1988hf,Moore:1989yh,Elitzur:1989nr}. 
Concretely, we place the 1+1d Ising CFT on a spacetime cylinder. 
In the bulk of the (solid) cylinder, we introduce a 2+1d $\mathbb{Z}_2$ gauge theory, and gauge the $\mathbb{Z}_2$ global symmetry of the 1+1d Ising CFT by the bulk gauge field.  
The original Ising CFT now becomes a gapless boundary condition for the 2+1d $\mathbb{Z}_2$ gauge theory \cite{Ho:2014vla,Freed:2018cec,Lin:2019kpn,Ji:2019ugf,Ji:2019jhk,Gaiotto:2020iye,Ji:2021esj,Kaidi:2022cpf, Freed:2022qnc,Lin:2022dhv,Chatterjee:2022jll,Lin:2023uvm,Choi:2023xjw}. 
This construction was recently discussed in the contexts of symmetry TFT in \cite{Gaiotto:2014kfa,Gaiotto:2020iye,Apruzzi:2021nmk,Freed:2022qnc,Moradi:2022lqp,Kaidi:2022cpf,Kaidi:2023maf} and of ``categorical symmetry" in  \cite{Ji:2019jhk,Kong:2020cie}.\footnote{The term ``categorical symmetry" is being abused in at least two contexts with different, but related, meanings.  The first use of the term  is the one mentioned in the main text which involves a bulk. The second one  is used as a synonym for ``non-invertible symmetry" in  many field theory discussions.  The ``categorical symmetry"  of \cite{Ji:2019jhk,Kong:2020cie} is the bulk TQFT of the ``categorical/non-invertible symmetry" of the boundary QFT in the field theory context. In the case of a  2+1d bulk/1+1d boundary, the former ``categorical symmetry"  is a modular tensor category that is  the Drinfeld center of the latter ``categorical symmetry", which is  a fusion category.}

\begin{figure}
\centering
\includegraphics[width=.45\textwidth]{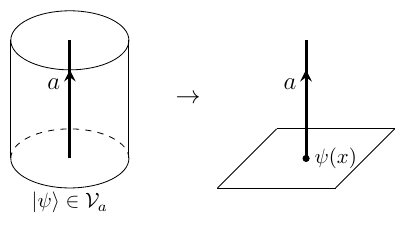}
\caption{Left: We have a 2+1d $\mathbb{Z}_2$ gauge theory in the bulk of the cylinder, with the gapless boundary condition corresponding to the Ising CFT imposed at the surface of the cylinder. There is an anyon line $a=1,e,m,f$ inserted in the middle of the cylinder.  If the vertical direction is time, then this gives the Hilbert space ${\cal V}_a$ of the $\mathbb{Z}_2$ gauge theory quantized on a spatial disk with the above gapless boundary condition and an anyon $a$ inserted at the origin of the disk.  Right: Via the cylinder-plane conformal map, a state $\ket{\psi}\in {\cal V}_a$ is mapped to a point operator confined to the 1+1d boundary that is attached to the anyon line $a$ extended into the bulk.}\label{fig:anyon}
\end{figure}

We quantize  the 2+1d $\mathbb{Z}_2$ gauge theory on a disk with the above gapless boundary condition and an anyon line $a$ insertion at the origin of the disk.  
We denote the Hilbert space of this 2+1d system by ${\cal V}_a$. 
Via the cylinder-plane conformal map, a state in ${\cal V}_a$ is mapped to a point operator (which is confined to the 1+1d boundary) that is attached to the anyon line $a$ extended into the bulk. See Figure \ref{fig:anyon}.

The states of the original 1+1d system in \eqref{CHpm} and \eqref{CHetapm} are now mapped to the Hilbert space ${\cal V}_a$. For instance, the $\mathbb{Z}_2$-even local operators (such as $1,\varepsilon$) of the Ising CFT are not affected by the coupling to the $\mathbb{Z}_2$ gauge theory, and they remain local operators confined to the boundary. They correspond to the states in ${\cal V}_1$. In contrast, the $\mathbb{Z}_2$-odd local operators (such as $\sigma$) are now attached to the $\mathbb{Z}_2$ Wilson line $e$, and belongs to  ${\cal V}_e$. 
We have the following identification between the 1+1d states in the (un)twisted Hilbert spaces $\CH^\pm,\CH_\eta^\pm$, and the 2+1d states in the Hilbert spaces ${\cal V}_a$ \cite{Ho:2014vla,Lin:2019kpn}:
\ie
\CH^+ = {\cal V}_1\,,\qquad \CH^- = {\cal V}_e\,,\qquad \CH_\eta^+ = {\cal V}_m \,,\qquad \CH_\eta^- = {\cal V}_f\,.
\fe
Note that the conformal spins $h-\bar h$ in each sector match with the topological spin of the corresponding anyons modulo integers.

What happens to the non-invertible symmetry $\CD$ when we couple the Ising CFT to the 2+1d $\mathbb{Z}_2$ gauge theory? 
It becomes the end locus of the $\mathbb{Z}_2^\text{em}$ surface defect  on the 1+1d boundary \cite{Ho:2014vla,Freed:2018cec,Ji:2019ugf}. This invertible condensation defect $\CC_f$ of the 2+1d $\mathbb{Z}_2$ gauge theory implements the  $\mathbb{Z}_2^\text{em}$ 0-form global symmetry that exchanges the $e$ and $m$ anyons. 
In this sense, the electromagnetic symmetry of the 2+1d $\mathbb{Z}_2$ gauge theory is identified with the Kramers-Wannier duality defect of the 1+1d Ising CFT.  
Indeed, the $e\leftrightarrow m$ exchange in the bulk swaps the order state $\ket{\sigma}\in {\cal V}_e$ and the disorder state $\ket{\mu}\in {\cal V}_m$.

\section{Half gauging}\label{sec:half}

In this section we review the half gauging construction of non-invertible symmetries in \cite{Choi:2021kmx,Choi:2022zal}.
We start with a QFT $\cal T$ in $d$ spacetime dimensions with a non-anomalous $\bZ_N^{(q)}$ $q$-form global symmetry, generated by the topological operator $\eta$.  
Let ${\cal L}_{\cal T}[A]$ be the Lagrangian of $\cal T$ coupled to the $(q+1)$-form background gauge field $A$.  
Gauging $\bZ_N^{(q)}$ gives another QFT, ${\cal T}/\bZ_N^{(q)}$, whose Lagrangian can be written as $\cal T$ coupled to a $\bZ_N$ gauge theory:\footnote{There are generally different ways of gauging a global symmetry, which are related by the discrete torsion.  In this section we  make a particular choice here. See \cite{Choi:2022zal,Choi:2022jqy} for examples of other choices of the discrete torsion.}
\ie\label{TZN}
{\cal L}_{ {\cal T}/\bZ_N^{(q)}} = {\cal L}_{\cal T} [a] + {iN\over 2\pi } bda
\fe
where $a, b$ are $(q+1)$- and $(d-q-2)$-form gauge fields, respectively.  
The gauged theory ${\cal T}/\bZ_N^{(q)}$ has a dual $(d-q-2)$-form global symmetry $\widehat{\bZ_N}^{(d-q-2)}$ \cite{Gaiotto:2014kfa,Tachikawa:2017gyf}, whose generators are the topological Wilson operators of $a$:
\ie\label{eta}
\hat\eta = \exp\left( i \oint a \right)\,.
\fe

\subsection{Topological boundary conditions of discrete gauge theory}

The last term of \eqref{TZN} 
\ie\label{BF}
 {iN\over 2\pi } bda
\fe
 is the Lagrangian for a  $\bZ_N$   gauge theory of $a$ in $d$ spacetime dimensions  \cite{Maldacena:2001ss,Banks:2010zn,Kapustin:2014gua}. 
 It is also known as the BF theory, with ``F" stands for the field strength $da$. 
 The equation of motion for $b$ enforces $a$ to be a $\bZ_N$ gauge field.  
(For $d=3$, $q=0$, $N=2$, it reduces to the 2+1d $\bZ_2$ gauge theory we discussed in \eqref{Z2Lag}.) 
Dually, \eqref{BF} can be viewed as  a $\bZ_N$   gauge theory of $b$. The equation of motion for $a$ enforces $b$ to be $\bZ_N$-valued.

To proceed, we need to analyze  \eqref{BF} by itself without coupling to the QFT $\cal T$.  
We are particularly interested in two boundary conditions of the BF theory:
\ie\label{topbc}
&\text{Dirichlet b.c.:}~~~ a | = 0\,,\\
&\text{Neumann b.c.:}~~~ b | = 0\,,
\fe 
where $|$ stands for the restriction of the differential form to the boundary. 
The notions of ``Dirichlet" and ``Neumann" here are conventional, and depend on whether we view \eqref{BF} as a gauge theory of $a$ or of $b$. Here we take the former perspective for reasons that will become clear later.  
These boundary conditions are topological in the sense that infinitesimal deformation of the boundary locus does not affect any  correlation function.

One intuitive way to understand the topological nature of, say, the Dirichlet boundary condition is the following. 
The equation of motion for $b$ implies that $a$ is a flat gauge field, and therefore infinitesimal deformations of the locus where $a|=0$ do not change the correlation functions. 
This is similar to the usual argument for the topological nature of the Wilson lines in Chern-Simons theory.  
For $q=0$ and $d=2$, this argument can be made precise on the lattice  using a discrete version of the BF theory from  \cite{Gorantla:2021svj}.
See Section 5.1.3 of \cite{Choi:2021kmx}.

In $d>3$, the BF theory has infinitely many topological (simple) boundary conditions, since one can always stack a decoupled $d-1$-dimensional TQFT to construct another one  (see Section \ref{sec:space}). 
When $d=3$, the space of topological boundary conditions of a given 2+1d TQFT  are under better control. This is because all nontrivial 1+1d TQFTs (without symmetry) have multiple topological local operators. Thus, stacking a decoupled 1+1d TQFT with a simple topological boundary does not give a simple boundary. 
See \cite{1995PhRvL..74.2090H,Bravyi:1998sy,Kapustin:2010hk,2012CMaPh.313..351K,Wang:2012am,2013PhRvX...3b1009L,2013PhRvB..88x1103B,2014PhRvB..89l5307K,Hung:2014tba,Lan:2014uaa,Kaidi:2021gbs} for simple, topological  boundary conditions of 2+1d TQFTs, and \cite{2010arXiv1009.2117D,2011arXiv1109.5558D,Fuchs:2012dt,Kong:2019byq,Kong:2019cuu,Freed:2020qfy} for  discussions in the context of modular tensor category and fully extended TQFT.
In particular, when $d=3, q=0, N=2$, \eqref{topbc} are the low energy limits of the two gapped boundary conditions of the toric code in \cite{Bravyi:1998sy}.

\subsection{Gauging in half of the spacetime}

Having discussed the topological boundary conditions of the BF theory itself \eqref{BF}, we now return to the coupled system of $\cal T$ and the BF theory in \eqref{TZN}. 
While the Neumann boundary condition $b|=0$ is no longer topological in the presence of the coupling, the Dirichlet boundary  $a|=0$ remains topological in \eqref{TZN} because the equation of motion for $b$ is implies $a$ is flat.

\begin{figure}[h!]
\centering
\includegraphics[width=.4\textwidth]{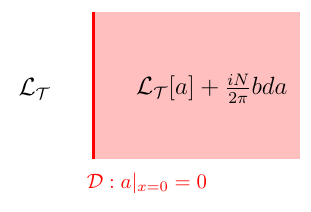}
\caption{We gauge the $q$-form $\bZ_N^{(q)}$ global symmetry of the QFT $\cal T$ only in half of the spacetime $x>0$. More specifically, we couple $\cal T$ to the BF theory in $x>0$ and impose a topological Dirichlet boundary condition $a| =0$ at $x=0$. This half gauging gives a topological interface $\CD$ between two QFTs, ${\cal T}$ and ${\cal T}/\bZ_N^{(q)}$. }\label{fig:halfgauging}
\end{figure}

Equipped with this topological  boundary condition for \eqref{TZN}, we can make precise what it means to gauge in only half of the spacetime, say, $x>0$. 
We start with the QFT $\cal T$ in the entire spacetime. All the degrees of freedom for $\cal T$ are continuous across $x=0$. 
Next, we introduce the BF theory \eqref{BF} only in half of the spacetime $x>0$, and impose  the topological Dirichlet boundary condition $a|_{x=0}=0$ at the interface. 
This construction gives a topological interface $\CD$ between $\cal T$ and ${\cal T}/\bZ_N^{(q)}$. 
See Figure \ref{fig:halfgauging}. 
We again refer the readers to  Section 5.1.3 of \cite{Choi:2021kmx} for a rigorous proof that the interface $\CD$ is topological.

The statement is more generally true for a QFT $\cal T$ with  a discrete $q$-form global symmetry $G^{(q)}$.  There is a topological interface $\CD$ between  $\cal T$ and its gauged version ${\cal T}/G^{(q)}$. 
The orientation reversal interface $\overline{\CD}$ can be similarly defined by gauging in the other half of the spacetime. 
It then follows from Figure \ref{fig:DDC} that
\ie\label{DDC2}
\CD (M) \times \overline{\CD}(M) = \CC(M)
= {1\over N} \sum_{S \in H_2( M,G)} \eta (S)\,,
\fe
where $\CC$ is the condensation defect from higher gauging $G^{(q)}$ along the codimension-1 submanifold  $M:x=0$. 
Furthermore, since the gauge field $a$ is set to be zero on the interface, we have
\ie
\CD \times \hat\eta =\CD\,,
\fe
where $\hat\eta=e^{i \oint a}$ is the generator for the dual $\widehat{G}^{(d-q-2)}$ global symmetry in ${\cal T}/G^{(q)}$.\footnote{Here $\widehat{G} = \text{Hom}(G,U(1))$ is defined as the Pontryagin dual of the finite group $G$. When $G$ is an abelian finite group, which is necessarily the case if $q\ge 1$, then $G\simeq \widehat{G}$. When $q=0$ and $G$ is a non-abelian finite group, then $\widehat{G}=\text{Rep}(G)$, which is a non-invertible fusion category. One can also generalize the above construction by replacing $G^{(q)}$ with a non-invertible $q$-form global symmetry, but we do not pursue it here. }

So far,  $\CD$ is an interface between two systems, not a defect in a single system, at least not yet. 
We now consider the special case when there is an isomorphism between the original QFT $\cal T$ and the gauged theory ${\cal T}/\bZ_N^{(q)}$:
\ie\label{iso}
&{\cal T} \simeq {\cal T}/\bZ_N^{(q)}\,,\\
&\bZ_N^{(q)}\simeq \widehat{\bZ_N}^{(d-q-2)}\,.
\fe
Importantly, the isomorphism   identifies the original $\bZ_N^{(q)}$ $q$-form global symmetry of $\cal T$ with the dual $(d-q-2)$-form global symmetry $\widehat{\bZ_N}^{(d-q-2)}$ in ${\cal T}/\bZ_N^{(q)}$.  Of course, \eqref{iso} is possible only if
\ie
q= {d-2\over2}\,.
\fe

The isomorphism \eqref{iso} does not hold for every QFT ${\cal T}$. 
For instance, the $c=1$ compact boson CFT is different from the orbifold $S^1/\bZ_2$ CFT. 
Similarly, the $SU(2)$ Yang-Mills theory differs from the $SU(2)/\bZ_2^{(1)}=SO(3)$ Yang-Mills theory in its line operator spectrum \cite{Kapustin:2005py,Gaiotto:2010be,Aharony:2013hda}.  
Nonetheless, it does happen for some special QFTs and we will provide examples below. 

If there is such an isomorphism \eqref{iso}, we can compose the isomorphism with the topological interface to obtain a topological defect   in a single   quantum system.  
As an abuse of notations, we also denote this defect as $\CD$, which will be called the duality defect. 
The fusion algebra of $\CD$ and the $\bZ_N^{(q)}$ symmetry operator $\eta$ is then given by
\ie\label{dualitynoninv}
&\CD (M) \times \overline{\CD}(M) = \CC(M)
= {1\over N} \sum_{S \in H_2( M,G)} \eta (S)\,,\\
&\eta\times \CD = \CD \times \eta =\CD\,,\qquad\eta^N=1\,,
\fe
where we have identified the $\bZ_N^{(q)}$ symmetry operator $\eta$ with its dual $\hat\eta$.  
The above algebra is generally non-invertible because $\CC$ is a nontrivial operator. We conclude that $\CD$ is a non-invertible global symmetry. 

Importantly, since the topological Dirichlet boundary condition is a well-defined boundary condition associated with a Hilbert space, so is the non-invertible defect from half gauging. This ensures that the non-invertible symmetry from half gauging obeys the operator/defect principle in Section \ref{sec:opdefect}.

\subsection{Examples of duality defects}

Below we give examples of non-invertible duality defects in diverse dimensions.

\subsubsection{1+1d Ising CFT and Kramers-Wannier duality}

We start with yet another construction of the non-invertible symmetry $\CD$ of the Ising CFT.  
In this case $d=2, q=0, N=2$. 
The isomorphism \eqref{iso} follows from the fact that the Ising CFT is isomorphic to its $\bZ_2$ orbifold,
\ie\label{Isingiso}
\text{Kramers-Wannier duality:}~~~\text{Ising CFT} \simeq \text{Ising CFT}/\bZ_2\,,
\fe
where the original $\bZ_2$ global symmetry is identified with the dual $\widehat{\bZ_2}$ global symmetry.  
This is a consequence of the Kramers-Wannier duality at the critical point.\footnote{Note that \eqref{Isingiso} is only true at the critical point.  Away from the critical point, Kramers-Wannier ``duality" is not a duality relating two equivalent descriptions of a single system. Rather, it is a map from the high temperature phase to the low temperature phase. See \cite{Kapustin:2014gua} for related discussions.} 
The isomorphism \eqref{Isingiso} maps the thermal operator $\varepsilon$ to $-\varepsilon$. 
In 1+1d, the duality defect $\CD$ is the same as its orientation reversal \cite{Choi:2021kmx}. The condensation operator  is simply twice the projection operator, i.e., $\CC= I+\eta$. 
The general non-invertible fusion algebra \eqref{dualitynoninv} reduces to that for the $\bZ_2$ Tambara-Yamagami category TY$_+$:
\ie
&\CD\times \CD = \CC=  I+\eta\,,\\
&\eta\times \CD = \CD\times \eta = \CD\,,\qquad \eta^2=1\,.
\fe

The half gauging construction also provides another explanation for the Euclidean process in Figure   \ref{fig:duality}. 
We start with the one involving the thermal operator $\varepsilon$. Since $\varepsilon$  is $\bZ_2$-even, it is not affected by the gauging. 
On the other hand, the isomorphism \eqref{Isingiso}  maps $\varepsilon \to -\varepsilon$. Combining the two steps,  we reproduce Figure \ref{fig:duality}(a). 
Next, consider the process involving the order operator $\sigma$.  Since $\sigma$ is $\bZ_2$-odd, it is no longer a local operator after the gauging. Rather, it  becomes the disorder operator $\mu$ attached to the end of the  $\bZ_2$ line $\eta$. 
This explains Figure \ref{fig:duality}(b).

\subsubsection{1+1d $c=1$ compact boson and T-duality}\label{sec:c=1}

The Lagrangian of the 1+1d $c=1$ compact boson CFT at  radius $R$ is
\ie
{\cal L} = {R^2\over 4\pi }\partial_\mu \phi\partial^\mu \phi\,,~~~~~\phi\sim \phi+2\pi\,.
\fe
It has a $U(1)^\text{m}$ momentum  global symmetry, a $U(1)^\text{w}$ winding    global symmetry, and also a $\bZ_2$ symmetry $\phi\to -\phi$. 
Together, the internal,  invertible 0-form symmetry group at a generic radius is $(U(1)^\text{m}\times U(1)^\text{w})\rtimes \bZ_2$.   
Gauging the $\bZ_N^\text{m}$ subgroup of $U(1)^\text{m}$ maps the compact boson CFT at radius $R$ to that at $R/N$. 
Similarly, gauging the $\bZ_N^\text{w}$ subgroup of $U(1)^\text{w}$ maps $R$ to $NR$. 

The compact boson CFT has the well-known T-duality, which states that the CFT as radius $R$ is equivalent to that at radius $1/R$, with momentum and winding operators exchanged:
\ie\label{Tduality}
\text{T-duality:}~~~R\sim {1\over R}  \,.
\fe

It follows that, when 
\ie
R = \sqrt{N}
\fe
for some positive integer $N$, the CFT is isomorphic to its $\bZ_N^\text{m}$ gauged version. Indeed,
\ie
R=\sqrt{N} \xrightarrow{\text{gauge~}\bZ_N^\text{m} } R={1\over \sqrt{N}} \xrightarrow{\text{T-duality}}  R= \sqrt{N} \,.
\fe
Given the isomorphism from the T-duality \eqref{Tduality}, we can half gauging the momentum $\bZ_N^\text{m}$ symmetry to construct a non-invertible symmetry in the compact boson CFT at $R=\sqrt{N}$.  
The explicit worldline Lagrangian for the resulting non-invertible defect is \cite{Choi:2021kmx}:
\ie
{i N\over 2\pi} \int_{x=0} \phi_\text{L} d\phi_\text{R}
\fe
where $\phi_\text{L,R}$ are the compact boson fields to the left and right of the defect at $x=0$.  
Other non-invertible global symmetries of the $c=1$ CFTs can be found in \cite{Fuchs:2007tx,Thorngren:2019iar,Thorngren:2021yso}.

\subsubsection{3+1d $U(1)$ Maxwell theory and S-duality}\label{sec:maxwell}

Next, we consider the 3+1d free $U(1)$ Maxwell gauge theory without matter fields:
\ie
{\cal L } = {1\over 2e^2}F\wedge \star F  + {i\theta\over 8\pi^2}F\wedge F
\fe
where $F=dA$ and $A$ is the dynamical $U(1)$ gauge field.  
We normalize our gauge fields so that $\oint F\in 2\pi \bZ$ on any closed 2-cycle in spacetime, i.e., the magnetic fluxes are quantized appropriately. 
The Maxwell theory is a 3+1d free CFT with a complex exactly marginal deformation
\ie
\tau  = {2\pi i \over e^2} +{\theta\over2\pi}\,.
\fe 
The exactly marginal deformation  $\tau$ does not change the local operator correlation function, but affects the line operator spectrum and the partition functions. 
(See, for instance, \cite{Witten:1995gf,Metlitski:2015yqa}.)

The internal global symmetry of the Maxwell theory at a generic $\tau$ is $(U(1)_\text{e}^{(1)} \times U(1)_\text{m}^{(1)} )\rtimes \bZ_2$, where the two $U(1)^{(1)}$'s are the electric and magnetic 1-form global symmetries \cite{Gaiotto:2014kfa}, and $\bZ_2$ is the charge conjugation symmetry which acts as $A\to -A$.\footnote{In contrast to the pure $U(1)_{2N}$ Chern-Simons theory, here the charge conjugation symmetry is not obtained from higher gauging and is not generated by a condensation operator. Indeed, the charge conjugation symmetry of the Maxwell theory acts nontrivially on the local operator such as $F=dA$. }  
The electric $U(1)^{(1)}_\text{e}$ 1-form symmetry shifts the gauge field by a flat connection, and acts on the Wilson lines. The symmetry operator is $\exp({\alpha\over e^2} \oint \star F )$. 
The magnetic $U(1)^{(1)}_\text{m}$  1-form symmetry acts on the 't Hooft lines. The symmetry operator is $\exp( {i\alpha\over 2\pi} \oint F)$. 

Gauging the $\bZ_N^{(1)}$ subgroup of the electric $U(1)^{(1)}_\text{e}$ 1-form global symmetry rescales the gauge field as $A\to A/N$, and therefore maps the theory from $\tau$ to $\tau/N^2$. Gauging a subgroup  of the magnetic $U(1)^{(1)}_\text{m}$ does the opposite. 

Famously, on spin manifolds, the Maxwell theory has the $SL(2,\bZ)$ duality, which states the following isomorphism between the theories at two different $\tau$'s:
\ie
\tau \simeq {a\tau+b\over c\tau+d}\,,~~~~ad-bc=1, ~a,b,c,d\in \bZ\,.
\fe
Below we focus on the S transformation:
\ie\label{Sduality}
\text{S-duality:}~~~~\tau\simeq -1/\tau\,.
\fe

Using the S-duality \eqref{Sduality}, we find that the Maxwell theory at
\ie
\tau  = i N
\fe
is isomorphic to the theory gauged by the electric $\bZ_N^{(1)}$ 1-form global symmetry. Indeed, we have
\ie
\tau=iN \xrightarrow{\text{gauge~}\bZ_N^{(1)} } \tau={i\over N} \xrightarrow{\text{S-duality}}  \tau=iN \,.
\fe
We can therefore apply half gauging to the Maxwell theory at $\tau=iN$ and obtain a non-invertible duality defect $\CD$ obeying the fusion algebra \eqref{dualitynoninv}, with $\eta= \exp({2\pi \over N e^2}\oint \star F)$ being the generator for the $\bZ_N^{(1)}$ subgroup of the electric $U(1)^{(1)}_\text{e}$ 1-form global symmetry.  
At the self-dual point $\tau=i$, the duality defect reduces to an invertible $\bZ_4$ symmetry \cite{Gaiotto:2008ak,Kapustin:2009av}. 
The explicit worldvolume of this duality defect is \cite{Choi:2021kmx}
\ie
{iN\over 2\pi} \int _{x=0} A_\text{L} dA_\text{R}
\fe
where $A_\text{L,R}$ are the dynamical gauge fields to the left and right of the defect at $x=0$.

\begin{figure}
\centering
\includegraphics[width=.4\textwidth]{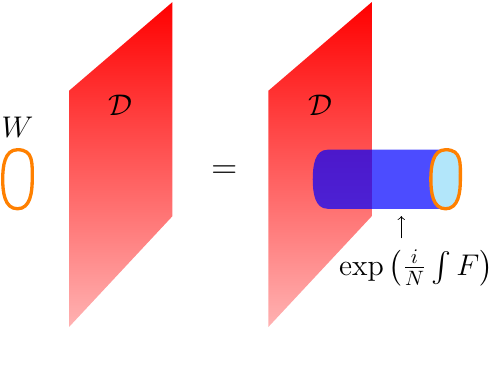}
\caption{The non-invertible symmetry $\CD$ of the 3+1d Maxwell gauge theory at $\tau = iN$. As we bring a minimally charged Wilson line $W=e^{i\oint_\gamma A}$ past the non-invertible defect $\CD$, it becomes the boundary line of a topological surface operator $\exp\left( {i\over N} \int F\right)$. The latter is a fractional Wilson line with charge $1/N$. 
This Euclidean process is the 3+1d counterpart of Figure \ref{fig:duality} in the 1+1d Ising model.  }\label{fig:maxwell}
\end{figure}

How does this non-invertible symmetry $\CD$ act on the operators?  
What are the analogs of the Euclidean processes in Section \ref{sec:euclidean}? 
As we bring a minimally charged (non-topological) Wilson line $W=e^{i \oint_\gamma A}$ from the left  past the duality defect $\CD$ to the right, it is gauged and 
 $A\to A/N$. Hence, the charge $+1$ Wilson line on the left side of the defect becomes a charge $1/N$ Wilson line on the right side:
\ie
\exp\left(i \oint_\gamma A\right) \to \exp\left( {i\over N}\oint_\gamma A\right) = \exp\left( {i\over N}\int_\Sigma F\right)\,,
\fe
where $\Sigma$ is the blue cylinder shown in Figure \ref{fig:maxwell}, whose boundary on the right is the curve $\gamma$. 
A fractionally charged Wilson line is not a genuine line operator; rather, it is the boundary of a topological surface $\exp\left( {i\over N}\int_\Sigma F\right)$. 
The fact that a genuine line operator $W$ becomes a boundary line attached to a topological surface is the hallmark of the non-invertible nature of $\CD$.
Under the isomorphism \eqref{Sduality}, this surface operator, which is the generator for the magnetic 1-form symmetry on the right, is identified as the generator of the $\bZ_N^{(1)}$ electric 1-form global symmetry on the left.

The non-invertible symmetries in the free Maxwell theory at $\tau = iN$ was first studied in \cite{Choi:2021kmx}.  
It was soon generalized to other locations in the space of $\tau$:
\begin{itemize}
\item Non-invertible triality defects at $\tau = N e^{2\pi i /3}$ for $N\in \mathbb{N}$ \cite{Choi:2022zal}.
\item Non-invertible time-reversal and $\mathsf{CP}$ symmetries at $\theta=\pi p/N$ for every gcd$(p,N)=1$ \cite{Choi:2022rfe}. They are generated by anti-linear operators satisfying $\mathsf{T}\times \mathsf{T}^\dagger=\CC_N$, where $\CC_N$ is the condensation defect. 
\item  In   \cite{Niro:2022ctq}, it was found that there is a non-invertible (linear) symmetry in the free Maxwell theory  for every rational values of $4\pi^2/e^4,\theta/2\pi$. More specifically, at
\ie
{\theta\over2\pi } = {N_R-N_L\over2}\,,~~~~{2\pi\over e^2} = \sqrt{N^2- \left({N_L+N_R\over2}\right)^2}\,,~~~~N,N_L,N_R\in \mathbb{Q},
\fe
the non-invertible symmetry rotates the field strength and its dual by an angle $\varphi$ with $\cos\varphi ={N_R+N_L\over2N}$, and acts non-invertibly on the line defects. The special case of $\theta=0$ was later analyzed in more detail in \cite{Cordova:2023ent}. 
\end{itemize}

\subsubsection{3+1d ${\cal N}=4$ super Yang-Mills theory and S-duality}\label{sec:N=4}

The non-invertible symmetries of the ${\cal N}=4$ super Yang-Mills theories  are explored in \cite{Kaidi:2021xfk,Choi:2022zal,Kaidi:2022uux}. 
  Below we focus on the simplest example of a non-invertible symmetry in the ${\cal N}=4$ $SU(2)$ super Yang-Mills theory at the self-dual point $\tau=i$ \cite{Kaidi:2021xfk}. 
  See also \cite{Choi:2022rfe} for the non-invertible time-reversal symmetries in this theory.

The ${\cal N}=4$ $SU(2)$ gauge theory has a complex exactly marginal deformation $\tau$. The S-duality is the statement that the $SU(2)$ theory at $\tau$ is equivalent to the $SO(3)$ theory at $-1/\tau$:
\ie\label{N=4S}
\text{S-duality}:~~SU(2)~\text{at}~\tau~~ ~\simeq~~ ~SO(3)~\text{at}~-1/\tau\,.
\fe

Since all the fields are in the adjoint representations, the $SU(2)$ theory has a $\bZ_2^{(1)}$ center 1-form global symmetry. Gauging the $\bZ_2^{(1)}$ center symmetry gives the $SO(3)$ theory, while preserving $\tau$.

Using the S-duality \eqref{N=4S}, we find that the $SU(2)$ theory is isomorphic to $SU(2)/\bZ_2^{(1)}$ at the self-dual point $\tau=i$:
\ie
SU(2)\text{~at~}\tau=i \xrightarrow{\text{gauge~}\bZ_2^{(1)} } SO(3)\text{~at~}\tau={i} \xrightarrow{\text{S-duality}}  SU(2)\text{~at~}\tau=i \,.
\fe
We can therefore half gauge the $\bZ_2^{(1)}$ symmetry of the $SU(2)$ theory at the self-dual point $\tau=i$ to produce a non-invertible duality defect.  
This is the usual S-duality defect of the ${\cal N}=4$ theory. 
The new point here is that while it acts as an invertible $\bZ_4$ global symmetry on the local operators, it acts non-invertibly on the line defects. Therefore, the S-duality defect of the ${\cal N}=4$ theory is non-invertible.

\section{Why do pions decay?}

Do non-invertible symmetries exist in Nature? 
The answer is a resounding yes. 
In this section we discuss non-invertible global symmetries in the 3+1d QED and QCD for the real world \cite{Choi:2022jqy,Cordova:2022ieu}. 
In the context of QCD, the neutral pion decay $\pi^0 \to 2\gamma$ is reinterpreted as a consequence of a non-invertible global symmetry \cite{Choi:2022jqy}.

\subsection{Is chiral symmetry a symmetry?}

Consider the 3+1d $U(1)$ QED with a single, massless, unit charge Dirac fermion $\Psi$:
\ie
{1\over 4e^2} F_{\mu\nu}F^{\mu\nu}  + i \bar\Psi (\partial_\mu - iA_\mu)\gamma^\mu \Psi\,,
\fe
where $A_\mu$ is the dynamical $U(1)$ gauge field. 
The gauge field is normalized so that the magnetic flux through any 2-cylce is quantized as $\oint F\in 2\pi \bZ$.
Classically, the Lagrangian has a chiral $U(1)_\text{A}$ symmetry
\ie\label{chiralrot}
U(1)_\text{A}:~~~\Psi \to e^{ i{\alpha \over2} \gamma_5 }\Psi\,,
\fe
where we normalize $\alpha$ so that $\alpha\sim \alpha+2\pi$. Note that when $\alpha=2\pi$, the $U(1)_A$ transformation corresponds to $\Psi\to -\Psi$, which is part of the gauge group.

Define the axial current as
\ie\label{anomcon}
j^\text{A}_\mu = \frac 12 \bar\Psi\gamma_5\gamma_\mu\Psi\,. 
\fe
The axial current obeys the anomalous conservation equation
\ie
d\star j^\text{A}  = {1\over 8\pi^2}F\wedge F\,.
\fe
Since the current $j^\text{A}_\mu$ is not conserved, the classical chiral symmetry $U(1)_\text{A}$ fails to be a global symmetry quantum mechanically. 
This is the famous Adler-Bell-Jackiw (ABJ) anomaly \cite{PhysRev.177.2426,Bell:1969ts}.  
The general structure of the anomaly was later analyzed in \cite{PhysRev.184.1848} and many other references.

However, this is not the end of the story.  
In some sense  $U(1)_\text{A}$ is still a symmetry in flat spacetime.  
For instance, the scattering amplitudes of electrons and positrons obey the helicity conservation law, which is a selection rule that follows from the chiral $U(1)_\text{A}$ symmetry.  
The traditional explanation is the following. 
The chiral $U(1)_\text{A}$ symmetry is broken by the instanton.  However, there is no $U(1)$ instanton in flat spacetime because $\pi_3(U(1))=0$. Therefore the chiral $U(1)_\text{A}$ symmetry is unbroken.  

Another way to see that there is a symmetry in flat spacetime is the following. 
In \cite{PhysRev.177.2426}, Adler defines the following operator in $\mathbb{R}^3$:
\ie
\hat U_\alpha(\mathbb{R}^3) = \exp\left[
i \alpha \int_{\mathbb{R}^3} 
\left( \star j^\text{A} - {1\over 8\pi^2}A dA\right)
\right]
\fe
The term $\star j^\text{A} - {1\over 8\pi^2}A dA$ in the parentheses is formally closed, but it is not gauge-invariant. 
Nonetheless, the operator $\hat U_\alpha(\mathbb{R}^3)$ is conserved and gauge-invariant, and leads to selection rules such as the helicity conservation in flat spacetime. 
(See also \cite{Harlow:2018tng} for a recent discussion.) 
 
However, if the 3-dimensional space $M$ has a nontrivial topology, then\footnote{Here and below we put operators such as  $\hat U_\alpha(M)$ that are  not gauge-invariant in quotation marks.}
\ie\label{hatU}
`` \, \hat U_\alpha (M) = \exp\left[ i\alpha\oint_M \left(
\star j^\text{A} -{1\over 8\pi^2 }A dA\right)\right]\, "
\fe
is no longer gauge-invariant. 
This is because the Chern-Simons action $\exp\left[ {i N\over 4\pi}\oint_M A dA\right]$ is only gauge-invariant if the level $N$ is an integer, and here the level is $\alpha\over 2\pi$.\footnote{We assume $M$ to be a spin manifold. One can generalize the discussion to spin$^c$ manifolds, but we do not pursue it here.}

One might argue that we are pretty comfortable in flat spacetime, so why bother? 
But the moment we have a magnetic monopole, it creates a nontrivial topology, and the chiral $U(1)_\text{A}$ symmetry is   violated. 
Indeed, it is well-known that helicity is not conserved in monopole scattering such as in the discussion of the Callan-Rubakov effect \cite{PhysRevD.15.2287,Callan:1982ah,Rubakov:1982fp}.

More conceptually, global symmetry is an intrinsic  property of a quantum system that does not depend on the duality frame. 
Sometimes, different ways of describing the same quantum system can look very different, but global symmetry helps us recognize and classify them.
Therefore, the global symmetry of a quantum system should be something inherent  to the quantum system. 
The fact that the chiral symmetry only exists in flat spacetime seems to be in tension with this philosophy.

So is the chiral $U(1)_\text{A}$ symmetry a global symmetry of the massless QED? 
The conventional answer is both yes and no.
Yes, in flat spacetime, because of the operator $\hat U_\alpha(\mathbb{R}^3)$. 
No, because   the symmetry is violated when we have monopoles or  nontrivial spacetime topologies.

Below we will provide an alternative viewpoint on the ABJ anomaly. 
We will see that the chiral symmetry is not completely broken by the ABJ anomaly; rather, it is resurrected as a non-invertible global symmetry.

\subsection{Fractional quantum Hall state cures the ABJ anomaly}\label{sec:FQH}

Let us be less ambitious, and focus on the case where the chiral rotation angle is a fraction:
\ie
\alpha  = {2\pi \over N}
\fe
where $N$ is any positive integer.   
The naive operator $\hat U_{2\pi \over N}(M)$ is still not gauge-invariant, because the Chern-Simons term
\ie\label{FQH1}
``  \, { i \over 4\pi N} \oint_M AdA \, "  
\fe
has a fractional level $1/N$.

Interestingly, the action \eqref{FQH1} makes an appearance in a completely different physical system: it is the effective response action for the fractional quantum Hall (FQH) state in 2+1d with filling fraction  $\nu=1/N$. 
In that context,  $A$ is a background gauge field, associated with the background magnetic field used in a table-top experiment, and $M$ is the spacetime manifold of the 2+1d system.  

However, as we stressed above, \eqref{FQH1} is not gauge invariant. 
How is it possible that a realistic physical system  is described by a gauge non-invariant action? 
Fortunately, there is a well-known fix to \eqref{FQH1}. 
The more precise,  gauge-invariant action for the FQH effect is
\ie\label{FQH2}
{iN\over 4\pi }ada +{i\over 2\pi } adA\,,
\fe
where we have introduced an additional dynamical $U(1)$ gauge field $a$.  
The new action \eqref{FQH2} is gauge-invariant because the levels are both properly quantized. 

Naively, one is tempted to integrate out $a$ in \eqref{FQH2} and find
\ie\label{illegal}
``\, a = - {A\over N}\, ."
\fe
Substituting this into \eqref{FQH2} returns \eqref{FQH1}. 
However, the manipulation is not globally correct because both $a$ and $A$ are properly normalized gauge fields with quantized magnetic fluxes, i.e., $\oint da , \oint dA \in 2\pi \bZ$. 
The substitution \eqref{illegal} does not respect the above quantization condition, and is therefore illegal. 
Indeed, we should not be able to relate a gauge non-invariant action \eqref{FQH1} to a gauge-invariant one \eqref{FQH2} by a valid manipulation.  
Having said that, the substitution \eqref{illegal} provides a heuristic understanding of the relation between them

In any case, in the context of the FQH effect, \eqref{FQH2} is the accurate, gauge-invariant effective action. 
Nonetheless, \eqref{FQH1} is still a very powerful and useful description that is valid for most of the local observables of the physical system.

We now return to the 3+1d QED. 
Motivated by the above discussion of the 2+1d FQH effect, we introduce a new operator  in QED \cite{Choi:2022jqy,Cordova:2022ieu}:
\ie\label{DQED}
\CD_{1\over N} (M) =  \int [Da]_M 
\exp\left[
 \oint_M \left(
{2\pi i\over N} \star j^\text{A} +  {iN\over 4\pi}ada + {i\over 2\pi }adA
\right)
\right] \,.
\fe
Importantly, $a$ is an auxiliary 1-form gauge field that only lives on the closed 3-dimensional  manifold $M$ on which  this operator is supported.  
It does not introduce any new asymptotic states in QED, and the bulk physics away from $M$ remains the same. 
$\CD_{1\over N}(M)$ can also be used as a defect when the 3-manifold $M$ extends in the time direction.

The new operator $\CD_{1\over N}$ is gauge-invariant because both Chern-Simons terms have properly quantized levels. 
It is also conserved, and more generally, topological.  
A heuristic way to understand this is to locally integrate out $a$ and use the anomalous conservation equation \eqref{anomcon}.  
More rigorously, the topological nature of $\CD_{1\over N}$ follows from half gauging, as we explain below.

We first recall that the free Maxwell theory has both the electric and the magnetic $U(1)^{(1)}$ 1-form symmetries. The coupling to the Dirac fermion breaks the electric 1-form symmetry, but preserves the magnetic one. 
It is shown in Section 2.3 of \cite{Choi:2022jqy} that gauging the $\bZ_N^{(1)}$ subgroup of the magnetic 1-form symmetry with a particular choice of the discrete torsion shifts the QED Lagrangian by a $\theta$-angle:
\ie\label{thetashift}
\theta \to  \theta- {2\pi \over N}\,.
\fe
However, the $\theta$-angle of the massless QED is not physically meaningful; it can be removed by a chiral rotation of the Dirac fermion \cite{PhysRevD.29.285}.  
Therefore, gauging the $\bZ_N^{(1)}$ magnetic 1-form global symmetry leaves QED invariant, up to an isomorphism from the chiral rotation. 
As shown in  \cite{Choi:2022jqy}, half gauging this $\bZ_N^{(1)}$ magnetic 1-form symmetry reproduces the \eqref{DQED}.  
By the general argument in Section \ref{sec:half}, we conclude that $\CD_{1\over N}$ is topological.

Thus, we have constructed a new gauge-invariant and conserved operator $\CD_{1\over N}$ that can be defined on any closed 3-manifold $M$. 
However, there is a price we pay: the operator is not invertible.  
Below we first show that the operator is non-unitary.  
Multiplying $\CD_{1\over N}$ with $\CD_{1\over N}^\dagger=  \CD_{- {1\over N}}$, we find
\ie
~&\CD_{1\over N} (M ) \times \CD_{1\over N}^\dagger(M)= 
\int [Da]_M [D\bar a]_M 
\exp\left[
\oint_M \left(
{iN\over 4\pi} ada - {iN\over 4\pi} \bar ad\bar a 
+{i\over 2\pi} (a-\bar a)dA
\right)
\right] \\
&= \CC_N(M)\,,
\fe 
where $\CC_N$  is the condensation operator from higher gauging the magnetic $\bZ_N^{(1)}$ 1-form symmetry (see Appendix B.2 of \cite{Choi:2022jqy} for more details). 
Since $\CD_{1\over N} \times\CD_{1\over N}^\dagger\neq1$, we conclude that $\CD_{1\over N}$ is not unitary. 
An intuitive way to understand that $\CD_{1\over N}$ is not unitary is that \eqref{DQED} takes the form of an integral of unitaries, which is not unitary. 

How do we show that $\CD_{1\over N}$ is a non-invertible operator/defect? 
This follows from the fact that the fractional quantum Hall state \eqref{FQH2} is a non-invertible topological phase in 2+1d, meaning that it cannot be trivialized by stacking. 
Since $\CD_{1\over N}$ contains a fractional quantum Hall state, it is also non-invertible. 
Indeed, when the space $M$ has a nontrivial magnetic flux, the operator $\CD_{1\over N}(M)$ has a kernel and is non-invertible \cite{Cordova:2022ieu}. 

More generally, we can resurrect the chiral rotation with angle $\alpha= 2\pi p/N$, where gcd$(p,N)=1$, to  a non-invertible operator $\CD_{p\over N}$:
\ie\label{DpN}
\CD_{p\over N}(M) =\exp\left[
\oint_M \left(
{2\pi i p\over N} \star j^\text{A}   +{\cal A}^{N,p}[dA/N]
\right)
\right]\,.
\fe
Here ${\cal A}^{N,p}[B]$ is the 2+1d minimal $\bZ_N$ TQFT \cite{Hsin:2018vcg} (see also \cite{Moore:1988qv,Bonderson:2007ci,Barkeshli:2014cna}) coupled to a $\bZ_N$ background 2-form gauge field $B$. 
It is the effective TQFT for the $\nu= p/N$ FQH state. 
The operator $\CD_{p\over N}$ can be constructed from half gauging the $\bZ_N^{(1)}$ magnetic 1-form symmetry with a more general discrete torsion. See \cite{Choi:2022jqy,Putrov:2022pua} for  the fusion algebras of these non-invertible operators.

We therefore conclude that the classical $U(1)_\text{A}$ chiral symmetry is not entirely broken by the ABJ anomaly.  
For every rational angle, it is resurrected as a non-invertible global symmetry $\CD_{p\over N}$ labeled by $p/N\in\mathbb{Q}/\bZ$.  
Intuitively,  the FQH state cures the ABJ anomaly. 
This is an infinite non-invertible symmetry labeled by the rational numbers, which are dense in the real line. 
However, it is not a continuous symmetry.

Let us compare the ABJ anomaly with other anomalies in 3+1d. 
Consider a perturbative QFT with a $U(1)_\text{global}$ global symmetry and a $U(1)_\text{gauge}$ symmetry.  
There are four possible one-loop triangle diagrams  shown in Table \ref{table:triangle}. 
\begin{itemize}
\item The first case is a triangle diagram involving  three global symmetry currents. It signals an 't Hooft anomaly for the global symmetry. 
There is nothing wrong about it. 
We have a perfectly well-defined, invertible $U(1)_\text{global}$ global symmetry in a well-defined QFT. You just can't gauge $U(1)_\text{global}$, but you don't have to.  
The B-L symmetry of the Standard Model is one such example. 
\item The second case is a diagram with two global symmetries and one gauge symmetry.  
It signals another kind of generalized global symmetry, the 2-group symmetry, where the 0- and 1-form global symmetries are mixed and form a nontrivial extension class \cite{Cordova:2018cvg}. 
\item The third case is a diagram with one gauge and two global symmetry currents. This is the ABJ anomaly discussed above, where the symmetry becomes non-invertible. 
\item The fourth case is when all three external legs are gauge symmetries, in which case the QFT is inconsistent. This is sometimes called a gauge anomaly, which should be canceled in any consistent QFT.
\end{itemize}

\begin{table}[h!]
\begin{align*}
\left.\begin{array}{|c|c|c|c|c|}
\hline \text{triangle} & U(1)_\text{global}&  U(1)_\text{gauge} & U(1)_\text{global} & U(1)_\text{gauge} \\
\text{diagrams}& U(1)_\text{global}~~U(1)_\text{global} &  U(1)_\text{global}~~U(1)_\text{global}  &  U(1)_\text{gauge}~~U(1)_\text{gauge}   &  U(1)_\text{gauge}~~U(1)_\text{gauge}   \\
\hline
\text{global }& \text{invertible}& \text{invertible}&\text{non-invertible}& \text{inconsistent} \\
\text{symmetry}&U(1)_\text{global}&\text{2-group}&~\CD_{p\over N}&\text{QFT}\\
\hline
 \end{array}\right.
\end{align*}
\caption{Consequences of  triangle diagrams in 3+1d involving a  $U(1)_\text{global}$ global symmetry and a $U(1)_\text{gauge}$ gauge symmetry.}\label{table:triangle}
\end{table}

\subsection{Selection rules, naturalness, and monopoles}\label{sec:monopole}

The action of the non-invertible symmetry $\CD_{p\over N}$ on the fields and operators can be most easily derived from the half gauging construction. 
Since the fermion fields  are unaffected by gauging the magnetic 1-form symmetry, $\CD_{p\over N}$ acts on  $\Psi$ as an invertible chiral rotation \eqref{chiralrot}.   
The non-invertible global symmetry thus gives an exact explanation for the helicity conservation of electron-positron scattering in massless QED.

In particular, the Dirac mass term $m\bar\Psi\Psi$ transforms by a phase under the non-invertible global symmetry $\CD_{p\over N}$. 
This gives another interpretation of naturalness  \cite{tHooft:1979rat} in massless QED. 
Given a global symmetry, a theory is said to be natural if all symmetry preserving terms in the Lagrangian have ${\cal O}(1)$ coefficients.  
Traditionally, the electron is said to be naturally massless in QED because of the classical chiral symmetry $U(1)_\text{A}$, which is only a symmetry in flat spacetime.  
With the new interpretation, we can now say that the massless QED is natural because of the non-invertible global symmetry ${\cal D}_{p\over N}$. 
See \cite{Choi:2022jqy,Cordova:2022ieu} for more discussions on naturalness and \cite{Cordova:2022fhg} for applications to neutrino physics.

\begin{figure}[t]
    \centering
    \begin{subfigure}[T]{.45\textwidth}
        \centering
        \includegraphics[width=0.9\linewidth]{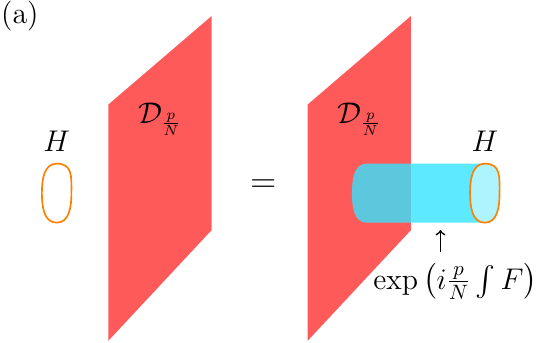}
    \end{subfigure}
    \hspace{0.5cm}
    \begin{subfigure}[T]{.45\textwidth}
        \centering
        \includegraphics[width=0.9\linewidth]{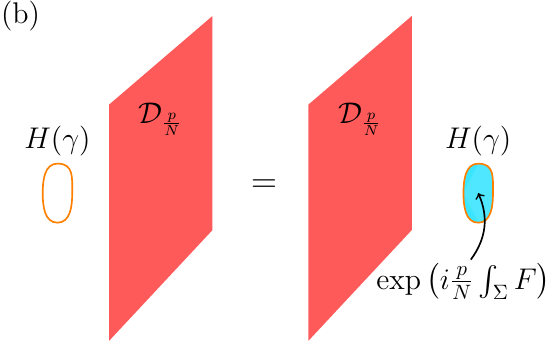}
    \end{subfigure}
    \caption{Action of the non-invertible symmetry  $\mathcal{D}_{\frac{p}{N}}$ on the 't Hooft line $H$.
    (a) As one sweeps the operator $\mathcal{D}_{\frac{p}{N}}$ past the 't Hooft line $H$, the latter is attached to the topological surface  $\exp \left( i \frac{p}{N} \int F \right)$.  
    (b) When the 't Hooft line is supported on a contractible loop $\gamma$ such that $\gamma= \partial \Sigma$ for a 2-dimensional disk $\Sigma$, the surface operator $\exp \left( i \frac{p}{N} \int F \right)$ can be deformed to be supported on $\Sigma$.
    The defect $\exp \left( i \frac{p}{N} \int_\Sigma F \right)$ can be thought of as a dyon with a fractional electric charge $p/N$.  }
    \label{fig:4d}
\end{figure}

While ${\cal D}_{p\over N}$ acts invertibly on the fermions, its non-invertible nature is  revealed when we consider monopoles. 
Consider a minimal 't Hooft line defect $H$, which is the worldline of an infinitely heavy, non-dynamical monopole.  
As $H$ goes to the other side of $\CD_{p\over N}$, it experiences the gauging of the magnetic $\bZ_N^{(1)}$ 1-form symmetry, and is no longer a genuine line defect. 
Rather, it is attached to the topological surface $\exp( {ip\over N}\int F)$ which generates the $\bZ_N^{(1)}$ magnetic 1-form symmetry.  
See Figure \ref{fig:4d}.

This Euclidean process is a consequence of the Witten effect \cite{Witten:1979ey}.  
As the 't Hooft line passes $\CD_{p\over N}$, the $\theta$-angle is shifted according to \eqref{thetashift}. 
By the Witten effect, a monopole of unit magnetic charge turns into a dyon with unit magnetic charge and a fractional electric charge $p/N$. 
The worldline of the dyon is not a genuine line defect, but the boundary line of a topological surface. This is precisely the defect $\exp( {ip\over N}\int F)$. 

We refer the readers to \cite{vanBeest:2023dbu} for further applications of non-invertible symmetries and higher groups in the context of monopole scattering. 
See \cite{Copetti:2023mcq} for more selection rules from the non-invertible symmetry.

\subsection{QCD and pion decay}\label{sec:pion}

We now return to the problem of pion decay, which was the motivation for the discovery of the ABJ anomaly. See the famous review \cite{tHooft:1986ooh}  for more detail.
Below we phrase the puzzle in the modern QCD language. 
Consider QCD below the electroweak scale where the $SU(2)\times U(1)_Y$ gauge group is broken to the electromagnetic gauge group $U(1)_\text{EM}$. 
We focus on the first generation of the up and down quarks in the massless limit. 
The classical QCD Lagrangian has the following symmetry in the massless quark limit:
\ie\label{chiralQCD}
U(1)_\text{A3}:~~\left(\begin{array}{c}u \\d\end{array}\right) \to 
e^{ i\alpha \gamma_5 \sigma_3}\left(\begin{array}{c}u \\d\end{array}\right) \,,
\fe
with $\alpha\sim \alpha+2\pi$.  
The axial current is $j^\text{A3} _\mu =\frac 12 \bar u \gamma_5 \gamma_\mu u-\frac 12 \bar d \gamma_5\gamma_\mu d$. 
The naive Goldstone boson associated with $U(1)_\text{A3}$ is the neutral pion $\pi^0$, which is shifted by the broken symmetry as 
\ie\label{SSB}
\pi^0 \to \pi^0 -2\alpha f_\pi\,,
\fe
 where $f_\pi \sim 92.4$ MeV is the pion decay constant.  
On the other hand, it was observed experimentally that the dominant decay channel for the neutral pion is into two photons. 
This decay channel would arise from a couple $\pi^0F\wedge F$ in the chiral Lagrangian. However, this term is not compatible with the spontaneously broken symmetry action \eqref{SSB}.  

The conventional resolution was that $U(1)_\text{A3}$ is broken by the ABJ anomaly with the abelian $U(1)_\text{EM}$ gauge group, and therefore $\pi^0$ is not really a Goldstone boson even in the massless quark limit. 
The coefficient of the term $\pi^0 F\wedge F$ was determined by matching the anomalous conservation law $d\star j^\text{A3} = {1\over 8\pi^2}F\wedge F$.

There is something slightly counterintuitive about this classic argument.  
Usually we celebrate for the discovery of a new global symmetry. 
But in the case of the ABJ anomaly, it is exactly the opposite. 
How can we match the absence of a global symmetry to derive a quantitative result that agrees with the experiment? 
Is it possible to reinterpret the result in terms of the presence of a generalized global symmetry?
This is where the non-invertible symmetry comes in to play \cite{Choi:2022jqy}.

By applying the same construction in QED, we find the similar non-invertible global symmetry $\CD_{p\over N}$  in QCD, with $j^\text{A}$ replaced by $j^\text{A3}$ in \eqref{DpN}. 
How does the IR chiral Lagrangian match this non-invertible global symmetry in QCD? 
The relevant terms in the chiral Lagrangian are
\ie
{\cal L}_\text{IR} = \frac 12 d\pi^0 \wedge \star d\pi^0 
+ i g \pi^0 F\wedge F +\cdots\,.
\fe
To determine the coupling constant $g$ in the chiral Lagrangian, we insert a non-invertible defect $\CD_{1\over N}$ at $x=0$. The total action is then
\ie
&\int_{x<0} \left( \frac 12 d\pi^0 \wedge \star d\pi^0 
+ i g \pi^0 F\wedge F\right)\\
& 
+\int_{x=0} \left( 
{2\pi i \over N }\star j^\text{A3}  +{iN\over 4\pi } ada +{i\over 2\pi} adA
\right)\\
&+\int_{x>0} \left( \frac 12 d\pi^0 \wedge \star d\pi^0 
+ i g \pi^0 F\wedge F \right)\,.
\fe
The equations of motion for $\pi^0,a,A$ have both a bulk part and a boundary part at $x=0$. The latter is
\ie
&\pi^0 \Big|_{x=0^+}  -\pi^0 \Big|_{x=0^- } = -{2\pi \over N}f_\pi\,,\\
&Nda + F=0\,,\\ 
&2i g\left(\pi^0 \Big|_{x=0^+}  -\pi^0 \Big|_{x=0^- } \right)F = {i\over 2\pi }da\,,
\fe
where we have used $j^\text{A3} = - i f_\pi d\pi^0+\cdots$ in Euclidean signature. These three boundary equations of motion are only consistent if
\ie
g= {1\over 8\pi^2f_\pi}\,,
\fe
which is the correct value compatible with the experiment.  
We conclude that the coupling 
\ie
{i\over 8\pi^2f_\pi}\pi^0 F\wedge F
\fe
 of the chiral Lagrangian can be determined from matching the non-invertible global symmetry in QCD.  
The non-invertible global symmetry $\CD_{p\over N}$ shifts the $\pi^0$ field as in \eqref{SSB} with $\alpha =2\pi p/N$, suggesting that $\pi^0$ can be viewed as a Goldstone boson for this infinite non-invertible global symmetry labeled by the rational numbers $\mathbb{Q}/\bZ$. (See \cite{GarciaEtxebarria:2022jky} for an alternative viewpoint.)

There are many generalizations of such non-invertible global symmetries to other physical systems, ranging from axions to M-theory  \cite{Damia:2022seq,Damia:2022bcd,Apruzzi:2022rei,Choi:2022rfe,Chen:2022cyw,Choi:2022fgx,Yokokura:2022alv,Garcia-Valdecasas:2023mis,Putrov:2023jqi,Yamamoto:2023uzq,Copetti:2023mcq,vanBeest:2023dbu}. 
However, there are also  limitations to this half gauging construction:
\begin{itemize}
\item When the rotation angle $\alpha/2\pi$ is not rational, there is no corresponding subgroup of the magnetic 1-form symmetry. Therefore we cannot apply our half gauging construction. Relatedly, there is no irrational quantum Hall state to cure the ABJ anomaly.\footnote{The authors of \cite{Karasik:2022kkq,GarciaEtxebarria:2022jky} define a non-invertible \textit{operator} for every chiral rotation angle $\alpha$ by directly integrating over the gauge orbit.  However, it is not entirely clear if their construction leads to  a \textit{defect} associated with a well-defined twisted Hilbert space. Relatedly, it is not clear if their operators have finite quantum dimensions on certain manifolds.  In contrast, the non-invertible symmetries from half gauging  manifestly obey the operator/defect principle since it arises from a well-defined Dirichlet boundary condition of the discrete gauge theory. }
\item If the gauge group is $SU(N)$, then there is no magnetic 1-form global symmetry, and we  cannot apply the half gauging construction to find a non-invertible symmetry in QCD. 
Relatedly, there is no fractional quantum Hall state for $SU(N)$ symmetry. In QCD, consider the classical chiral symmetry  that rotates the $u$ and $d$ quarks as $\left(\begin{array}{c}u \\d\end{array}\right) \to  e^{i \alpha\gamma_5} \left(\begin{array}{c}u \\d\end{array}\right)$,  without the $\sigma_3$ in \eqref{chiralQCD}. 
This symmetry is broken by the ABJ anomaly with the $SU(3)$ color gauge group, and is not resurrected as a non-invertible global symmetry.
Nonetheless, see \cite{Cordova:2022ieu} for non-invertible symmetries when the gauge group is $PSU(N)$.
\item The chiral symmetry in 1+1d QED also has a similar anomaly. However, there is no magnetic global symmetry in 1+1d, so again   the half gauging construction does not apply.
\end{itemize}

\section{Applications}\label{sec:application}

We have seen how these novel global symmetries offer a unified framework for understanding diverse physical phenomena.  
Some examples are the selection rules  in 1+1d CFT (Section \ref{sec:selection}), the helicity conservation law  in massless QED (Section \ref{sec:monopole}), the  pion decay (Section \ref{sec:pion}), the generalized Landau paradigm  for  non-abelian topological orders   (Section \ref{sec:interlude}), and more.

But the program of generalized global symmetries is more than just a cosmetic change of perspective. 
These symmetries lead to new implications for diverse physical systems. 
Below we briefly mention a few of these applications:
\begin{itemize}
\item Non-invertible global symmetries  can have generalized anomalies, which impose dynamical constraints on RG flows \cite{Chang:2018iay,Thorngren:2019iar,Komargodski:2020mxz,Thorngren:2021yso,Choi:2021kmx,Choi:2022zal,Apte:2022xtu,Kaidi:2023maf,Zhang:2023wlu,Choi:2023xjw,Sun:2023xxv}. 
In the cases of gauge theory, it provides an analytic obstruction to a trivially confining phase \cite{Choi:2021kmx,Choi:2022zal,Apte:2022xtu}.
\item The higher symmetry structure  of non-invertible symmetries (which is reminiscent of the higher group) results in lower bounds on the axion string tension and the monopole mass  in axion physics \cite{Choi:2022fgx}. 
\item They  inspire new phenomenological models in particle physics, such as in the context of neutrino physics  \cite{Cordova:2022fhg}.
\item   Non-invertible global symmetries consolidate conjectures in quantum gravity  \cite{Rudelius:2020orz,Heidenreich:2021xpr,Arias-Tamargo:2022nlf,Choi:2022fgx}.  
 \item In monopole scattering, there was a longstanding puzzle that certain outgoing states have fractional global symmetry quantum numbers \cite{Callan:1982ah,Rubakov:1982fp}. 
In  \cite{vanBeest:2023dbu}, the authors resolve
 the  puzzles by arguing that the outgoing states are attached to an invertible or non-invertible topological defect, which ends on the monopole. Since the outgoing states are in the twisted Hilbert space of a topological defect, they generally have fractional quantum numbers.
 \item Applications to the Weak Gravity Conjecture \cite{Cordova:2022rer,Choi:2022fgx}.
 \item Categorical generalizations of the Monster Moonshine \cite{Lin:2019hks}.
 \end{itemize}
Above is only a partial list of new results from  non-invertible symmetries, and is by no means comprehensive. 
  Below we elaborate on  a couple of these novel applications.

\subsection{Anomalies and constraints on renormalization group flows}\label{sec:anomaly}
  
The 't Hooft anomaly of an ordinary global symmetry is defined as the obstruction to gauging the global symmetry. 
It doesn't signal any pathology of the global symmetry or the quantum system; you simply cannot gauge it. 
 One important consequence of the 't Hooft anomaly of an ordinary global symmetry is that the low energy phase cannot be a trivially gapped phase, which is described by a trivial TQFT   with a unique ground state on any spatial manifold.  
 In 1+1d, it leaves one with the following two options: it is either a gapless phase described by a CFT, or a gapped phase with degenerate vacua described by a nontrivial TQFT. 
Either way,  there must be nontrivial degrees of freedom in the IR to match the 't Hooft anomalies of the UV. 
 This is the celebrated 't Hooft anomaly matching argument \cite{tHooft:1979rat}.

Sometimes a non-invertible global symmetry can also be gauged. 
The gauging is defined by inserting a mesh of topological defects on the spacetime manifold. 
In 1+1d, the gauging of a fusion category  has been systematically developed in \cite{Frohlich:2009gb,Carqueville:2012dk,Brunner:2013ota,Brunner:2013xna,Brunner:2014lua}, which are discussed more recently in \cite{Bhardwaj:2017xup,Thorngren:2019iar,Komargodski:2020mxz,Zhang:2023wlu,Choi:2023xjw}. 
We will not discuss the detail of this gauging procedure, but refer the readers to the references above.

Interestingly, some of the non-invertible symmetries  are intrinsically incompatible with a trivially gapped phase.  
This can be viewed as a generalized anomaly of the non-invertible symmetry.\footnote{For an ordinary invertible global symmetry, an 't Hooft anomaly can be equivalently defined as either the obstruction to gauging, or as the incompatibility with a trivially gapped phase. These two notions however bifurcate for non-invertible symmetries  \cite{Choi:2023xjw}. Here we adopt the definition that  a non-invertible symmetry is  anomalous if it is incompatible with a trivially gapped phase. Mathematically, an anomalous fusion category  does not admit a fiber functor.  When the fusion category is non-anomalous, there can be a symmetric gapped phase with a unique ground state, i.e., a non-invertible SPT phase \cite{Inamura:2021szw}.}  
The presence of such non-invertible symmetries in the UV can be used for a  generalized 't Hooft anomaly matching argument to rule out a trivially gapped phase in the IR. 

Below we discuss one particular application of this generalized anomaly for non-invertible symmetries in 1+1d CFT \cite{Chang:2018iay}. 
We will prove the following statement:   
\textit{Consider a CFT deformed by a relevant operator that preserves  a topological line $\CL$. If   $\langle\CL\rangle\notin\bZ_{\ge0}$, then the low energy phase cannot be gapped with a unique ground state.} 
In other words, a non-integral quantum dimension $\langle \CL\rangle$ (defined in \eqref{qdim}) is a sufficient condition for an 't Hooft anomaly of the non-invertible symmetry. 
However, it is not necessary. For instance, an anomalous, invertible global symmetry has $\langle\CL\rangle=1$ but is also incompatible with a trivially gapped phase. 
Nonetheless, since the quantum dimension of a line is a quantity that can be easily computed from the fusion algebra, this statement provides a quick diagnostic of the generalized anomaly.

We prove this statement by contradiction.   
Suppose $\langle \CL\rangle\notin\bZ_{\ge0}$ but the low energy phase is a trivially gapped phase.  
By definition, a trivially gapped phase is described by a trivial TQFT (with $c=0$) and a 1-dimensional Hilbert space, dim$\CH=1$. 
Let us compute the torus partition function in this trivial TQFT with the topological line $\CL$ inserted as an operator at a fixed time:
\ie
Z^\CL = \text{Tr}_\CH[ \, \CL \,  q^{L_0 -{c\over 24} } \bar q^{\bar L_0 -{c\over24}} \, ] = \langle \CL\rangle\,.
\fe
Since the low energy phase is by assumption a TQFT, we have $L_0=\bar L_0=0$ and $c=0$. 
The partition function is then just the eigenvalue of $\CL$ on the only state in $\CH$, which is nothing but the quantum dimension $\langle \CL\rangle$. 
In particular,  the partition function  is independent of the complex structure moduli $\tau$ of the spacetime torus.

In contrast, the torus partition function with $\CL$ inserted as a defect extended in the time direction computes the dimension of the $\CL$-twisted Hilbert space:
\ie
Z_\CL= \text{Tr}_{\CH_\CL} [q^{L_0 -{c\over 24} } \bar q^{\bar L_0 -{c\over24}}]  = \dim  \CH_\CL\,.
\fe

By  a modular S transformation, these two partition functions are equal $Z^\CL= Z_\CL$, which gives
\ie
\dim \CH_\CL = \langle\CL\rangle\,.
\fe
This is a contradiction if $\langle \CL\rangle$ is not a non-negative integer.  This completes the proof.

This statement leads to several new constraints on RG flows. 
For instance, in the tricritical Ising CFT of $c=7/10$, there is a Fibonacci fusion category satisfying the fusion rule $W^2 = I+W$. 
Its quantum dimension is $\langle W\rangle = {1+\sqrt{5}\over2}$.  
(This non-invertible symmetry can be realized microscopically by the golden chain \cite{Feiguin:2006ydp}.)
There is a relevant deformation $\sigma'$ of $(h,\bar h)=({7\over 16}, {7\over 16})$ in the tricritical Ising CFT that only preserves the $W$ line.  
In particular, the $\bZ_2$ symmetry is explicitly broken by $\sigma'$. 

What is the low energy phase of the tricritical Ising deformed by $\sigma'$?  
The statement proven above indicates that it cannot be a trivially gapped phase with a unique ground state. 
Indeed, it turns out that there are two degenerate ground states \cite{Zamolodchikov:1990xc,Ellem:1997vz,Fateev:1997yg}. 
Without knowing the non-invertible symmetry $W$, this is somewhat puzzling at first sight since the only ordinary global symmetry $\bZ_2$ is explicitly broken, so the two states cannot be interpreted as a spontaneously broken phase an ordinary global symmetry.  
It turns out that a unique gapped ground state is incompatible with $W$, and the two states should be viewed as a spontaneously broken phase for the non-invertible symmetry  \cite{Chang:2018iay}. 
See \cite{Huang:2021zvu} for a systematic construction of 1+1d TQFT describing the spontaneously broken phases of  fusion category symmetry.

There are further constraints one can derive from non-invertible symmetries. 
For example, the non-invertible line $\CD$ of the TY$_+$ fusion category has quantum dimension $\sqrt{2}$, and is thus anomalous. 
It was further shown in \cite{Chang:2018iay} that in a symmetric gapped phase with a TY$_+$ fusion category symmetry, the number of ground states has to be a multiple of 3. 
Similar constraints are also found on the lattice \cite{Aasen:2020jwb}.

Generalized anomalies of non-invertible symmetries that signal the incompatibility with a trivially gapped phase have also been studied in higher dimensions.  
See \cite{Choi:2021kmx,Choi:2022zal,Choi:2022rfe,Apte:2022xtu,Kaidi:2023maf} for  a partial list of references.  
These anomalies give an analytic obstruction to a trivially confining phase in gauge theory. 
For example, the non-invertible global symmetry in the 3+1d Maxwell theory (when viewed as a bosonic QFT) at $\tau=iN$ is compatible with a trivially gapped phase   only if each prime factor of $N$
is one modulo four \cite{Choi:2021kmx}.

\subsection{Axions}\label{sec:axion}

\subsubsection{Emergent symmetries}

Not all symmetries are on the same footing. 
Some non-invertible symmetries are subordinate to an underlying invertible one. 
 For instance, the fusion rule $\CD^2 =I+\eta$ \eqref{Isingcat} in the Ising CFT implies that the non-invertible symmetry $\CD$ cannot exist without the invertible $\bZ_2$ symmetry $\eta$. 
 As another example, the non-invertible symmetry $\CD_{p\over N}$ in QED is constructed from half gauging the invertible $\bZ_N^{(1)}$ magnetic 1-form symmetry, so it cannot exist independently without the latter.\footnote{In 1+1d, there are non-invertible symmetries that are unrelated to any invertible ones, such as the Fibonacci line $W$ with the fusion rule $W^2=I+W$.} 
 
A simple toy example of such a phenomenon is the following. 
Consider a $\bZ_4$ group generated by $g$, with $g^4=1$. 
The $\bZ_2$ \textit{quotient} (which is generated by $g$) cannot exist without the $\bZ_2$ \textit{subgroup} (which is generated by $g^2$).
A more advanced analogy is with the higher group symmetry, where a  global symmetry of a lower form degree cannot exist independently of another one with a higher form degree \cite{Kapustin:2013uxa,Tachikawa:2017gyf,Cordova:2018cvg,Benini:2018reh}.

This hierarchy between global symmetries, which we refer to as the \textit{higher symmetry structure},  has far-reaching physical implications in realistic quantum systems.  
Low energy EFTs typically have emergent global symmetries that are not present in the UV.  
These symmetries are approximate, and 
as we go up in energy, they are broken either by higher dimensional operators in the Lagrangian or by massive particles or strings. 
Suppose the emergent global symmetry of  an IR EFT has such a higher symmetry structure where one symmetry $G_1$ cannot exist without the other $G_2$.  
Let $E_1$ and $E_2$ be the energy scales above which $G_1$ and $G_2$ are respectively broken. 
The higher symmetry structure implies that we must have
\ie\label{inequality}
E_2 \gtrsim E_1\,.
\fe
Since the scales $E_i$ are only defined approximately, \eqref{inequality} is not a sharp inequality. It means that there cannot be a parametrically large range of energy scales where $G_2$ is broken  while $G_1$ is approximately preserved. 
Constraints of the form \eqref{inequality} from  higher groups have found applications in \cite{Cordova:2020tij,Brennan:2020ehu,Cordova:2022qtz}.

\subsubsection{Non-invertible 1-form symmetries}

Consider the 3+1d axion-Maxwell Lagrangian
\ie\label{axionmaxwell}
{\cal L} = {f^2\over 2} d\theta\wedge \star d\theta  +{1\over 2e^2} F\wedge \star F - {i  K \over 8\pi^2}\theta F\wedge F\,,
\fe
where $A$ is the dynamical $U(1)$ gauge field and $F=dA$. 
Here $\theta(x)\sim \theta(x)+2\pi$ is the  axion field, which is  a  dynamical compact scalar field. 
The axion-photon coupling constant $K$ is quantized to be an integer as required by the compactness of the axion field $\theta(x)$ and the gauge symmetry of $A$. 
See \cite{Reece:2023czb} for a recent review of axion physics and in particular for the quantization of the level $K$.

The generalized global symmetry of this axion model is extremely rich. 
When $|K|>1$, there is an invertible  higher group symmetry \cite{Hidaka:2020iaz,Hidaka:2020izy,Brennan:2020ehu,Brennan:2023kpw}. 
(The higher group symmetry in the axion-Yang-Mills theory was studied in \cite{Seiberg:2018ntt,Cordova:2019uob,Brennan:2020ehu}.)
It was later found in  \cite{Choi:2022fgx}  that the axion-Maxwell theory has a non-invertible 1-form global symmetry even for $|K|=1$, in addition to the non-invertible shift symmetry $\CD_{p\over N}$ similar to the one in Section \ref{sec:pion}. 
Below we discuss this non-invertible 1-form symmetry in \cite{Choi:2022fgx} and its consequences in the $K=1$ case.

What is the Gauss law in axion physics? 
The equation of motion of $A$ reads
\ie\label{anomgauss}
- {i\over e^2} d\star F = {1\over 4\pi^2} d\theta \wedge F\,.
\fe
The non-vanishing of the righthand side means that the Gauss law in the axion-Maxwell theory is anomalous \cite{Fischler:1983sc}.  
The naive Gauss law surface operator
\ie
Q =  - {i\over e^2}\oint_\Sigma \star F
\fe
is not conserved or topological because $\star F$ is not a closed 2-form. 
Here $\Sigma$ is a closed 2-cycle in spacetime.  
Instead, one can consider the following operator
\ie
``\, Q_\text{Page} =  \oint_\Sigma \left( -{i\over e^2} \star F- {1\over 4\pi^2 }\theta dA\right) \,."
\fe
However, it is not invariant under $\theta\sim \theta+2\pi$, which should be viewed as a gauge symmetry.  
(Following the convention in  Section \ref{sec:FQH},   we put this gauge non-invariant operator $Q_\text{Page}$ in quotation marks.)   
This gauge non-invariant operator is known as the ``Page charge" in the supergravity literature \cite{Page:1983mke,Marolf:2000cb}.  
To conclude, there is no gauge-invariant, quantized, and conserved electric charge in axion physics \cite{Kogan:1993yw,Fukuda:2020imw}. 
In the language of higher-form symmetry, it means that there is no invertible electric 1-form  global symmetry in the axion-Maxwell theory.

We  can understand the absence of a conventional Gauss law through the following process. 
We start with an axion string defect, which is a 2-dimensional defect in spacetime, around which the axion field winds once $\theta(x)\to \theta(x)+2\pi$.  
It is charged under a $U(1)^{(2)}$ winding 2-form global symmetry. 
 The axion string defect can be  viewed as the worldsheet of an infinitely heavy axion string, or a limit of the axion domain wall.  
Now we bring a monopole of electromagnetic charge $(q,m) = (0,1)$ around this axion string defect.  
As the monopole goes around the axion string, it experiences the Witten effect $\theta\to \theta+2\pi$, and becomes a dyon of charge $(q,m) = (1,1)$.   
The net effect is that the monopole gains one unit of the electric charge.

One can take the perspective that there is simply no notion of conserved electric charge whatsoever in axion physics. 
This is however too pessimistic. 
We know that the electric charge is only violated in a very specific way when there are axion strings and monopoles. 
Is there  a modified conservation law for axions?

As the readers might have already anticipated, we can ``cure" the anomaly \eqref{anomgauss} by a TQFT following the strategy in Section \ref{sec:FQH}. 
More precisely, for every coprime integers $p,N$, we define a new operator \cite{Choi:2022fgx,Yokokura:2022alv}:\footnote{The superscript $(1)$ is to remind us that this is a (non-invertible) 1-form symmetry operator supported on a 2-surface in the 4-dimensional spacetime.}
\ie
\CD_{p\over N}^{(1) } (\Sigma)  
=  \int [D\phi Dc]_\Sigma
\exp\left[
\oint_\Sigma \left(
{2\pi  p\over Ne^2 }\star F + { iN\over 2\pi }\phi dc +{ip \over 2\pi }\theta dc+{i\over 2\pi }\phi dA
\right)
\right]\,,
\fe
where $\phi\sim \phi+2\pi$ is a compact scalar field and $c$ is a dynamical 1-form  gauge field, both of them living only on the 2-surface $\Sigma$. 
The $ { iN\over 2\pi }\phi dc$ stands for a 1+1d $\bZ_N$ gauge theory that we introduce to cure the anomalous Gauss law. 
This gauge-invariant operator $\CD_{p\over N}^{(1)}$ is intuitively  
 related to the gauge non-invariant one $\exp (2\pi i p Q_\text{Page}/N)$ by integrating out $\phi$ to obtain $c = -A/N$. However, this manipulation is not correct globally, and $\CD^{(1)}_{p\over N}$ is the precise, gauge-invariant expression. 
Furthermore, this operator is topological by a generalization of the half gauging construction, the half higher gauging.  
It involves gauging the invertible 1-form magnetic and 2-form winding global symmetries along half of a higher codimensional manifold  \cite{Choi:2022fgx} . 
 At the end of the day, we still do not have a conserved charge, but we do have a conserved and gauge-invariant symmetry operator $\CD_{p\over N}^{(1)}$ labeled by the rational numbers $p/N\in \mathbb{Q}/\bZ$.
 
 The only price we pay is that $\CD^{(1)}_{p\over N}$ is not invertible because of the 1+1d TQFT living on it. 
The non-invertible nature can be seen by its action on a dyonic line $H_{q,m}$ with charge $(q,m)$ as
 \ie
 \CD_{p\over N}^{(1)} \cdot H_{q,m} = \begin{cases}
 0 \,,~~~~~~~~~~~~~~~~~~~~~\text{if}~m\neq 0\text{~mod}~N\,,\\
  e^{2\pi i p/N} H_{q,m} \,,~~~~\text{if}~m= 0\text{~mod}~N\,.
 \end{cases}
 \fe
 where $\cdot$ stands for the action of encircling the surface operator $\CD^{(1)}_{p\over N}$ around the dyon in space. 
The selection rule from $\CD^{(1)}_{p/N}$ states that the electric charge is only conserved modulo $m$ in the presence of a charge $m$ monopole. This is precisely the modified Gauss law  we were looking for. 

We conclude that a dynamical axion turns the conventional electric $U(1)^{(1)}$ 1-form global symmetry  of the free Maxwell theory into a non-invertible electric 1-form global symmetry $\CD^{(1)}_{p\over N}$ labeled by the rational numbers.

\subsubsection{Bounds on axion string tension and monopole mass}

Finally, we discuss the universal inequality from the non-invertible 1-form global symmetry $\CD^{(1)}_{p\over N}$.  
As mentioned above, this non-invertible symmetry is constructed by gauging the winding 2-form symmetry and the magnetic 1-form symmetry.

As we try to UV complete the axion-Maxwell theory, we bring in the dynamical electric and magnetic particles, as well as the dynamical axion string. 
They break the non-invertible 1-form symmetry, the invertible magnetic  1-form symmetry, and the winding 2-form symmetry, respectively. 
However, since the non-invertible symmetry cannot exist without its invertible parents, these symmetries have to be broken in the right order. 
We therefore find the following universal inequalities \cite{Choi:2022fgx,Choi:2023pdp}:
\ie
~&m_\text{electric} \lesssim m_\text{magnetic} \,,\\
&m_\text{electric} \lesssim \sqrt{T} \,.
\fe
where $m_\text{electric}, m_\text{magnetic}$ are the masses of the lightest dynamical electrically and magnetically charged particles, respectively, and $T$ is the tension of the dynamical axion string.  
Interestingly, we find that the non-invertible symmetries give rise to lower bounds on the string tension and the monopole mass. 
These generalize the  inequalities from the invertible higher groups in \cite{Brennan:2020ehu}.

In particular, this inequality implies that the lightest electrically charged particle has to be lighter than the lightest magnetic one. 
 At the level of the slogan, one can say that the electron has to be lighter than the monopole when there is an axion.\footnote{Note that here we assume the axion couples to the gauge field as in \eqref{axionmaxwell}, which breaks the electromagnetic duality of the free Maxwell theory. Hence the electron and the monopole are on different footings.}

\subsection{Conjectures in quantum gravity}

There are two famous conjectures in quantum gravity: the completeness hypothesis and the no  global symmetry conjecture. 
The former states that any gauge theory coupled to quantum gravity must have objects charged under every finite dimensional irreducible representation of the gauge group \cite{Polchinski:2003bq},  and the latter states that there cannot be exact global symmetries in quantum gravity \cite{Misner:1957mt,Banks:2010zn}.   
However, in the context of QFT, the two conjectures are not equivalent. 
For instance, non-abelian finite group gauge theory with certain matter spectrum satisfies one but not the other \cite{Harlow:2018tng}.

The inclusion of non-invertible global symmetries consolidate these two conjectures. 
It is  argued in  \cite{Rudelius:2020orz}  that there cannot be any invertible or non-invertible global symmetries in quantum gravity. 
In fact, since we are instructed to sum over the topologies of spacetime in quantum gravity, it is not even clear how one would even define a generalized global symmetry in terms of a topological operator/defect supported on some fixed nontrivial cycles. 
By extending the no global symmetry conjecture to the absence of  invertible and non-invertible global symmetries in quantum gravity, the two conjectures become equivalent under mild assumptions  \cite{Rudelius:2020orz,Heidenreich:2021xpr}.

There is another application to quantum gravity from  the non-invertible 1-form symmetry of the axions  in Section \ref{sec:axion}. 
In \cite{Heidenreich:2020tzg}, a puzzle was raised when we embed the axion-Maxwell theory \eqref{axionmaxwell} into a full-fledged quantum gravity.  
In all UV complete theories with a $U(1)$ gauge theory coupled to an axion, there are always other electrically charged particles. 
Who ordered these particles? 
Naively, following the no global symmetry conjecture, we need these electrically charged particles to break the  electric $U(1)^{(1)}$ 1-form global symmetry. 
However, the axion already does that for us. 
The electrically charged particles seem to have lost their purposes in life. 

This is when the non-invertible 1-form global symmetry comes to rescue \cite{Choi:2022fgx}.  
As discussed in Section \ref{sec:axion}, the axion does not completely break the electric $U(1)^{(1)}$ 1-form global symmetry. Rather, the rational part of $U(1)^{(1)}$ is resurrected as a non-invertible 1-form symmetry. 
Assuming the no generalized global symmetry conjecture, this non-invertible 1-form global symmetry has to be  broken eventually, which is precisely what the electric particles do. 
This gives another  argument why the no global symmetry conjecture should be extended to the absence of invertible and non-invertible global symmetries.

\section{Conclusions}\label{sec:conclusion}

We have come a long way from the non-invertible symmetry in the Ising model, to that in QED.  
These new non-invertible symmetries explain and unify different physical phenomena. 
For instance, despite  the apparent differences between the Ising model and QED, the two systems actually have similar  symmetry structures (see Table \ref{table:IsingQED}). 
More generally, these new symmetries resolve longstanding puzzles, have new implications for strongly-coupled quantum systems, lead to new notions of naturalness, and even inspire new physical models.

 \begin{table}[h!]
\begin{align*}
\left.\begin{array}{|c|c|c|}\hline
  & \text{1+1d Ising lattice model} & \text{3+1d QED} \\\hline \text{non-invertible} & ~~\text{Kramers-Wannier duality sym.} &~~ \text{non-invertible chiral sym.}\\
\text{symmetry}&\mD=e^{- {2\pi i N\over 8}} \times&
\CD_{1\over N} = \int Da  \times\\
& \prod_{j=1}^{N-1} \left(
{1+iX_j \over \sqrt{2}} {1+iZ_jZ_{j+1}\over \sqrt{2}} \right)
{1+iX_N\over \sqrt{2}} {1+\eta\over2} & 
\exp\left[
 \oint \left(
{2\pi i\over N} \star j^\text{A} +  {iN\over 4\pi}ada + {i\over 2\pi }adA
\right)
\right] \\
\hline \text{invertible} & \text{0-form sym.} & \text{1-form magnetic  sym.} \\
\text{symmetry}&\bZ_2^{(0)}:~\eta =\prod_j X_j&\bZ_N^{(1)}:~ \eta = \exp({i\over N} \oint F)\\
\hline  \text{action on}& \text{order}~\sigma & \text{'t Hooft line} \\
\text{observables}&\updownarrow& \updownarrow\\
 & \text{disorder}~\mu & \text{dyon} \\\hline \end{array}\right.
\end{align*}
\caption{Comparison between the non-invertible symmetries in the 1+1d Ising lattice model (Section \ref{sec:TFIM}) and 3+1d QED (Section \ref{sec:FQH}).  (Note that the $N$ for the Ising model is the number of lattice sites, while the $N$ for QED is related to the chiral rotation angle $\alpha=2\pi/N$.) }\label{table:IsingQED}
\end{table}

There are several exciting future directions for the applications of non-invertible symmetries.  
\begin{itemize}
\item In higher than 1+1d, most of the non-invertible (0-form) symmetries are constructed from gauging, including the ordinary gauging of a non-anomalous subgroup, higher gauging, half gauging, and other generalizations. Are there non-invertible symmetries that have nothing to do with gauging? In 1+1d, the Fibonacci fusion category is one such example.
\item The general mathematical language for these generalized global symmetries are encoded in the formalism of higher fusion category, which is a subject under development by the mathematicians.  It has become increasingly clear that while group theory is the language for symmetry in quantum mechanics, category theory is the appropriate language  in QFT.
\item How do we gauge non-invertible symmetries in general dimensions? What are the most general anomalies? Is there a mathematical classification? How do we systematically use them to constrain the phase diagram of strongly-coupled quantum systems? In particular, non-invertible symmetries have been used to prove (de)confinement in 1+1d \cite{Komargodski:2020mxz}. Can we learn anything about confinement in higher dimensions? See \cite{Nguyen:2021yld,Choi:2021kmx,Choi:2022zal,Apte:2022xtu,Sun:2023xxv} for discussions along this line. 

\item  While there is a large class of lattice models realizing general non-invertible symmetries in low dimensions, the constructions are rather involved. What kinds of non-invertible symmetries exist in the conventional lattice models, such as  those with a tensor product Hilbert space?  What's the relation between the symmetries on the lattice and in the continuum? Can we generalize the Lieb-Schultz-Mattis-type theorems in \cite{Seiberg:2024gek} to other non-invertible symmetries?
 \item New symmetries lead to new notions of naturalness. Can they be used to address various hierarchy problems in particle physics?  
\item What more can we learn about monopoles and axions from non-invertible symmetries? 
\item Do they lead to new selection rules in scattering amplitudes in addition to the helicity conservation?
\item Non-invertible global symmetries of the string worldsheet CFT implies non-invertible gauge symmetries in spacetime.  It would be interesting to explore the consequences of these novel gauge symmetries from the string theory point of view.
\item Can we derive other universal constraints from the breaking of the non-invertible symmetries in quantum gravity? 
\end{itemize}
These open problems span over topics in different disciplines, including formal quantum field theory, particle phenomenology, condensed matter physics, quantum information theory, quantum gravity, string theory, and mathematics. 
 They  require new ideas and techniques that can bridge the gaps between different disciplines. This is what makes this subject so exciting and rewarding.  I am hopeful that this field will keep advancing, with even more surprises and discoveries in the near future.

\section*{Acknowledgements}

I would like to  thank    Ken Intriligator, Ibou Bah, and the   local organizers, Oliver DeWolfe, Ethan Neil, Tom DeGrand  for organizing and running the TASI 2023 summer school ``Aspects of Symmetry" together.  
I also thank the TASI students for the stimulating questions and discussions. 
I am grateful to Yichul Choi and Yunqin Zheng for comments on a draft.

The material presented here is based on what I've learned over the years from working with many collaborators on non-invertible symmetries.  
I would especially like to thank Chi-Ming Chang, Yichul Choi, Clay Cordova, Po-Shen Hsin, Ho Tat Lam, Ying-Hsuan Lin,   Sahand Seifnashri,  Nati Seiberg, Yifan Wang, and Xi Yin for the fruitful and close collaborations on this subject. 
I am also grateful to Lakshya Bhardwaj, Meng Cheng, 
Michele Del Zotto, Dan Freed, Daniel Harlow, Wenjie Ji,  Theo Johnson-Freyd, Justin Kaidi,  Zohar Komargodski, John McGreevy, Greg Moore, Kantaro Ohmori,  Brandon Rayhaun,  Matt Reece, Kostis Roumpedakis, Tom Rudelius, Yaman Sanghavi, Yuji Tachikawa,  Ryan Thorngren, Xiao-Gang Wen, and Yunqin Zheng   for many stimulating and insightful discussions throughout the years.

This work  was supported in part by NSF grant PHY-2210182. 
I thank Harvard University for its hospitality during the course of this work.

\bibliographystyle{JHEP}
\bibliography{ref}

\end{document}

%% file: symbound.tikzstyles
\tikzstyle{redrectangle}=[fill=red, draw=none, shape=rectangle]

\tikzstyle{dashedline}=[dashed, -, thick]
\tikzstyle{blueline}=[-, draw=blue, thick]
\tikzstyle{arrowline}=[<-, thick]
\tikzstyle{normalline}=[-, thick]
\tikzstyle{fillline}=[-, thick, fill=red, fill opacity=.5]
\tikzstyle{bluefillline}=[-, fill opacity=.5, fill=blue, thick]
\tikzstyle{purplefillline}=[-, fill opacity=.5, fill=purple, thick]